\documentclass[a4paper]{article}
\pdfoutput=1
\PassOptionsToPackage{
backend=biber,
language=auto,
style=numeric-comp,
sorting=none,
maxbibnames=3,
natbib=true
}{biblatex}
\usepackage{biblatex}

\usepackage{silence}
\usepackage{jcappub}
\usepackage{float}

\usepackage[T1]{fontenc}

\usepackage{dcolumn}
    \newcolumntype{d}[1]{D{.}{\cdot}{#1}}
    \newcolumntype{.}{D{.}{.}{-2}}
    \newcolumntype{,}{D{,}{,}{-3}}

\usepackage{booktabs}

\usepackage{soul}
\usepackage{hhline}
\usepackage{arydshln}
\usepackage{physics}

\usepackage{booktabs}
\usepackage[stable]{footmisc}
\usepackage[table,dvipsnames]{xcolor}
\usepackage{rotating}

\usepackage{lipsum}

\def\setsymbol#1#2{\expandafter\def\csname #1\endcsname{#2}}
\def\getsymbol#1{\csname #1\endcsname}

\newbox\tablebox    \newdimen\tablewidth
\def\leaderfil{\leaders\hbox to 5pt{\hss.\hss}\hfil}
\def\endPlancktable{\tablewidth=\columnwidth 
    $$\hss\copy\tablebox\hss$$
    \vskip-\lastskip\vskip -2pt}

\def\tablenote#1 #2\par{\begingroup \parindent=0.8em
    \abovedisplayshortskip=0pt\belowdisplayshortskip=0pt
    \noindent
    $$\hss\vbox{\hsize\tablewidth \hangindent=\parindent \hangafter=1 \noindent
    \hbox to \parindent{$^#1$\hss}\strut#2\strut\par}\hss$$
    \endgroup}
\def\doubleline{\vskip 3pt\hrule \vskip 1.5pt \hrule \vskip 5pt}

\def\NHUNIT{\ifmmode {\rm \,cm^{-2}} \else $\rm \,cm^{-2}$ \fi} 

\def\muKcmb{\ifmmode \,\mu$K$_{\rm CMB}$\else \,$\mu$K$_{\rm CMB}$\fi}

\newcommand{\sign}{\text{sign}}

\newcommand{\As}{A_{\rm s}}

\newcommand{\ns}{n_{\rm s}}

\newcommand{\hnot}{H_{\rm 0}}

\newcommand{\OmegaM}{\ifmmode\Omega_{\rm M}\else $\Omega_{\rm M}$\fi}

\providecommand{\Planck}{\textit{Planck}}

\providecommand{\text}[1]{\rm{#1}}

\newcommand{\Mpc}{\text{Mpc}}

\renewcommand{\d}{\text{d}}

\newcommand{\Hunit}{\,\text{km}\,\text{s}^{-1}\,\Mpc^{-1}}

\providecommand{\bl}{$B_\ell$}
\providecommand{\Npp}{$N_{pp^\prime}^{\rm cov}$}
\providecommand{\regn}{$\sigma_{\rm reg}$}
\providecommand{\apo}{FWHM${}_{\rm apo}$}
\providecommand{\deltal}{$\Delta\ell$}
\providecommand{\lswitch}{$\ell_{\rm switch}$}
\providecommand{\lmaxabs}{$\ell_{\rm max}^{\rm abs}$}
\providecommand{\Nll}{$N_{\ell\ell^\prime}^{\rm cov}$}
\providecommand{\Nval}{$N_{\rm val}$}

\providecommand{\Omb}{\Omega_{\mathrm{b}}h^2}
\providecommand{\Omc}{\Omega_{\mathrm{c}}h^2}

\newcommand{\begm}{\begin{pmatrix}}
\newcommand{\enm}{\end{pmatrix}}

\newcommand{\threejz}[3]{{\begm #1 & #2 & #3 \\ 0 & 0 & 0 \enm}}

\providecommand{\Tr}{\text{Tr}}

\def\pmb#1{\setbox0=\hbox{#1}%
    \kern-.025em\copy0\kern-\wd0
    \kern.05em\copy0\kern-\wd0
    \kern-.025em\raise.0433em\box0}

\def\p2Y{\;_2Y}
\def\m2Y{\;_{-2}Y}
\def\beglet{
  \addtocounter{equation}{1}%
  \setcounter{parentequation}{\value{equation}}%
  \setcounter{equation}{0}%
  \def\theequation{\arabic{parentequation}\alph{equation}}%
  \ignorespaces
}
\def\endlet{
  \setcounter{equation}{\value{parentequation}}%
  \def\theequation{\arabic{equation}}%
}
\providecommand{\beglet}{\begin{subequations}}
\providecommand{\endlet}{\end{subequations}}

\newcommand{\mksym}[1]{\ifmmode {\rm #1}\else #1\fi}

\providecommand{\text}[1]{\rm{#1}}

\renewcommand{\d}{\text{d}}

\providecommand{\Omb}{\Omega_{\mathrm{b}}}
\providecommand{\Omc}{\Omega_{\mathrm{c}}}

\newcommand\ba{\begin{eqnarray}}
\newcommand\ea{\end{eqnarray}}
\newcommand\bea{\begin{eqnarray}}
\newcommand\eea{\end{eqnarray}}

\newcommand\be{\begin{equation}}
\newcommand\ee{\end{equation}}

\newcommand{\Nside}{\ensuremath{N_{\mathrm{side}}}}

\providecommand{\Tr}{\text{Tr}}

\appto{\bibsetup}{\sloppy} 

\title{\boldmath Validation of harmonic-space likelihood methods for \textit{LiteBIRD} $B$-mode polarisation analysis}

\author[1,2]{G.\,Galloni,}
\author[1,2]{P.\,Campeti,}
\author[3]{C.\,Gimeno-Amo,}
\author[4,5]{A.\,Carones,}
\author[6,7]{A.\,Rizzieri,}
\author[8,9]{Y.\,Chinone,}
\author[4,5,10]{N.\,Krachmalnicoff,}
\author[3]{E.\,Martínez-González,}
\author[1,2,11]{L.\,Pagano,}
\author[12]{D.\,Adak,}
% ...
\affiliation[1]{Dipartimento di Fisica e Scienze della Terra, Università di Ferrara, Via Saragat 1, 44122 Ferrara, Italy}
\affiliation[2]{INFN Sezione di Ferrara, Via Saragat 1, 44122 Ferrara, Italy}
\affiliation[3]{Instituto de Fisica de Cantabria (IFCA, CSIC-UC), Avenida los Castros SN, 39005, Santander, Spain}
\affiliation[4]{International School for Advanced Studies (SISSA), Via Bonomea 265, 34136, Trieste, Italy}
\affiliation[5]{INFN Sezione di Trieste, via Valerio 2, 34127 Trieste, Italy}
\affiliation[6]{Department of Physics, University of Oxford, Denys Wilkinson Building, Keble Road, Oxford OX1 3RH, UK}
\affiliation[7]{Université Paris Cité, CNRS, Astroparticule et Cosmologie, F-75013 Paris, France}
\affiliation[8]{International Center for Quantum-field Measurement Systems for Studies of the Universe and Particles (QUP), High Energy Accelerator Research Organization (KEK), Tsukuba, Ibaraki 305-0801, Japan}
\affiliation[9]{Kavli Institute for the Physics and Mathematics of the Universe (Kavli IPMU, WPI), UTIAS, The University of Tokyo, Kashiwa, Chiba 277-8583, Japan}
\affiliation[10]{IFPU, Via Beirut, 2, 34151 Grignano, Trieste, Italy}
\affiliation[11]{Université Paris-Saclay, CNRS, Institut d’Astrophysique Spatiale, 91405, Orsay, France}
\affiliation[12]{Instituto de Astrofísica de Canarias, E-38200 La Laguna, Tenerife, Canary Islands, Spain}
% ...

\emailAdd{giacomo.galloni@unife.it}

\abstract{The \textit{LiteBIRD} satellite mission aims to measure the tensor-to-scalar ratio $r$ with unprecedented sensitivity through observations of the cosmic microwave background (CMB) $B$-mode polarisation. Robust constraints on $r$ require accurate likelihood modelling across a wide range of angular scales. In this work, we validate harmonic-space likelihood methods against pixel-based benchmarks using 20\,000 end-to-end simulations that include CMB, instrumental noise, and Galactic foreground residuals processed through a realistic component-separation pipeline. We compare a debiased pixel-based likelihood, which serves as our benchmark on large scales ($\ell\leq32$), with an offset Hamimeche \& Lewis (oHL) likelihood applicable across the full multipole range. We systematically explore the space of analysis settings, identifying the number of simulations used for covariance estimation and the choice of beam window function as the most critical factors. Over the full multipole range considered ($\ell\leq128$), where the benchmark combines the pixel-based likelihood at $\ell\leq32$ with a harmonic one at smaller scales, the 68\% upper bound on $r$ from the oHL likelihood is $19$--$29\%$ larger than the benchmark one, depending on the beam window function. This demonstrates its viability as a computationally efficient baseline for \textit{LiteBIRD} data analysis and cosmological inference from the polarisation $B$-mode field.}

\begin{document}
\ActivateWarningFilters[warn]

\maketitle

\flushbottom

\section{Introduction}\label{sec:introduction}

The detection of primordial gravitational waves generated during cosmic inflation remains one of the most compelling targets in observational cosmology. These tensor perturbations would leave a distinctive imprint on the polarisation of the cosmic microwave background (CMB), producing a curl-like pattern known as $B$-mode polarisation on large angular scales \cite{guzzettiGravitationalWavesInflation2016, seljakLineofSightIntegrationApproach1996, seljakSignatureGravityWaves1997, huWanderingBackgroundCMB1995}. The amplitude of this signal is parametrised by the tensor-to-scalar ratio $r$, which directly probes the energy scale of inflation. Current upper bounds place $r < 0.032$ at 95\% confidence level when assuming slow-roll single field inflation \cite{tristram2022ImprovedLimitsTensortoscalarratio} and $r < 0.028$ when marginalising over a non slow-roll spectral tilt\footnote{Note that the results of ref.~\cite{tristram2022ImprovedLimitsTensortoscalarratio} are obtained at the pivot scale for tensor perturbations of $0.05 \Mpc^{-1}$, while those from refs.~\cite{galloniUpdatedConstraintsAmplitude2022, galloni2024RobustConstraintsTensorPerturbations} are obtained at $0.01 \Mpc^{-1}$.} \cite{galloniUpdatedConstraintsAmplitude2022, galloni2024RobustConstraintsTensorPerturbations}, leaving ample room for detection by next-generation experiments.

\textit{LiteBIRD} (Lite satellite for the study of $B$-mode polarisation and Inflation from cosmic background Radiation Detection) is a JAXA-led satellite mission designed to observe the full sky in polarisation across 12 frequency bands spanning 40–402 GHz. Its targets are a total uncertainty $\delta_r < 0.002$ on the tensor-to-scalar ratio, which includes statistical noise, instrumental systematics, foreground residuals and margins, and a statistical uncertainty $\delta_r < 0.001$, which includes foreground residuals and lensing \cite{LiteBIRD:2026gdf}. Meeting these ambitious targets requires not only exquisite control of instrumental systematics but also a robust data analysis pipeline capable of extracting the faint primordial signal from astrophysical foregrounds and noise. The path from multifrequency sky maps to constraints on $r$ involves several stages: component separation to isolate the CMB from Galactic emission, power spectrum estimation to compress the map-level information, and likelihood evaluation to infer cosmological parameters.

Throughout this work, we adopt the instrumental configuration and sensitivity of the \textit{LiteBIRD} Phase-A1 study \cite{litebirdcollaboration2022ProbingCosmicInflationLiteBIRDa}, commonly referred to as the ``PTEP configuration'', which observes the sky in 15 frequency bands and targets a total $\delta_r < 0.001$. We note that the mission design has since been revised \cite{LiteBIRD:2026gdf}, so that the absolute values of the forecast uncertainties reported here should be taken indicatively. A different configuration would in principle modify the constraining power on the tensor-to-scalar ratio, and a change in the signal-to-noise ratio (SNR) could also change how well a given approximation of the likelihood describes the truth. However, during the revision of the mission design, we applied the same pipeline to different instrument configurations, and in all cases the conclusions of this work held. Although we do not report these tests here, they indicate that adopting the final configuration would only change the numbers quoted in this work, not its conclusions. The same holds for the value of $r$. All the simulations of this work have $r=0$, which is also the fiducial model of the oHL, but during the same revision we also ran the pipeline on simulations with $r\neq0$, and the agreement between the oHL and the benchmark held.

The likelihood function is the central statistical object that connects measurements to parameter inference. For CMB analysis, a plethora of different approaches has been investigated \cite{smoot1992cosmicBackgroundExplorer, smootCOBEObservationsResults1999, Gorski:1994uu, Gorski:1994ye, Gorski:1996cf, Gorski:1996ti, Jewell:2008hg, Taylor:2007sj, Eriksen:2004ss, aghanimPlanck2018Results2020a, de_Belsunce_2021, deBelsunce:2022yll}. Here, we explore two broad classes of likelihood, each with distinct advantages and limitations.

Pixel-based (PB) likelihoods operate directly on the map and naturally capture the full covariance structure of the data, including correlations induced by partial sky coverage and inhomogeneous noise. Under the assumption of Gaussianity, the PB likelihood takes a well-known form involving the determinant and inverse of the pixel-pixel covariance matrix \cite{gerbinoLikelihoodMethodsCMB2020a}. However, this approach becomes computationally prohibitive at high resolution, since the covariance matrix scales as the square of the number of pixels. In practice, PB likelihoods are therefore restricted to low-resolution maps targeting the largest angular scales.

Harmonic-based likelihoods circumvent this limitation by working with angular power spectra rather than maps, dramatically reducing the dimensionality of the problem. The Hamimeche \& Lewis (HL) approximation \cite{hamimecheLikelihoodAnalysisCMB2008a} provides an accurate description of the non-Gaussian distribution of power spectrum estimates on a cut sky while accounting for correlations between multipoles through a covariance matrix estimated from simulations. This approach has been widely adopted, including by the \textit{Planck} \cite{aghanimPlanck2018Results2020a, ade2016Planck2015ResultsXX, tristram2022ImprovedLimitsTensortoscalarratio} and the BICEP/Keck array \cite{Ade_2014, collaboration2021BICEPKeckXIIIImproved} collaborations, and can be efficiently evaluated across a broad range of angular scales. Nevertheless, the accuracy of harmonic likelihoods at the largest scales—where the number of modes is small and the distribution of power spectrum estimates deviates most strongly from Gaussianity—warrants careful validation against pixel-based methods \cite{Galloni:2025eqk}.

\begin{figure}[t]
    \centering
    \includegraphics[width=\textwidth]{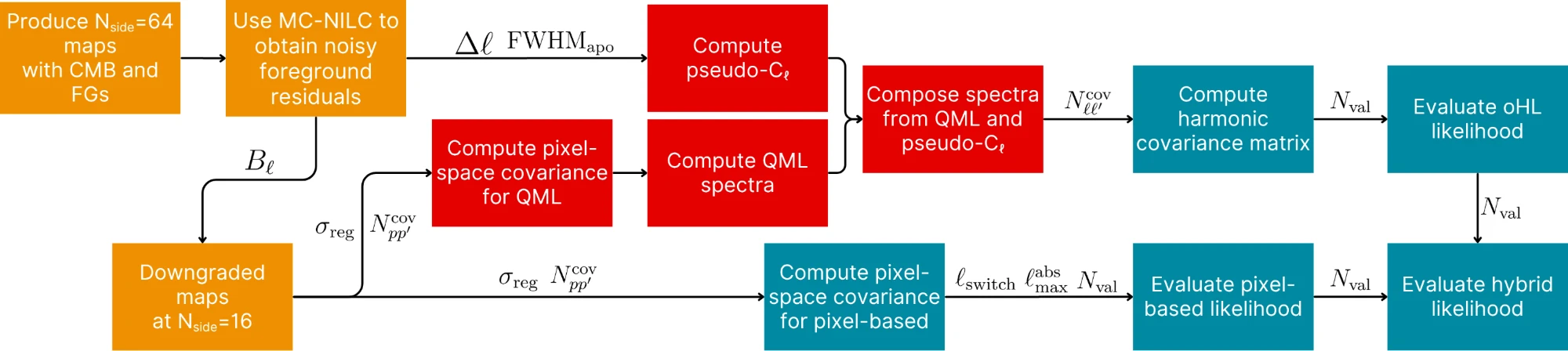}
    \caption{Complete pipeline of this work. The simulation, power spectrum estimation, and likelihood layers are highlighted in yellow, red, and blue, respectively. The analysis settings needed for each step are reported on the connecting arrows; see the following sections for details.}
    \label{fig:pipe}
\end{figure}

In this paper, we perform a comprehensive validation of harmonic-space likelihood methods for \textit{LiteBIRD}, using the pixel-based likelihood as a benchmark. The complete analysis pipeline is illustrated in \autoref{fig:pipe}. Starting from simulated multifrequency sky maps containing CMB, instrumental noise, and Galactic foregrounds, we apply the Multi-Clustering Needlet Internal Linear Combination (MC-NILC) component-separation algorithm \cite{caronesMultiClusteringNeedletILCCMB2023} to obtain foreground-cleaned maps with residual contamination (yellow block). Indeed, MC-NILC has proven to give excellent results when applied to $B$-mode cosmological inference, becoming one of the most promising techniques for future searches for primordial gravitational waves. The cleaned maps are fed into two parallel tracks for power spectrum estimation (red block): a quadratic maximum likelihood (QML) estimator operating on low-resolution maps for large scales ($\ell \leq 32$) \cite{Tegmark:1994we, tegmark1997HowMeasureCMBpower, Bond:1998zw, Efstathiou:2003dj}, and the pseudo-$C_\ell$ formalism via \texttt{\texttt{NaMaster}} \footnote{Publicly available at \url{https://github.com/LSSTDESC/NaMaster}.} \cite{hivonMASTERCMBAnisotropy2002, wandeltPseudoC_lMethodCosmic2001, alonsoUnifiedPseudoC_ellFramework2019} for smaller scales. Finally, we evaluate and compare three likelihood approaches (blue block): a debiased pixel-based likelihood serving as our large-scale benchmark (computed from two independent codes for robustness), an offset Hamimeche \& Lewis (oHL) likelihood \cite{mangilli2015LargescaleCMBTemperaturepolarization} applicable across the full multipole range, and a hybrid combination of the two. Each stage depends on a set of analysis settings—indicated along the connecting arrows in \autoref{fig:pipe}—whose impact on the final constraints we systematically explore.

Our main result is that, over the multipole range $\ell\leq128$, the 68\% upper bound on $r$ from the oHL likelihood is $19$--$29\%$ larger than that of the benchmark, depending on the beam window function. This validates the use of computationally efficient harmonic-space methods for \textit{LiteBIRD} parameter estimation while quantifying the residual differences that arise from the likelihood approximation.

The paper is organised as follows. Section~\ref{sec:sims} describes the simulation production pipeline. Section~\ref{sec:compsep} summarises the MC-NILC component-separation method. Section~\ref{sec:pse} presents the power spectrum estimation algorithms. Section~\ref{sec:likelihood} introduces the pixel-based, harmonic-based, and hybrid likelihood approaches. Section~\ref{sec:settings} collects the analysis settings explored throughout the pipeline. Section~\ref{sec:results} reports our results, and we present our conclusions in \autoref{sec:conclusions}. Appendix~\ref{app: settings} provides additional details on the exploration of the analysis settings, and appendix~\ref{app: PBLik} describes the \texttt{PBLik} code.

\section{Simulation production}\label{sec:sims}

Below, we summarise the workflow used to build the multifrequency sky simulations that serve as input to the component-separation pipeline discussed in \autoref{sec:compsep}. The configuration is identical to that adopted in section 5.2 of ref.~\cite{litebirdcollaboration2022ProbingCosmicInflationLiteBIRDa}, to which the reader is referred for full technical details. Following the structure of ref.~\cite{litebirdcollaboration2022ProbingCosmicInflationLiteBIRDa}, we first introduce the reference skies and then outline how the simulated maps are produced in each of \textit{LiteBIRD}'s frequency bands.

\subsection{Reference skies}
All input CMB and Galactic foreground maps are generated on a \texttt{HEALPix} \cite{healpix2005} grid with \Nside$=512$.

\paragraph{CMB maps.} We draw Gaussian realisations from the lensed CMB power spectra that correspond to these $\Lambda$CDM parameters $\{\As, \ns, \Omb, \Omc, \hnot, \tau\} = \{2.12\times10^{-9}, 0.9651, 0.02237, \\ 0.1201, 67.32~\Hunit, 0.06\}$ \cite{planckcollaboration2020Planck2018ResultsVI, Pagano_2020, Genesini:2026lmg, rosenberg2022CMBPowerSpectraCosmological, Tristram:2023haj}, and we set the tensor-to-scalar ratio to $r=0$. Hence, the only $B$-mode signal in our simulations is the one generated by gravitational lensing, i.e., by the deflection of CMB photons by the intervening large-scale mass distribution, which converts part of the $E$ modes into $B$ modes \cite{Zaldarriaga:1998ar, Lewis:2006fu}.

\paragraph{Galactic foregrounds.} Foreground templates are produced with the \texttt{PySM} \cite{Thorne:2016ifb, Zonca:2021row} framework using its default \texttt{d1s1} model, which combines thermal-dust and synchrotron emission \cite{Planck:2015mvg, Remazeilles:2014mba, WMAP:2012fli}. The dust signal is modelled as a modified blackbody whose temperature and spectral index vary across the sky \cite{Planck:2015mvg}, whereas the polarised synchrotron component follows a power law with a spatially varying spectral index \cite{Miville-Deschenes:2008lza}. The thermal dust spectral parameters and reference template are derived from a map-based parametric fit to the \textit{Planck} 2015 data using the \texttt{Commander} component-separation approach \cite{Planck:2015mvg}. The synchrotron emission model is instead based on analyses of low-frequency observations, including the Haslam 408\,MHz survey and the \textit{WMAP} 7-year data at 23\,GHz \cite{Miville-Deschenes:2008lza}. We note that during the course of this work, a new set of improved foreground concordance models was released \cite{Panexp_2025}. A dedicated validation of the likelihood framework using these updated models is deferred to future work. Nevertheless, we do not expect their inclusion to qualitatively affect the main conclusions of the present analysis.

\subsection{Multifrequency input maps}
Each foreground template is passed through \textit{LiteBIRD}'s instrument model (IMo): the spectra are integrated over idealised bandpasses (top-hat in intensity units), co-added (including CMB maps), smoothed with circular Gaussian beams, and finally combined with white-noise realisations. Bandpass width, beam full-width half-maximum (FWHM), and noise level for each of the 15 \textit{LiteBIRD} channels are identical to those listed in table 3 of ref.~\cite{litebirdcollaboration2022ProbingCosmicInflationLiteBIRDa}, reported for convenience in \autoref{tab:lb_sensitivity}.

We generate 20\,000 independent realisations for every \textit{LiteBIRD} band, each containing the CMB signal, instrumental noise, and Galactic foregrounds. These ensembles constitute the data sets on which the component-separation method of \autoref{sec:compsep} is applied.

Table~\ref{tab:lb_sensitivity} lists the key sensitivity parameters of the 15 \textit{LiteBIRD} observational bands. Columns provide, in order: the telescope, i.e. the Low-, Mid- or High-Frequency Telescope (LFT, MFT, HFT), or ``comb.'' when values are combined, band identifier, central frequency (GHz), bandwidth (GHz) and the corresponding fractional width in parentheses, beam FWHM (arcmin), number of detectors, array NET ($\mu$K$\,\sqrt{\rm s}$), and map-level sensitivity ($\mu$K-arcmin). Throughout this paper, $\mu$K denotes $\mu$K$_{\rm CMB}$, i.e. thermodynamic CMB temperature units. For the LFT 68-, 78-, and 89-GHz bands, two values are quoted—those in parentheses correspond to the two detector-pixel options of 16 mm and 32 mm, respectively.

Since the goal of this analysis is to validate the cosmological likelihood for large-scale $B$ modes, the input simulations are restricted to polarisation-only maps, expressed in terms of the standard Stokes parameters $Q$ and $U$. This choice also motivates the inclusion of only synchrotron and thermal dust emission among the foreground components, as these are the two dominant astrophysical sources of polarised radiation in the microwave sky. This selection is further justified by the fact that our simulations contain no systematic effects able to leak the temperature signal into polarisation.

\begin{table}[ht]
	\begingroup
	\caption{\textit{LiteBIRD} instrumental specifics. The values are taken from table 3 of ref.~\cite{litebirdcollaboration2022ProbingCosmicInflationLiteBIRDa}.}
	\label{tab:lb_sensitivity}
	\nointerlineskip
	\footnotesize
	\setbox\tablebox=\vbox{
		\newdimen\digitwidth
		\setbox0=\hbox{\rm 0}
		\digitwidth=\wd0
		\catcode`*=\active
		\def*{\kern\digitwidth}
		\newdimen\signwidth
		\setbox0=\hbox{+}
		\signwidth=\wd0
		\catcode`!=\active
		\def!{\kern\signwidth}
		\halign{
			\hfil#\hfil\tabskip=10pt&
			\hfil#\hfil\tabskip=10pt&
			\hfil#\hfil\tabskip=10pt&
			\hfil#\hfil\tabskip=10pt&
			\hfil#\hfil\tabskip=10pt&
			\hfil#\hfil\tabskip=10pt&
			\hfil#\hfil\tabskip=10pt&
			\hfil#\hfil\tabskip=10pt&
			\hfil#\hfil\tabskip=0pt\cr
			\noalign{\doubleline}
			\omit\hfil  \hfil& ID \hfil&  $\nu$   \hfil& $\delta\nu$ [GHz] \hfil& Beam size \hfil&  No. of  \hfil& NET$_{\rm arr}$ \hfil& Sensitivity \cr
                  & & [GHz] & ($\delta\nu/\nu$) & [arcmin]  & detectors &
 [$\mu$K$\sqrt{\rm s}$] & [$\mu{\rm K}$-arcmin] \cr
			\noalign{\vskip 3pt\hrule\vskip 5pt}
LFT	      	& 1			& 40 		& 12 (0.30)	 	& 70.5	 	    		& 48			& 18.50				  	& 37.42 \cr 		
\noalign{\vskip 3pt\hrule\vskip 5pt}
LFT	      	& 2			& 50 		& 15 (0.30)	 	& 58.5					& 24			& 16.54				  	& 33.46 \cr 	
\noalign{\vskip 3pt\hrule\vskip 5pt}
LFT	      	& 3			& 60 		& 14 (0.23)	 	& 51.1					& 48			& 10.54				  	& 21.31 \cr	
\noalign{\vskip 3pt\hrule\vskip 5pt}
LFT	      	& 4			& 68 		& 16 (0.23)	 	& (41.6, 47.1)			& (144, 24)		& (9.84, 15.70)		  	& (19.91, 31.77) \cr
comb.	&			&			&				&							&				& 8.34					& 16.87 \cr 			
\noalign{\vskip 3pt\hrule\vskip 5pt}
LFT	      	& 5			& 78 		& 18 (0.23)	 	& (36.9, 43.8)			& (144, 48)		& (7.69, 9.46)			& (15.55, 19.13) \cr
comb.	&			&			&				&							&				& 5.97					& 12.07 \cr
\noalign{\vskip 3pt\hrule\vskip 5pt}
LFT	      	& 6			& 89 		& 20 (0.23)	 	& (33.0, 41.5)			& (144, 24)		& (6.07, 14.22)			& (12.28, 28.77) \cr
comb.	&			&			&				&							&				& 5.58					& 11.30 \cr
\noalign{\vskip 3pt\hrule\vskip 5pt}
LFT/  		& 7			& 100 		& 23 (0.23)		& 30.2/					& 144/			& 5.11/					& 10.34 \cr
MFT			&			&			& 				& 37.8 		 			& 366 			& 4.19 					& 8.48 \cr
comb.	&			&			&				&							&				& 3.24					& 6.56 \cr			
\noalign{\vskip 3pt\hrule\vskip 5pt}
LFT/		& 8			& 119 		& 36 (0.30)		& 26.3/					& 144/			& 3.8/					& 7.69 \cr
MFT			&			&			&				& 33.6					& 488			& 2.82					& 5.70 \cr
comb.	&			&			&				&							&				& 2.26					& 4.58 \cr
\noalign{\vskip 3pt\hrule\vskip 5pt}
LFT/		& 9			& 140 		& 42 (0.30)		& 23.7/					& 144/			& 3.58/					& 7.25 \cr
MFT			&			&			&				& 30.8					& 366			& 3.16					& 6.38 \cr
comb.	&			&			&				&							&				& 2.37					& 4.79 \cr	
\noalign{\vskip 3pt\hrule\vskip 5pt}
MFT	      	& 10		& 166		& 50 (0.30)		& 28.9					& 488			& 2.75					& 5.57 \cr
\noalign{\vskip 3pt\hrule\vskip 5pt}
MFT/   		& 11		& 195		& 59 (0.30)		& 28.0/					& 366/			& 3.48/ 				& 7.05 \cr
HFT			&			&			&				& 28.6					& 254			& 5.19					& 10.50 \cr	
comb.	&			&			&				&							&				& 2.89					& 5.85 \cr
\noalign{\vskip 3pt\hrule\vskip 5pt}
HFT	      	& 12		& 235		& 71 (0.30)		& 24.7					& 254			& 5.34					& 10.79 \cr
\noalign{\vskip 3pt\hrule\vskip 5pt}
HFT	      	& 13		& 280		& 84 (0.30)		& 22.5					& 254			& 6.82					& 13.80 \cr
\noalign{\vskip 3pt\hrule\vskip 5pt}
HFT	      	& 14		& 337		& 101 (0.30)	& 20.9					& 254			& 10.85					& 21.95 \cr
\noalign{\vskip 3pt\hrule\vskip 5pt}
HFT	      	& 15		& 402		& 92 (0.23)		& 17.9					& 338			& 23.45					& 47.45 \cr
\noalign{\vskip 3pt\hrule\vskip 5pt}
Total   	&			&			&				&						& 4508			&						& 2.16 \cr
			\noalign{\vskip 5pt\hrule\vskip 3pt}}}
	\endPlancktable
	\endgroup
\end{table}

\section{Component separation with MC-NILC}\label{sec:compsep}

The simulated \textit{LiteBIRD} multifrequency data sets are processed using a component-separation pipeline designed to reconstruct a cleaned estimate of the underlying CMB polarisation signal, with reduced contamination from astrophysical foregrounds and instrumental noise.
Prior to applying the component-separation pipeline, all input maps are smoothed to a common angular resolution, as required by most map-based component-separation methods. Since our analysis targets the extraction of cosmological information from the largest angular scales, we adopt the lowest angular resolution of the simulated \textit{LiteBIRD}-like experiment, corresponding to a Gaussian beam with $\mathrm{FWHM}=70^{\prime}$, without impacting the results of the analysis. The maximum multipole included in the component-separation run is set to $\ell_{\mathrm{max}} = 150$ to avoid any pixelisation effect approaching 3\Nside.

\begin{figure}
    \centering
    \includegraphics[width=.5\linewidth]{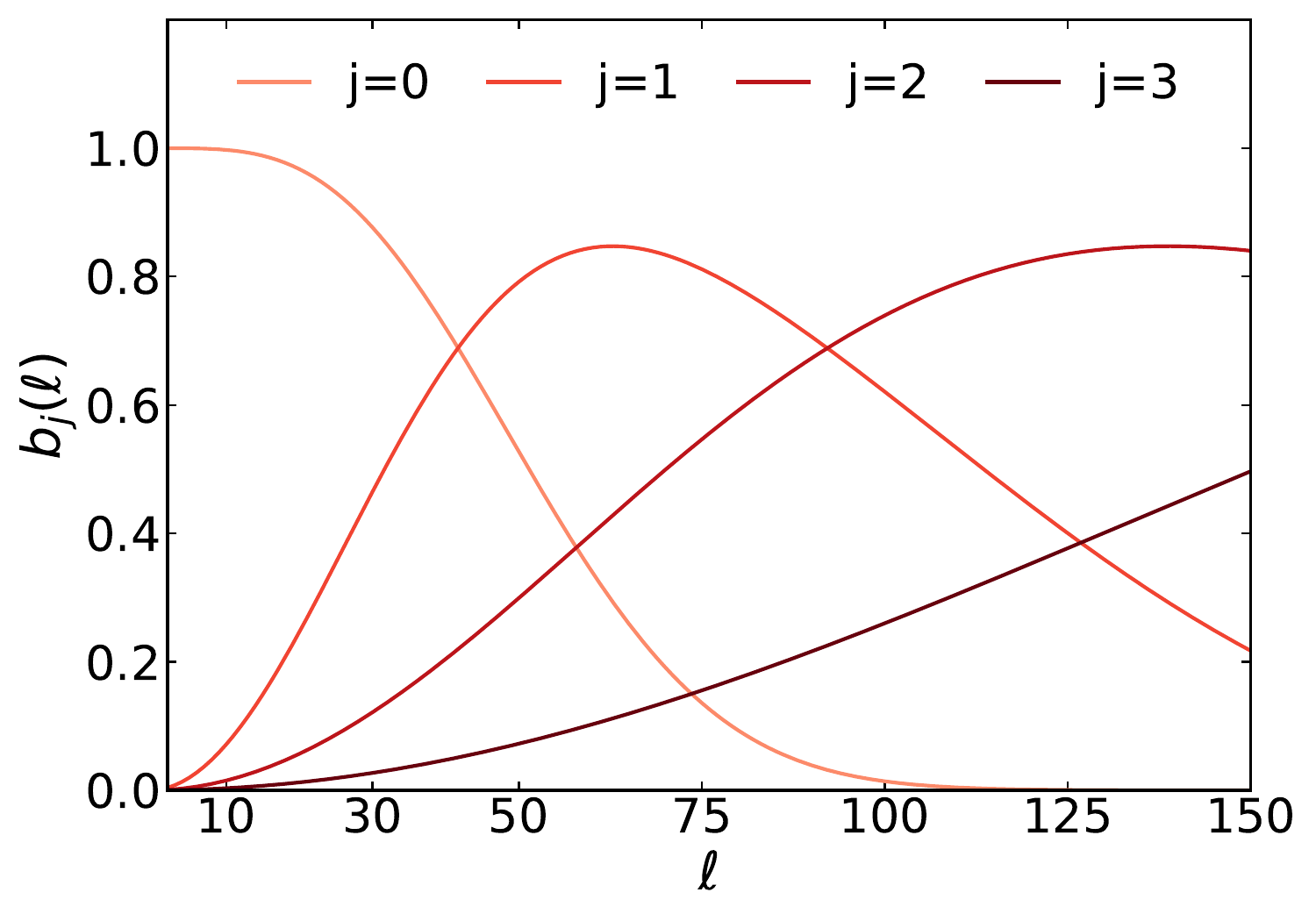}
    \caption{Configuration of needlet harmonic bands used for component separation in this work. Each colour represents a different needlet scale, with bandpass filters $b_{j}(\ell)$ plotted as a function of multipole $\ell$.}
    \label{fig:needlets}
\end{figure}

We employ a tailored implementation of a minimum-variance component-separation method. Since minimum-variance techniques are formulated for scalar fields, the input spin-$2$ polarisation signal—described by the Stokes parameter maps $Q$ and $U$—is first decomposed into its scalar components, namely the $E$ and $B$ modes, via a full-sky harmonic transformation. This procedure yields two independent multifrequency data sets corresponding to the $E$- and $B$-mode polarisation fields. Since the outputs needed for the subsequent cosmological likelihood validation refer to $B$ modes, component separation is uniquely applied to the data set of full-sky reconstructed $B$-mode maps. 

The component-separation pipeline adopted in this analysis is implemented in needlet space~\cite{Baldi2006,Marinucci2008}. The needlet framework acts as a set of localised wavelet filters, enabling the isolation of data modes over specific ranges of angular scales. Component separation is performed independently at each needlet scale, and the final cleaned $B$-mode map in pixel space is obtained by applying an inverse needlet transformation to the scale-dependent component-separated solutions. This procedure is repeated for all simulations. The adopted needlet functions correspond to Gaussian wavelets, also known as Mexican needlets \cite{mexican}, and are shown in figure~\ref{fig:needlets}. The bands of figure~\ref{fig:needlets} are the harmonic filters that define these scales. In particular, the needlet map $B^{(j)}_\nu(\hat{n})$ of the frequency channel $\nu$ is obtained by weighting the harmonic coefficients of the corresponding $B$-mode map by $b_j(\ell)$ and transforming back to pixel space. In practice, at each needlet scale $j$, a cleaned CMB solution $\tilde{B}^{(j)}_{\mathrm{CMB}}$ is obtained by linearly combining the multifrequency $B$-mode needlet maps $B^{(j)}_\nu(\hat{n})$ using frequency-dependent weights:
\begin{equation}
    \tilde{B}^{(j)}_{\mathrm{CMB}}(\hat{n}) =
    \sum_{\nu} w^{(j)}_{\nu}(\hat{n})\, B^{(j)}_{\nu}(\hat{n}) .
\end{equation}
The optimal set of weights is determined according to the Internal Linear Combination (ILC) criterion \cite{ILC,NILC}, by minimising the variance of the reconstructed map while preserving the CMB signal. This is enforced by requiring the weights to satisfy the constraint
\begin{equation}
    \sum_{\nu} w^{(j)}_{\nu}(\hat{n})\, A^{\mathrm{CMB}}_{\nu} = 1 ,
\end{equation}
where $A^{\mathrm{CMB}}_{\nu}$ denotes the CMB spectral energy distribution (SED) evaluated in the adopted units. Under this constraint, the weights are uniquely determined and can be written as
\begin{equation}
    \mathbf{w}^{(j)}(\hat{n}) =
    \left(\mathbf{A}^{\sf T}_{\mathrm{CMB}} \mathbf{C}_{(j)}^{-1}(\hat{n}) \mathbf{A}_{\mathrm{CMB}}\right)^{-1}
    \mathbf{A}^{\sf T}_{\mathrm{CMB}} \mathbf{C}_{(j)}^{-1}(\hat{n}) ,
    \label{eq:NILC}
\end{equation}
where $\mathbf{w}^{(j)}$ and $\mathbf{A}_{\mathrm{CMB}}$ are vectors of dimension equal to the number of observed frequency channels, collecting the multifrequency weights and the CMB SED, respectively. Here, $\mathbf{C}_{(j)}(\hat{n})$ denotes the local multifrequency covariance matrix estimated at needlet scale $j$:
\begin{equation}
    C^{\nu\nu'}_{(j)}(\hat{n}) =
    \left\langle
    B^{(j)}_{\nu}(\hat{n})\, B^{(j)}_{\nu'}(\hat{n})
    \right\rangle_{\mathcal{D}^{(j)}_{\hat{n}}} ,
\end{equation}
where $\mathcal{D}^{(j)}_{\hat{n}}$ denotes the domain over which the ensemble average is computed. This domain depends on both the sky direction $\hat{n}$ and the considered needlet scale $j$.

The choice of $\mathcal{D}^{(j)}_{\hat{n}}$, and therefore the practical implementation of the minimum-variance linear combination, depends on the needlet scale. At the largest angular scales (corresponding to the first needlet band), variance minimisation is performed over distinct patches of the sky. In this regime, the component-separation strategy corresponds to the Multi-Clustering Needlet Internal Linear Combination (MC-NILC) \cite{caronesMultiClusteringNeedletILCCMB2023}.
Each patch is constructed by grouping together sky pixels for which the $B$-mode foreground emission is expected to exhibit a similar spectral behaviour. Under this condition, the multifrequency covariance of the astrophysical signal can be described by a reduced number of independent eigenmodes.
In practice, the spatial variability of the foreground spectral properties is estimated blindly by computing the ratio
$\tilde{B}_{\mathrm{fgrs}}^{\mathrm{hf}} / \tilde{B}_{\mathrm{fgrs}}^{\mathrm{cf}}$
of two $B$-mode foreground templates, one obtained at a high-frequency channel and the other at a central frequency within the data set; in what follows, we refer to this ratio as the foreground ``tracer''. These templates are derived by applying the Generalised NILC (GNILC) pipeline \cite{GNILC_intro,GNILC} to a simulated observation containing CMB, foregrounds, and noise, which provides, at each observed frequency channel, a cleaned estimate of the foreground $B$-mode signal with reduced contamination from CMB and instrumental noise. The tracer is computed once from a single such realisation and held fixed across the whole simulation set, so that the patch definition, and hence the component-separation weights, do not vary from realisation to realisation.
The only implicit assumption underlying the GNILC application, and therefore the identification of the sky patches, is the prior knowledge of the two-point correlation function of the nuisance components (CMB and noise), which can be reliably predicted.
We stress that holding the tracer fixed makes the component-separation stage idealised and not fully realistic. In a real analysis, the tracer would be re-estimated from the data at hand; thus, it would fluctuate from realisation to realisation, together with the patch definition and the weights that follow from it. We keep it fixed here because the target of this work is the cosmological likelihood and not the component separation, and we want to avoid entangling the two. Dropping this assumption, along with the analogous one on the masking strategy discussed below, is left to the dedicated $B$-mode component-separation study in preparation (\textit{LiteBIRD} Collaboration, in prep.).

At the remaining needlet scales, a standard Needlet Internal Linear Combination (NILC) approach \cite{NILC} is adopted. In this case, the domain $\mathcal{D}^{(j)}_{\hat{n}}$ corresponds to a Gaussian kernel centred on the pixel $\hat{n}$. As a result, the data covariance is estimated locally at each pixel from a weighted average of the cross-products of the needlet maps. 

Although MC-NILC is expected to significantly reduce foreground residuals with respect to the standard NILC approach, it is applied only at the largest angular scales. This choice is motivated by the fact that a reliable and realistic reconstruction of the tracer defined above is currently available only on those scales.

For each \textit{LiteBIRD} simulated data set, in addition to the component-separated CMB $B$-mode solution, a set of ancillary products is provided, namely maps of foreground and noise residuals. These residual maps are obtained by linearly combining the foreground-only and noise-only multifrequency data sets using the same component-separation weights adopted for the CMB reconstruction. We note that these products, in particular the foreground residual maps, will not be available in real data analyses.

The last step before proceeding with the cosmological likelihood is to define a proper masking strategy that mitigates the impact of component-separation foreground residuals in the subsequent cosmological inference. In this study, we adopt the same procedure followed in \cite{litebirdcollaboration2022ProbingCosmicInflationLiteBIRDa}, simply extended to a $B$-mode solution. The bulk of the mask is the \textit{Planck} Galactic intensity mask, which retains $60\%$ of the sky (\texttt{GAL60}) \cite{planckcollaboration2016Planck2015ResultsI, planckcollaboration2015ExplanatorySupplementPlanck}. On top of this, an additional $10\%$ of the sky is masked by thresholding the map of the average foreground residuals, which has been previously smoothed with a Gaussian beam with FWHM=$2^\circ$. We note that this masking procedure is not fully applicable to real data, although it is not expected to have an impact on this study. Indeed, a different mask would certainly modify the constraining power on the tensor-to-scalar ratio, but it will not change whether a harmonic approximation of the likelihood is a good description of the truth. Given that this is the scope of this analysis, the choice of the mask becomes irrelevant. We also emphasise that the component-separation step of this work is based on the publicly available \texttt{broom} package\footnote{\url{https://github.com/alecarones/broom}}, described in Ref.~\cite{caronesDebiasingCosmologicalParameters2025}.

\section{Power spectrum estimation}\label{sec:pse}

\subsection{The quadratic maximum likelihood approach}\label{sec:qml}

The quadratic maximum likelihood (QML) estimator is the best estimator of the angular power spectrum. In fact, it is both unbiased $\langle\hat{C}_\ell\rangle = {C}_\ell$ and optimal, since its covariance matrix saturates the Cramér-Rao bound when computed on the true model \cite{tegmarkMethodExtractingMaximum1996, tegmarkHowMeasureCMB2001}. Thus, the QML estimator offers an efficient way to compress the information contained in a map to reduce the computational cost of working in real space with minimal information loss.

If we have a map $\mathbf{m}$ of the observed $B$-mode sky \footnote{Here, we consider $B$ modes as a scalar field, analogous to CMB temperature.}, its (real-space) covariance matrix reads:

\begin{equation}
\begin{aligned}
    \mathbf{C} & \equiv \expval{\mathbf{m}\mathbf{m}^{\sf T}} \\
    & = \mathbf{N} + \sum_\ell \mathbf{P}^\ell C_\ell\ , 
\end{aligned}
\label{eq:real_space_cov}
\end{equation}
where the elements of the matrix $\mathbf{P}^\ell$, which depend on two directions in the sky $\hat{n}_i, \hat{n}_j$ through the Legendre polynomials $P_\ell$, are defined as
\begin{equation}
    P^\ell_{ij} \equiv \frac{2\ell+1}{4\pi}P_\ell(\hat{n}_i\cdot\hat{n}_j)
\label{eq:matrix_legendre}
\end{equation}
and $\mathbf{N}$ is the noise covariance matrix. This matrix is a key ingredient of the computation, since obtaining it accurately from simulations is not trivial.

Under the assumption that the pixels are distributed according to a multivariate Gaussian distribution with a mean of zero and covariance $\mathbf{C}$, we can write the likelihood as (see \autoref{sec:pixel-based} for further details) \cite{hamimecheLikelihoodAnalysisCMB2008a, gerbinoLikelihoodMethodsCMB2020a}
\begin{equation}
    -2\ln\mathcal{L} = \ln\det\mathbf{C} + \mathbf{m}^{\sf T}\mathbf{C}^{-1}\mathbf{m} + \text{const.}\ ,
\label{eq:qml_like}
\end{equation}
where the constant collects the normalisation of the Gaussian and does not depend on the power spectrum $C_\ell$.

The Fisher matrix for this likelihood, which saturates the Cramér-Rao inequality, reads
\begin{equation}
    F_{\ell\ell^\prime} \equiv -\expval{\frac{\partial^2\ln\mathcal{L}}{\partial C_\ell \partial C_{\ell^\prime}}} = \frac{1}{2}\Tr\qty[\mathbf{C}^{-1}\mathbf{P}^\ell\mathbf{C}^{-1}\mathbf{P}^{\ell^\prime}]\ .
\label{eq:qml_fisher}
\end{equation}
It can be shown that the QML estimator can be written as \cite{tegmarkHowMeasureCMB2001}
\begin{equation}
    \hat{C}_\ell = \mathbf{m}^{\sf T}\mathbf{E}^\ell\mathbf{m} - \Tr\qty[\mathbf{E}^\ell\mathbf{N}]\ ,
\label{eq:estimated_cl_qml}
\end{equation}
where the operator $\mathbf{E}^\ell$ is
\begin{equation}
    \mathbf{E}^\ell = \frac{1}{2}\sum_{\ell^\prime}\qty(\mathbf{F}^{-1})_{\ell\ell^\prime}\,\mathbf{C}^{-1}\mathbf{P}^{\ell^\prime}\mathbf{C}^{-1}\ .
\label{eq:E_ell_operator}
\end{equation}
The second term of eq.~\eqref{eq:estimated_cl_qml} is the noise bias,
\begin{equation}
    N_\ell = \Tr\qty[\mathbf{E}^\ell\mathbf{N}]\ ,
\label{eq:qml_noise_bias}
\end{equation}
whose subtraction makes the estimator unbiased, $\expval{\hat{C}_\ell} = C_\ell$, with covariance ${\rm Cov}(\hat{C}_\ell, \hat{C}_{\ell^\prime}) = \qty(\mathbf{F}^{-1})_{\ell\ell^\prime}$. Note that the normalisation in eq.~\eqref{eq:E_ell_operator} involves the full inverse Fisher matrix and not only its diagonal. Hence, in the language of ref.~\cite{tegmarkHowMeasureCMB2001}, the estimator we use is the deconvolved one, in which the mode coupling induced by the mask is undone at the price of anti-correlating neighbouring multipoles. We will come back to this in \autoref{sec:negative_cls}.

We emphasise again that this procedure is performed in real space and depends on the pixel-covariance matrix of the map. While the signal component can be computed analytically as a sum of Legendre polynomials weighted by the fiducial power spectrum, the noise contribution is typically estimated with Monte-Carlo realisations of the expected noise. This is exactly what we do: leveraging the large number of simulations produced, we study the dependence of the results on the number of simulations dedicated to describing the pixel covariance \Npp. We stress that the same matrix $\mathbf{N}$ enters both the total covariance of eq.~\eqref{eq:real_space_cov} and the bias term of eq.~\eqref{eq:qml_noise_bias}.

The noise bias $N_\ell$ of eq.~\eqref{eq:qml_noise_bias} is the projection onto multipole $\ell$ of the autocorrelation of the noise and foreground residuals encoded in $\mathbf{N}$; it is a property of the data set, and the \Npp~simulations only provide the means to estimate it. In our case, $\mathbf{N}$ is built from the MC-NILC residual maps, so it describes the scatter of the noise residuals and of the foreground residuals, including any cross-covariance between the two, since neither is constant from simulation to simulation (see \autoref{sec:compsep}). On the other hand, $\mathbf{N}$ does not capture the part of the foreground residuals that is common to all simulations. This part is a systematic offset in the mean of the recovered spectra. We denote its QML projection by $F_\ell$.

We explored two ways of estimating this quantity: one can obtain $F_\ell$ from the total spectra $\hat{C}_\ell$ by subtracting $N_\ell$ and the fiducial power spectrum $C_\ell^{\rm fid}$. Alternatively, one can run the QML procedure on noisy foreground residual maps and subtract $N_\ell$ from the spectra. Given that the QML is unbiased and the CMB is statistically independent of foreground residuals, we expect these two estimates of $F_\ell$ to be equivalent. We verified this explicitly. Over the multipole range where the QML estimates are used, $\ell\leq32$, the two estimates of $F_\ell$ agree to a median of $1.8\%$, with individual multipoles differing by at most $15\%$. Both are means over the simulation set, so their accuracy is ultimately determined by its size. Whether this limit is relevant has to be assessed at the level of $r$: since all our simulations are generated at $r=0$, any residual wrong subtraction propagates into a non-zero maximum a posteriori (MAP) estimate of $r$, which the accuracy criterion of eq.~\eqref{eq:screen_accuracy} bounds for every configuration we explore.

\subsubsection{QML implementation}\label{sec:qml_implementation}

To obtain the low resolution maps for the QML, we start from the recovered CMB maps from MC-NILC (see \autoref{sec:compsep}), which contain the CMB and noisy foreground residuals at \Nside$=64$. Thus, we downgrade them to a lower resolution of \Nside$=16$ after smoothing them with an aggressive beam $B_\ell$ to limit the harmonic content and avoid sharp cuts \cite{sullivanMethodsCMBMap2025}. Note that when including a beam and a pixel window function $w_\ell$, eq.~\eqref{eq:real_space_cov} is recast to
\begin{equation}
    \mathbf{C} = \mathbf{N} + \sum_\ell \mathbf{P}^\ell B^2_\ell w_\ell^2 C_\ell\ .
\end{equation}
The beam \bl~used for this operation, the number of simulations \Npp~from which $\mathbf{N}$ is estimated, and the regularisation noise \regn~added to it are all analysis settings, which we define in \autoref{sec:settings} together with the ranges we explore.

\begin{figure}[t]
    \centering
    \includegraphics[width=.5\textwidth]{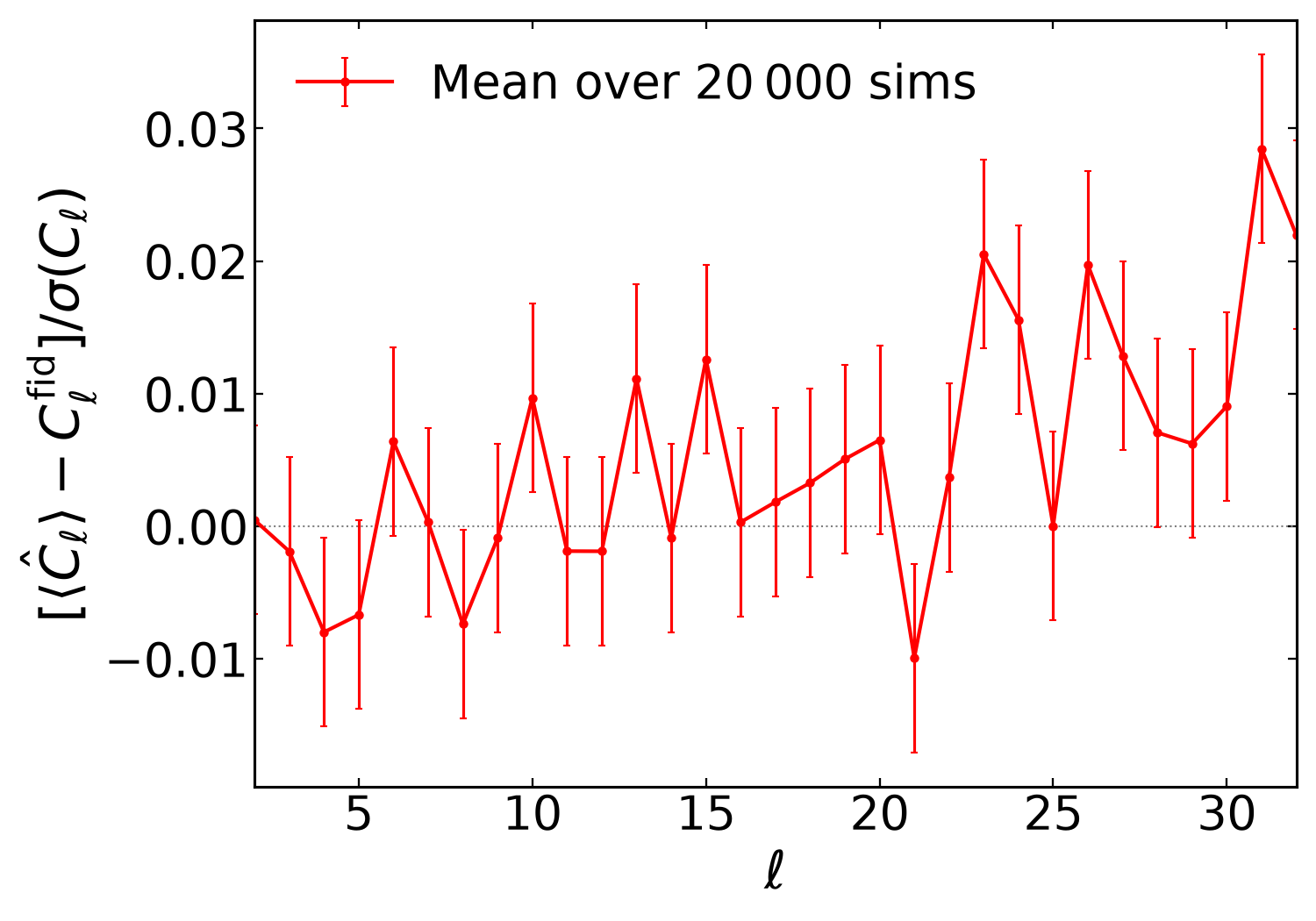}
    \caption{Validation of the QML estimator on CMB-alone maps. Points show the difference between the mean recovered spectrum over 20\,000 simulations and the input fiducial, normalised by the standard deviation $\sigma(C_\ell)$ of a single realisation. The error bars are the Monte Carlo uncertainty on the mean, which in these units is $1/\sqrt{N}$. The residual never exceeds $2.9\%$ of $\sigma(C_\ell)$.}
    \label{fig:validation_spectra}
\end{figure}
To perform the QML estimation, we use the publicly available code \texttt{pse\_qml} \footnote{\url{https://baltig.infn.it/cosmology_ferrara/lowell-likelihood-analysis/-/tree/master/pse_qml}.}. Together with the angular power spectrum of the input map $\hat{C}_\ell$, it provides the noise bias $N_\ell$ (see eq.~\eqref{eq:qml_noise_bias}). In terms of performance, it takes a fraction of a second to run the QML algebra described above on one map. Thus, processing all our twenty thousand simulations takes $\order{10}$ minutes.

In \autoref{fig:validation_spectra}, we show the validation of the angular power spectra from 20\,000 simulations as obtained from our QML algorithm. In particular, we perform the QML estimation on CMB-alone maps, assuming small white noise to regularise the eigenvalues of the pixel-space covariance matrix. The mean of these spectra is compared with the scatter of the mean; thus, this represents a very stringent test since the standard deviation of a single spectrum is 140 times larger than the uncertainty shown in the figure.

The residual is consistent with zero over most of the range, but it drifts slightly positive towards the upper end of the band because of power outside the modelled range. Indeed, the signal covariance of the estimator is built over $\ell\leq4\Nside$, while the input maps are downgraded from a higher resolution and thus retain some power above that range, which the beam \bl~suppresses but does not remove. The mask then couples such unmodelled power into the estimated band, preferentially at its upper end.

\begin{figure}[t]
    \centering
    \includegraphics[width=.5\textwidth]{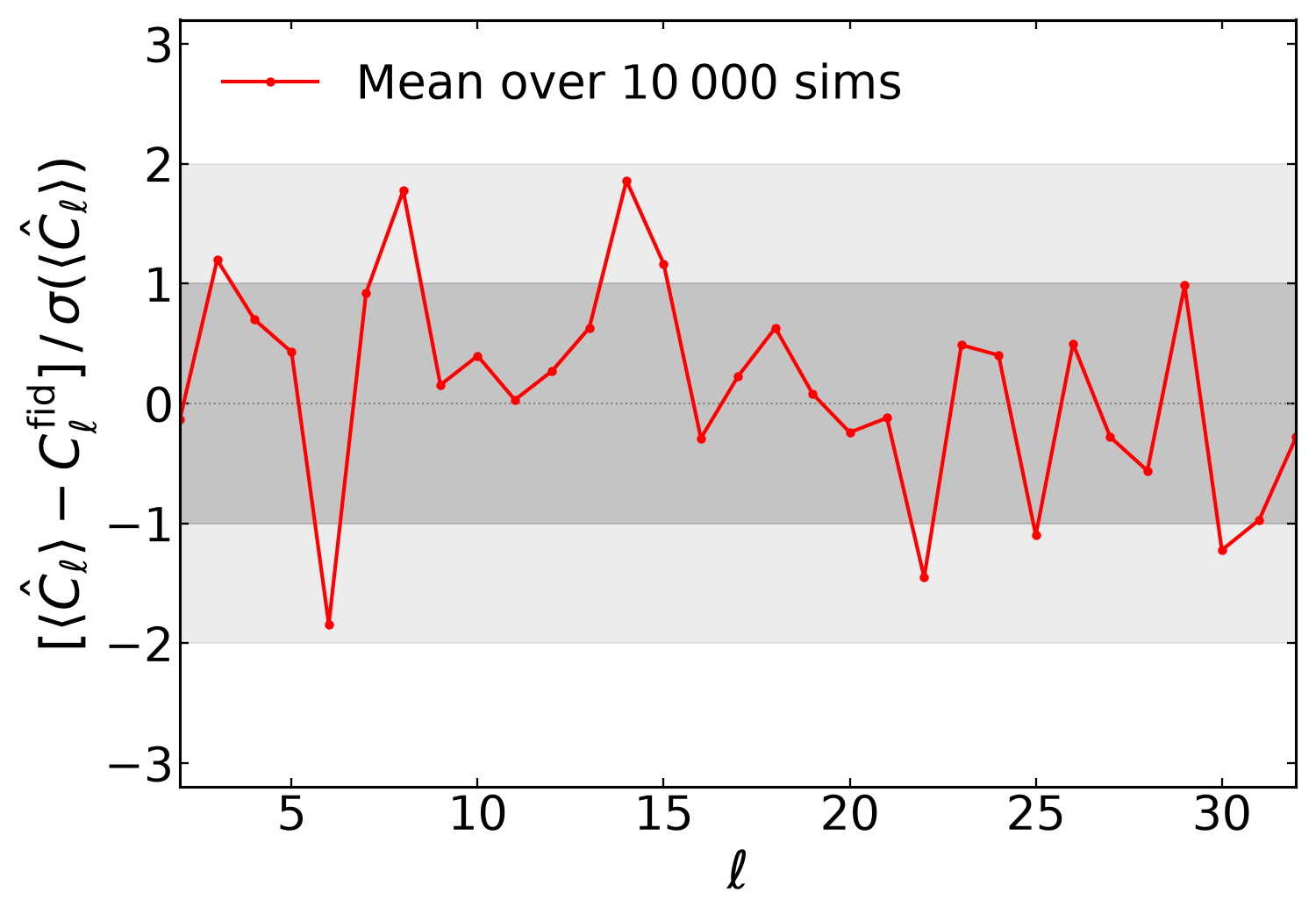}
    \caption{Test of the bias of the QML estimator when no sky power lies above the range covered by the signal covariance. Points show the difference between the mean recovered spectrum over 10\,000 band-limited simulations and the input fiducial, normalised by the uncertainty on the mean, with the dark and light bands marking 1 and 2 times this uncertainty. The results are obtained with \texttt{CosmoForge} \citep{CosmoForge}.}
    \label{fig:band_validation}
\end{figure}

To isolate this effect, we run a dedicated Monte Carlo in which the input sky is band-limited, so that no power lies above the modelled range. In this case, the estimator is unbiased at every multipole, as shown in \autoref{fig:band_validation}, with $\chi^2=23.4$ for $N_{\rm dof}=31$ over $2\leq\ell\leq32$ and no multipole deviating by more than $1.9$ times the uncertainty on the mean. This test is performed with \texttt{CosmoForge} \citep{CosmoForge}, which in this configuration reproduces the covariance products of \texttt{pse\_qml} to $10^{-11}$--$10^{-13}$ and its QML outputs to about $10^{-7}$, so the drift comes from the input maps. As for $F_\ell$, any residual effect of this drift on the MAP at $r=0$ is bounded by the accuracy criterion of eq.~\eqref{eq:screen_accuracy}.

\subsection[The pseudo-\texorpdfstring{$C_\ell$}{Cl} formalism]{The pseudo-\texorpdfstring{$\boldsymbol{C_\ell}$}{Cl} formalism}\label{sec:NaMaster}

On small scales, where the QML procedure becomes prohibitive, one typically applies the pseudo-$C_\ell$ formalism to extract the angular power spectra. Here, we recall the main concepts \cite{gerbinoLikelihoodMethodsCMB2020a}. Decomposing the pixel-space map using the spherical harmonics $Y_{\ell m}$ as a basis, we can write \cite{bondStatisticsCosmicBackground1987}
\begin{equation}
    a_{\ell m} = \int d\hat{n} \Theta(\hat{n}) Y^*_{\ell m}(\hat{n})\ ,
\end{equation}
where the $a_{\ell m}$ are the coefficients of this decomposition and $\Theta(\hat{n})$ is some field on the celestial sphere (e.g., the CMB temperature anisotropy $\Delta T/T(\hat{n})$). In the standard cosmological picture, these coefficients are Gaussian distributed with a mean of zero. Thus, their statistic is entirely determined by their variance—the angular power spectrum $C_\ell$:
\begin{equation}
    \expval{a_{\ell m} a^*_{\ell^\prime m^\prime}} = \delta_{\ell\ell^\prime}\delta_{mm^\prime}C_\ell\ .
\end{equation}
Thus, an observed map $\mathbf{m}$ can be converted to an observed set of coefficients $a_{\ell m}$, whose angular power spectrum can be estimated with
\begin{equation}
    \hat{C}_\ell = \frac{1}{2\ell+1}\sum_{m=-\ell}^\ell a_{\ell m} a^*_{\ell m}\ .
\end{equation}

However, in a realistic analysis, only a fraction of the sky would be available, introducing a series of complications. Given a generic mask $W(\hat{n})$ of the sky, with values from 0 to 1, we can decompose it into spherical harmonics $Y_{\ell m}(\hat{n})$ with \cite{hivonMASTERCMBAnisotropy2002, wandeltPseudoC_lMethodCosmic2001, brownCosmicMicrowaveBackground2005}
\begin{equation}
    w_{\ell m} = \int d\hat{n} W(\hat{n}) Y^*_{\ell m}(\hat{n})\ .
\end{equation}
Here, $w_{\ell m}$ are the coefficients of such a decomposition, with their angular power spectrum defined as 
\begin{equation}
    \mathcal{W}_{\ell} = \frac{1}{2\ell+1}\sum_m\abs{w_{\ell m}}^2\ .
\end{equation}
Thus, in such a context, the spherical harmonic decomposition of our field $\Theta(\hat{n})$ reads
\begin{equation}
\begin{aligned}
    \Tilde{a}_{\ell m} &= \int d\hat{n} \Theta(\hat{n}) W(\hat{n}) Y^*_{\ell m}(\hat{n}) \\
    &= \sum_{\ell^\prime m^\prime} a_{\ell^\prime m^\prime} \int d\hat{n} Y_{\ell^\prime m^\prime}(\hat{n}) W(\hat{n}) Y^*_{\ell m}(\hat{n}) \\
    &= \sum_{\ell^\prime m^\prime} a_{\ell^\prime m^\prime} K_{\ell m \ell^\prime m^\prime}(W) \ ,
\end{aligned}
\end{equation}
where $K_{\ell m \ell^\prime m^\prime}(W)$ is the coupling kernel between modes, depending on the geometry of the mask. Indeed, $\Tilde{a}_{\ell m}$ are still Gaussian random variables (sums of the ``full-sky'' $a_{\ell m}$); however, the spherical harmonic basis is no longer orthonormal on a partial sky, giving rise to coupling between $\ell-\ell^\prime$ multipoles.

This allows us to define the pseudo-$C_\ell$s as
\begin{equation}
    \tilde{C}_\ell = \frac{1}{2\ell+1}\sum_{m=-\ell}^\ell \tilde{a}_{\ell m} \tilde{a}^*_{\ell m}\ ,
\end{equation}
which, in the ensemble average, are linked to the ``full-sky'' $C_\ell$s through
\begin{equation}
\begin{aligned}
    \expval{\tilde{C}_{\ell}} &= \frac{1}{2\ell+1}\sum_{m=-\ell}^{\ell} \expval{\tilde{a}_{\ell m} \tilde{a}^*_{\ell m}} \\
    &= \sum_{\ell^\prime} \mathcal{K}_{\ell\ell^\prime}\expval{C_{\ell^\prime}}\ ,
\end{aligned}
\end{equation}
where the coupling matrix is defined as
\begin{equation}
    \mathcal{K}_{\ell\ell^\prime} = \frac{2\ell^\prime+1}{4\pi}\sum_{\ell^{\prime\prime}}\mathcal{W}_{\ell^{\prime\prime}}
    \threejz{\ell}{\ell^\prime}{\ell^{\prime\prime}}^2\ .
\end{equation}
Finally, our estimate of the angular power spectrum on the cut-sky is then
\begin{equation}
    \hat{C}_\ell = \sum_{\ell^\prime} \mathcal{K}_{\ell\ell^\prime}^{-1}\tilde{C}_{\ell^\prime}\ .
\end{equation}

In practice, the pseudo-$C_\ell$ formalism is publicly available via \texttt{NaMaster} \cite{alonsoUnifiedPseudoC_ellFramework2019}. In performing this computation, we must choose whether and how to apodise the mask and whether and how we want to bin the power spectra. Both are analysis choices, and we collect them in \autoref{sec:settings} together with the other settings.

\subsection{The recovered spectra}\label{sec:mean_spectra_sec}

\begin{figure}[t]
    \centering
    \includegraphics[width=.71\textwidth]{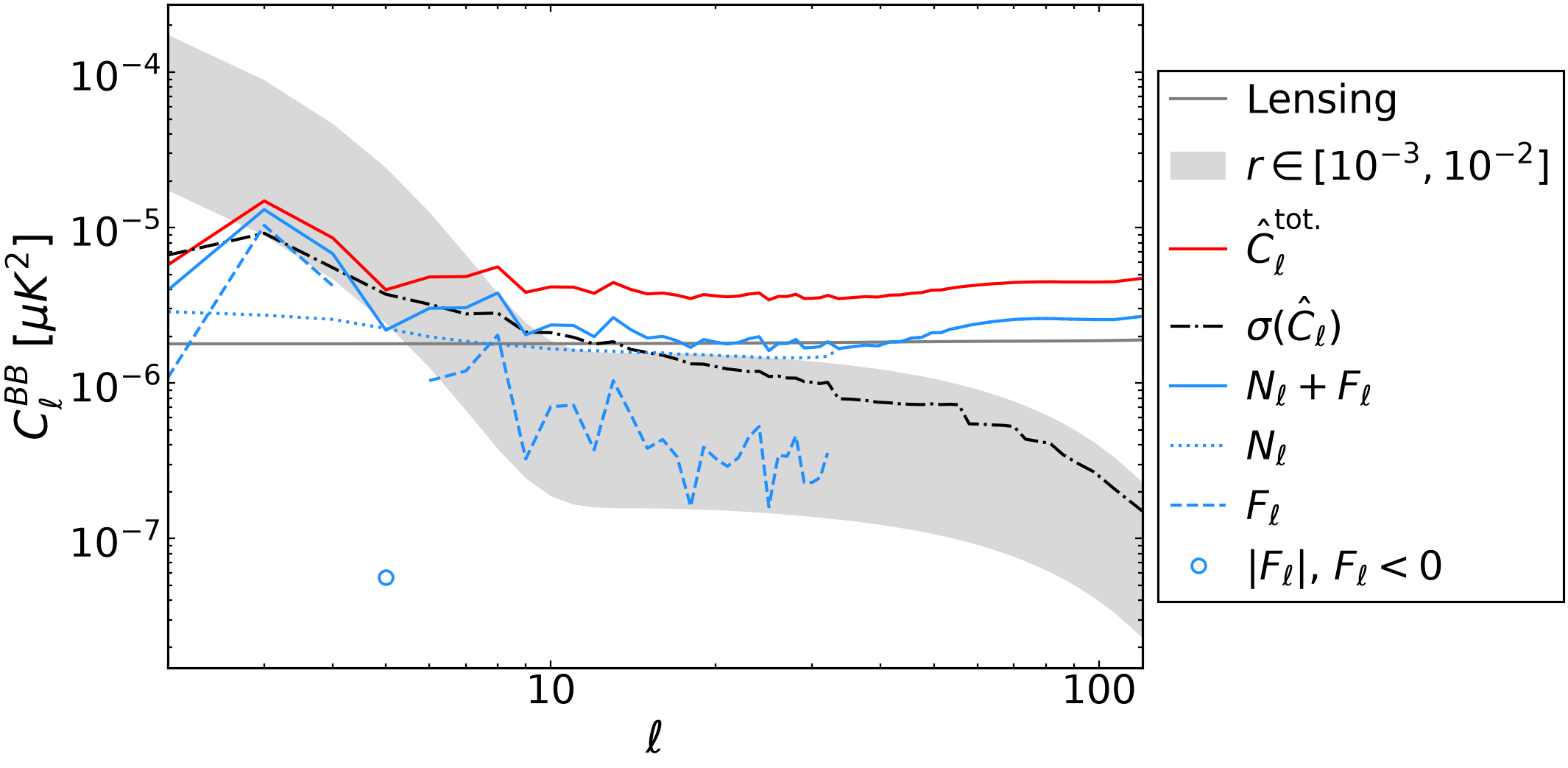}
    \caption{$B$-mode power spectrum. The solid blue line shows the average total bias spectrum from noisy foreground residuals, as a sum of the statistical $N_\ell$ and systematic $F_\ell$ contributions (shown in dotted and dashed, respectively). Instead, the solid red line is the average total power spectrum, thus containing CMB and the total bias. The black dot-dashed line is the standard deviation of $\hat{C}_\ell$ over the 20\,000 simulations; since $N_\ell$ and $F_\ell$ are the same for every simulation, this is also the scatter of the debiased spectrum. $F_\ell$ is estimated only below the binning transition at $\ell=32$ and is negative only at $\ell=5$, where it cannot be drawn on a logarithmic axis, so the open circle marks $|F_\ell|$ at that multipole. As a reference we plot the input lensing power spectrum and a range of $BB$ spectra for $r\in [10^{-3}, 10^{-2}]$.}
    \label{fig:mean_spectra}
\end{figure}

Before moving to the likelihood, we show in \autoref{fig:mean_spectra} an example of the angular power spectra recovered with the estimators of this section, since they determine which multipoles carry most of the constraining power on $r$ and motivate several of the choices of the following sections.
The noise bias from foreground residuals appears almost white: at large scales, the systematic contribution has its most relevant impact; while at small scales, we find an increase in the noise bias, typical of MC-NILC reconstruction \cite{carones2023MultiClusteringNeedletILCCMBBmodes, caronesDebiasingCosmologicalParameters2025}. The systematic contribution is positive at every multipole of $\ell\leq32$ except $\ell=5$, where it is $-5.6\times10^{-8}\,\mu{\rm K}^2$ and cannot be shown on the logarithmic axis of \autoref{fig:mean_spectra}. This value lies $2.1$ standard deviations of the Monte Carlo error on the mean below zero; among the $31$ multipoles of $\ell\leq32$, a single excursion of this size is expected by chance.

It is also clear from the figure that the largest scales will carry most of the constraining power on primordial $B$ modes. In fact, the dot-dashed line quantifies the uncertainty per multipole, which a signal of $r=10^{-3}$ exceeds at $\ell=2$ by a factor of $2.6$ and falls below it from $\ell=3$ onwards. Summing in quadrature the resulting signal-to-noise ratio over $2\leq\ell\leq128$, we find that $88\%$ of the total comes from $\ell\leq5$ and $91\%$ from $\ell\leq10$.

\section{Likelihood analysis}\label{sec:likelihood}

In a parameter-estimation problem, the likelihood function $ \mathcal{L}(\theta | D) = P(D | \theta) $ is the central quantity that connects a statistical model (parametrised by $\theta$) to the actual observations $D$. Throughout this work, we write the likelihood as $\mathcal{L}(\theta|D)$, placing its variable argument, $\theta$, to the left of the vertical bar. Regardless of this notation, it is always computed from the probability distribution of the data given the parameters, $P(D|\theta)$. Once the data are observed and thus fixed, $ \mathcal{L}(\theta | D) $ is viewed as a function of $\theta$: it quantifies how plausible different parameter values are in light of the data. Because $\theta$ is not a random variable in this context, $ \mathcal{L}(\theta | D) $ is not a probability distribution over $\theta$—hence, there is no requirement that it integrates to unity.

Through Bayes' theorem, the likelihood is linked to the posterior distribution $P(\theta|D)$, which is typically the final goal of such an analysis, since it tells what the most probable values of the parameters of interest are (viewed as random variables in a Bayesian framework).
Here, we explore two different incarnations of this. In particular, we will assume our data to be the reconstructed CMB map in a pixel-based likelihood (see \autoref{sec:pixel-based}) or the angular power spectra extracted from the same map in a harmonic-based likelihood (see \autoref{sec:harmonic-based}). On top of this, we will also study a hybrid approach, trying to combine these two frameworks in \autoref{sec:hybrid}.

\subsection{Pixel-based}\label{sec:pixel-based}

A first, natural choice is to take the data to be the map itself, $D = \mathbf{m}$, which defines the so-called pixel-based (PB) likelihood. Indeed, inference performed directly on maps has a long history in CMB analysis, in forms considerably more elaborate than the one used here \cite{jewellMarkovChainMonte2009, taylorFastOptimalCMB2008, eriksenPowerSpectrumEstimation2004a, bull2015CMBGIBBSSAMPLERLOCALIZED, galloway2023BeyondPlanckIIICommander3, hajian2007EfficientCosmologicalParameterEstimation}.

Statistical isotropy implies that the ensemble mean of the CMB anisotropy field vanishes, so the expected value of each pixel of the CMB anisotropy map is zero. Its statistics are then entirely described by the covariance of the pixels, which serves as our model: $\theta = \mathbf{C}$, a function of the cosmological parameters.

Actually, we have already introduced the PB likelihood when describing the QML algebra in eq.~\eqref{eq:qml_like}, which reads
\begin{equation}
    -2\ln\mathcal{L}(\mathbf{C}|\mathbf{m}) = \ln\det\mathbf{C} + \mathbf{m}^{\sf T}\mathbf{C}^{-1}\mathbf{m} + \text{const.}\ .
\label{eq:pb_like}
\end{equation}

The covariance matrix of eq.~\eqref{eq:real_space_cov} is divided into signal and noise contributions, namely $\mathbf{S}(C_\ell) = \sum_\ell \mathbf{P}^\ell C_\ell$ and $\mathbf{N}$. The form of the signal term, an expansion in Legendre polynomials with the $C_\ell$ as coefficients, follows from statistical isotropy alone, while Gaussianity ensures that the covariance is a complete statistical description of the field. Through $\mathbf{S}(C_\ell)$, the cosmological parameters are propagated to real space. In particular, the signal covariance matrix is built from the angular power spectrum up to \lmaxabs. However, only part of these ($\ell \leq$ \lswitch) varies as a function of the cosmological parameters. This allows for the proper description of the harmonic content of a map without encountering ``border effects''. Both \lmaxabs~and \lswitch~are analysis settings, and the values we explore are listed in \autoref{sec:settings}.

$\mathbf{N}$ is the covariance of the experimental noise and of the other contributions to the sky map, such as the foreground residuals. As mentioned above, it is imperative to have a good estimate of $\mathbf{N}$ to obtain a reliable PB likelihood. Given that it depends explicitly on the real space covariance, this likelihood is particularly prone to biases and distortions when the noise covariance is estimated with few simulations. For this reason, similarly to the QML, the PB depends on \Npp~and on the regularisation noise \regn, both of which we vary over the ranges given in \autoref{sec:settings}.

The foreground residuals contain a systematic component that does not change across simulations. Thus, this is not captured by $\mathbf{N}$ and biases the estimation of parameters. Thus, in the following, we will debias the PB by subtracting this systematic contribution from the maps. The ``systematic map'' we subtract from the maps is simply computed as the mean map from the same \Npp~simulations, since the statistical part averages to zero while the systematic part never changes. 

The PB computation needs $\order{20}$ minutes to be evaluated in our simulations, which is slightly more involved with respect to the QML procedure. To perform it, we used two different codes: the \texttt{parameter\_estimation} algorithm \footnote{\url{https://baltig.infn.it/cosmology_ferrara/lowell-likelihood-analysis/-/tree/master/parameter_estimation}} and the \texttt{PBLik} code \footnote{\url{https://github.com/ChristianGim/PBLik}} (see appendix~\ref{app: PBLik}), allowing us to compare the two to ensure robustness. The results shown in the following have been obtained using the former algorithm. We also verified that \texttt{PICSLike}, the pixel-based likelihood of \texttt{CosmoForge} \citep{CosmoForge}, recovers the results of \texttt{parameter\_estimation} to machine precision.

\subsection{Harmonic-based}\label{sec:harmonic-based} \label{sec:harmonic_based}

In \autoref{sec:NaMaster}, we have already shown how to compress the information of a map into harmonic-space. Also, we mentioned that the $a_{\ell m}$s are Gaussian distributed quantities. Thus, their harmonic likelihood can be written as \cite{hamimecheLikelihoodAnalysisCMB2008a}
\begin{equation}
    -2\ln\mathcal{L}(C_\ell|a_{\ell m}) = \sum_{m=-\ell}^\ell \qty[a^*_{\ell m} C_\ell^{-1} a_{\ell m} + \ln |2\pi C_\ell|]\ .
\label{eq:alm_like}
\end{equation}
Equation~\eqref{eq:alm_like} is the likelihood of a single multipole, and the total likelihood is the sum of such terms over $\ell$; the same holds for the expressions that follow. This can be recast as a likelihood on $\hat{C}_\ell$, i.e., an inverted Wishart distribution:
\begin{equation}
    -2\ln\mathcal{L}(C_\ell|\hat{C}_\ell) = \sum_\ell \qty(2\ell+1)\qty[\frac{\hat{C}_\ell}{C_\ell} - \ln\frac{\hat{C}_\ell}{C_\ell} - 1] + \text{const.}\ .
\label{eq:wishart_like}
\end{equation}

Here, the observed spectra $\hat{C}_\ell$ contain contributions from the CMB lensing power spectrum ($r=0$) and noisy foreground residuals. Thus, the theoretical spectra are obtained as
\begin{equation}
    C_\ell = C_\ell^{\rm prim}(r) + C_\ell^{\rm lens} + N_\ell + F_\ell\ , \label{eq:model_cl}
\end{equation}
which, for the sake of notation, we will not write out again: in the following, the theoretical spectra always include the lensing term and the two bias terms of this equation. Note that eq.~\eqref{eq:alm_like} is no longer valid on the cut-sky; as a consequence, eq.~\eqref{eq:wishart_like} is no longer exact, and approximations must be introduced.
Here, we consider the Hamimeche \& Lewis (HL) likelihood \cite{hamimecheLikelihoodAnalysisCMB2008a}, which is largely adopted in the literature to address this problem. In a full sky context, eq.~\eqref{eq:wishart_like} can be rewritten as
\begin{equation}
    -2\ln\mathcal{L}(C_\ell|\hat{C}_\ell) = \sum_{\ell\ell^\prime} X_{\ell}^{\sf T} \eval{M^{-1}_{\ell \ell^\prime}}_{\rm fid} X_{\ell^\prime}\ ,
\label{eq:hl_likelihood}
\end{equation}
where $M^{-1}_{\ell \ell^\prime}$ is the inverse covariance matrix of $C_\ell$, evaluated for a fiducial model $C_\ell^{\rm fid}$. Then, $X_{\ell}$ is a function of 
$C_\ell$
\begin{equation}
    X_{\ell} = C_\ell^{\rm fid} \sign\qty(\frac{\hat{C}_\ell}{C_\ell} - 1)\sqrt{2\qty(\frac{\hat{C}_\ell}{C_\ell} - \ln \frac{\hat{C}_\ell}{C_\ell} - 1)} = C_\ell^{\rm fid} g\qty(\frac{\hat{C}_\ell}{C_\ell})\ ,
\label{eq:X_vec_hl}
\end{equation}
where we define the function $g$. Thus, on the complete celestial sphere, this is also an exact likelihood. On top of this, on a cut sky, it allows one to naturally account for couplings between multipoles while considering the non-Gaussian nature of $C_\ell$. Note, however, that $M_{\ell \ell^\prime}$ needs to be estimated with simulations; thus, we will study the dependence of the HL likelihood on the number of simulations used \Nll. In addition, we will explore the Sellentin \& Heavens (SH) correction \cite{sellentin2016ParameterInferenceEstimated} and the Hartlap factor \cite{hartlap2007WhyYourModel} as means to alleviate the effect of low \Nll. The SH correction marginalises over the estimation error of $M_{\ell\ell^\prime}$, replacing the Gaussian $\chi^2$ with $N\ln\qty[1+\chi^2/(N-1)]$, where $N$ stands for \Nll. The Hartlap factor instead rescales the inverse of the estimated covariance by $(N-p-2)/(N-1)$, where $p$ is the size of the data vector.

Being defined in harmonic space, the HL likelihood is very fast to compute and can be easily implemented at arbitrary resolutions, i.e., multipole ranges, with an impact on computation time only when the data vector becomes very large. Thus, we want to assess the accuracy of this approach with respect to the pixel-based method, which acts as our benchmark for the largest scales.

By choosing this likelihood instead of the PB, the complexity of the analysis has essentially been transferred to the power spectrum estimation step (see \autoref{sec:qml}). Aside from the need to estimate the real-space covariance matrix mentioned above, this hides another often overlooked issue: the fact that power spectra estimated on a cut-sky may be negative \cite{Galloni:2025eqk, tegmark1997HowMeasureCMBpower}.

\subsubsection[Dealing with negative values of \texorpdfstring{$\hat{C}_\ell$}{Cl}]{Dealing with negative values of \texorpdfstring{$\boldsymbol{\hat{C}_\ell}$}{Cl}}\label{sec:negative_cls}

Despite all the spectra considered in this work being auto-correlations, the estimates produced by the algorithms of \autoref{sec:qml} are not guaranteed to be positive. Note that the raw quadratic form built by the QML, $\mathbf{m}^{\sf T}\mathbf{C}^{-1}\mathbf{P}^\ell\mathbf{C}^{-1}\mathbf{m}$, is always non-negative \cite{tegmark1997HowMeasureCMBpower}. Negative values arise instead after the deconvolution of eq.~\eqref{eq:estimated_cl_qml}, since the normalisation by $\mathbf{F}^{-1}$ makes each bandpower a linear combination of positive and negative quantities. For the mask used here, every off-diagonal element of $\mathbf{F}^{-1}$ is negative, with the strongest coupling between $\ell=2$ and $\ell=4$. In our simulations, we measure this on the spectra that enter the likelihood, i.e., the deconvolved spectra before the subtraction of the noise bias, which is instead included in the model of eq.~\eqref{eq:model_cl}. With the CosWin (GausWin 640$^\prime$), $1.3\%$ ($0.85\%$) of all the $(\ell,\,{\rm simulation})$ pairs at $\ell\leq32$ are negative, rising to $17\%$ ($9\%$) at $\ell = 2$, while the corresponding raw quadratic estimates are positive in $100\%$ of the realisations at every multipole, as expected. The negative values concentrate at the largest scales, with $79\%$ ($87\%$) of them at $\ell\leq5$. The same mechanism operates for the pseudo-$C_\ell$ estimates at $\ell>32$, with the mode-decoupling matrix $\mathcal{K}_{\ell\ell^\prime}^{-1}$ playing the role of $\mathbf{F}^{-1}$, although in our simulations the spectra entering the likelihood at those multipoles remain positive. Negative spectra have to be dealt with, since they cause a series of problems when applied to eq.~\eqref{eq:X_vec_hl}.

A similar problem was faced in ref.~\cite{mangilli2015LargescaleCMBTemperaturepolarization} when dealing with an HL likelihood defined on cross-spectra, which, by their nature, can be negative. To solve this issue, an offset was introduced in that work to mimic the distribution of an auto-correlation. In particular, this offset was chosen to ensure that 99\% of the simulated $C_\ell$ values are positive. In particular, eq.~\eqref{eq:X_vec_hl} can be recast to
\begin{equation}
    X_{\ell} = \qty(C_\ell^{\rm fid} + O_\ell) g\qty(\frac{\hat{C}_\ell + O_\ell}{C_\ell + O_\ell})\ ,
\label{eq:X_vec_ohl}
\end{equation}
where $O_\ell$ is the offset, and $g$ is substituted by
\begin{equation}
    g(x) \to \sign(x) g(\abs{x})\ .
\end{equation}

Although we are in a different situation, where negative spectra are artefacts of our estimation, we still introduce a similar quantity to regularise the likelihood computation. A related approach is followed by ref.~\cite{Galloni:2025eqk}, where the authors study a new HL-based approximation in a multi-field scenario, showing good accuracy with respect to the PB likelihood. From here on, we will refer to this likelihood as offset Hamimeche \& Lewis (oHL).

\subsection{Hybrid approach}\label{sec:hybrid}

We also consider a hybrid approach to the likelihood. For instance, to obtain a benchmark likelihood over the full angular range, we want to combine the PB likelihood with the oHL at small scales. Splitting the multipole range in this way, with a pixel-based likelihood at low $\ell$ and a harmonic-space one above the transition, is standard practice. Indeed, the \Planck~2015 baseline likelihood is built exactly along these lines \cite{planckcollaboration2016Planck2015ResultsXI}.

\begin{figure}[t]
    \centering
    \includegraphics[width=.5\textwidth]{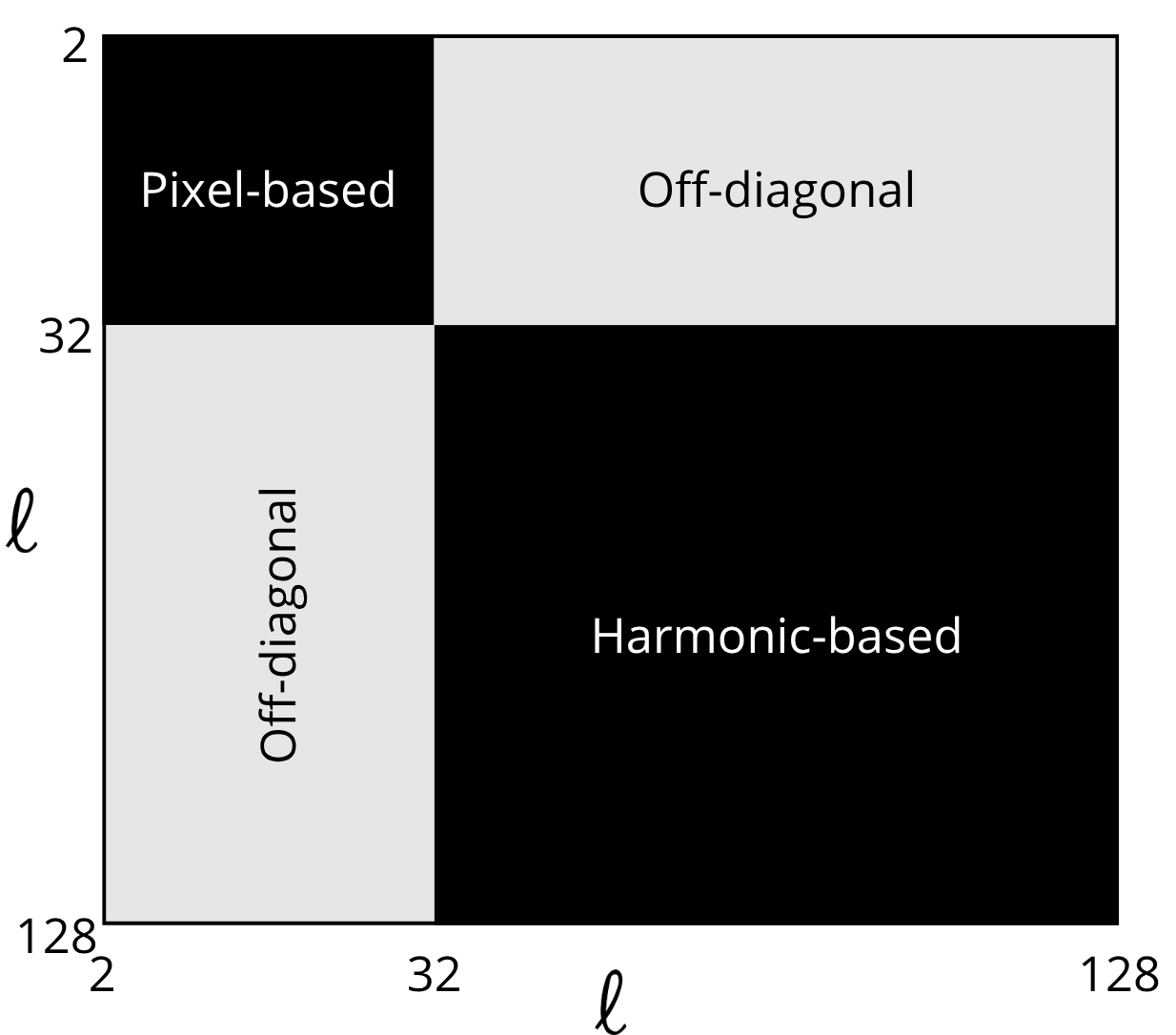}
    \caption{Block structure of the hybrid likelihood over the full multipole range $2\leq\ell\leq128$. The large scales, $2\leq\ell\leq32$, are described by the pixel-based likelihood and the small scales, $33\leq\ell\leq128$, by the harmonic-based one, while the off-diagonal blocks represent the correlation between the two ranges.}
    \label{fig:hybrid_chi2_blocks}
\end{figure}

In order to do this, one could simply sum the $\chi^2$ contributions of the two likelihoods, provided that the PB is limited to the large scales ($\ell \leq 32$) and the HL to the small ones ($33\leq\ell\leq128$); however, in principle, one needs to account for their correlation (see \autoref{fig:hybrid_chi2_blocks}).
Before listing the prescriptions that we test, we introduce the notation for the various $\chi^2$ blocks, which differ in the covariance they are evaluated with. Let $\mathbf{X}$ be the vector of eq.~\eqref{eq:X_vec_ohl} over the full range $2\leq\ell\leq128$, partitioned into a large-scale block $\mathbf{X}_{\rm lo}$ ($\ell\leq32$, built from the QML spectra) and a small-scale one $\mathbf{X}_{\rm hi}$ ($33\leq\ell\leq128$, built from the pseudo-$C_\ell$ spectra). We denote by $\mathbf{M}$ the corresponding harmonic covariance, estimated from simulations as in eq.~\eqref{eq:hl_likelihood}. Then
\begin{equation}
    \chi^{2}_{\rm HL, \ell\leq128} = \mathbf{X}^{\sf T} \eval{\mathbf{M}^{-1}}_{\rm fid} \mathbf{X}
\label{eq:chi2_full_range}
\end{equation}
is the full-range oHL, whereas $\chi^{2}_{\rm HL, \ell\leq32}$ and $\chi^{2}_{\rm HL, 33\leq\ell\leq128}$ denote the same quadratic form restricted to one block and evaluated with the covariance of that block alone, i.e., $\eval{\mathbf{M}_{\rm lo\,lo}^{-1}}_{\rm fid}$ and $\eval{\mathbf{M}_{\rm hi\,hi}^{-1}}_{\rm fid}$, respectively. Each of them is the likelihood one would write for that multipole range in isolation. The cross-term is instead the off-diagonal block of the inverse of the full-range covariance, sandwiched between the two halves of the data vector,
\begin{equation}
    \chi^{2}_{\rm HL, off} = \mathbf{X}_{\rm hi}^{\sf T} \eval{\qty(\mathbf{M}^{-1})_{\rm hi\,lo}}_{\rm fid} \mathbf{X}_{\rm lo}\ .
\label{eq:chi2_offdiag}
\end{equation}
Note that $\chi^{2}_{\rm HL, off}$ is a cross-term between the two blocks and can take either sign. With this notation, we test two ways of combining the two likelihoods.
\begin{enumerate}
    \item The full likelihood $\mathcal{L}(C_\ell|\hat{C}_\ell)$ can be decomposed using conditional probability in such a way that
    \begin{equation}
        \chi^{2} = \chi^{2}_{\ell\leq32} + \chi^{2}_{33\leq\ell\leq128|\ell\leq32}\ ,
    \end{equation}
    where the first term on the right hand side is the $\chi^2$ contribution computed on the large scales, and the second term is the small-scale one conditioned on the large scale part. In our case, the large scale part given by PB, while the second by the oHL (which we condition), hence the name ``hybrid''. For a Gaussian, the conditional term is the difference between the full-range $\chi^2$ and the marginal one of the conditioning block. Hence, in practice we compute $\chi^{2}_{\rm PB}$, $\chi^{2}_{\rm HL, \ell\leq128}$ (which intrinsically contains the off-diagonal terms), and $\chi^{2}_{\rm HL, \ell\leq32}$, and we compose them as
    \begin{equation}
        \chi^{2}_{\rm hyb} = \chi^{2}_{\rm PB} + \chi^{2}_{\rm HL, \ell\leq128} - \chi^{2}_{\rm HL, \ell\leq32}\ ,
    \end{equation}
    where the subtraction removes the large-scale information already carried by $\chi^{2}_{\rm PB}$.
    \item After computing $\chi^{2}_{\rm PB}$ and $\chi^{2}_{\rm HL, 33\leq\ell\leq128}$ as independent contributions, we compute the cross-term in harmonic space $\chi^{2}_{\rm HL, off}$ of eq.~\eqref{eq:chi2_offdiag} and sum it twice, once for each of the two symmetric off-diagonal blocks of \autoref{fig:hybrid_chi2_blocks}, obtaining
    \begin{equation}
        \chi^{2}_{\rm hyb} = \chi^{2}_{\rm PB} + \chi^{2}_{\rm HL, 33\leq\ell\leq128} + 2 \chi^{2}_{\rm HL, off}\ .
    \label{eq:chi2_hyb_sum}
    \end{equation}
\end{enumerate}
The two prescriptions differ only in how they carry over the correlation between the two ranges: the second keeps each block marginal and adds the coupling explicitly, while in the first the full-range inverse covariance accounts for it. Unless stated otherwise, all the hybrid results presented in \autoref{sec:results} are obtained with eq.~\eqref{eq:chi2_hyb_sum}, but we verified that the two methods give almost identical results. Finally, we also verify the role played by the correlation between large and small scales by neglecting the off-diagonal terms altogether, i.e., by dropping $2\chi^{2}_{\rm HL, off}$ from eq.~\eqref{eq:chi2_hyb_sum} and treating the two ranges as independent.

\subsection{Details on posterior evaluation}\label{sec:details_post}

Given that we are ultimately interested in estimating the tensor-to-scalar ratio $r$, we use it as a proxy for the accuracy of the likelihood description. The theoretical spectra entering the likelihoods are computed with \texttt{CAMB} \cite{lewis2000EfficientComputationCMB}, including lensing and assuming the same $\Lambda$CDM parameters used to generate the simulations (\autoref{sec:sims}). The tensor tilt is fixed by the slow-roll consistency relation, $n_{\rm t} = -r/8$ \cite{liddleCOBEGravitationalWaves1992, guzzettiGravitationalWavesInflation2016}, and both the scalar and the tensor spectra are normalised at the pivot scale $k_\ast = 0.05\,\Mpc^{-1}$. In particular, for each simulation, we evaluate the likelihoods on a grid of $r$ values spanning $\qty[0.0, 0.005]$ with 400 steps, which we over-sample to $10^4$ points through spline interpolation to avoid grid-resolution effects.

Assuming a uniform prior on $r$ over the range of this grid, the posterior is proportional to the likelihood, so that the likelihood sampled on the grid, once normalised, gives the posterior directly. We repeat this process for \Nval~simulations, which constitute our validation set. The order of the two operations that follow affects the results quantitatively: we first average the $\chi^2$ over the validation set, and only then convert it into a single posterior,
\begin{equation}
    P(r) \propto \exp\qty[-\frac{1}{2}\expval{\chi^2(r)}_{\text{\Nval}}]\ ,
\label{eq:mean_chi2_posterior}
\end{equation}
from which the MAP and the credible intervals defined below are derived. Equivalently, eq.~\eqref{eq:mean_chi2_posterior} is the geometric mean of the \Nval~individual posteriors, and it describes the uncertainty attainable in a typical realisation. Taking instead the arithmetic mean of the \Nval~posteriors, or of the \Nval~individual credible intervals, gives a different result, and for our default configurations the former would inflate by roughly $40\%$ the uncertainty $\delta_r$ that we introduce in the following. All the numbers quoted in this work follow eq.~\eqref{eq:mean_chi2_posterior}. Finally, we compute the MAP estimate of $r$, the 68\% and 95\% credible intervals using the highest-density interval (HDI) prescription\footnote{Given a posterior $P(\theta|D)$, the 68\% HDI is the shortest interval $\qty[\theta_{\rm low}, \theta_{\rm high}]$ such that
\begin{equation}
    \int_{\theta_{\rm low}}^{\theta_{\rm high}} P(\theta|D)\,\d \theta = 0.68\ , \qq{with} P(\theta_{\rm low}|D) = P(\theta_{\rm high}|D)\ .
\label{eq:hdi}
\end{equation}
The interval becomes automatically one-tailed if one of the two edges crosses a prior boundary.}, along with the first and second moments of the distribution. 
Among these, we use the upper edge of the 68\% HDI as our measure of the uncertainty on the tensor-to-scalar ratio, named $\delta_r$, and we use it consistently throughout this work to compare our results. Since the grid starts at $r=0$, which is our fiducial value, almost all our intervals end up being one-tailed; thus, $\delta_r$ is both the upper edge and the width of the interval, and it coincides with the 68\% upper bound on $r$. The lower edge of the interval departs from zero only in a few cases, which we flag where they occur; for example, the strongly unconverged configurations at the smallest \Npp~and \Nll~shown in appendix~\ref{app: settings}. In those cases, $\delta_r$ remains the upper edge of the interval, and a non-vanishing $r_{\rm low}$ is itself a useful indicator that a configuration has not converged.

\section{Analysis settings}\label{sec:settings}

Besides the choice of estimator and likelihood, the pipeline described so far requires a number of numerical and methodological choices at every stage, which we collectively call ``analysis settings'' and which are indicated on the connecting arrows of \autoref{fig:pipe}. Having introduced all the ingredients, in this section we gather these choices in a single place, in the order in which they enter the pipeline, and we report the ranges we explore.

These settings are not all of the same nature, so we group them into three classes, which we use throughout the rest of this work:
\begin{enumerate}
    \item[(i)] freely optimisable settings, which can be changed at negligible computational cost and thus tuned at will, namely the beam \bl, the regularisation noise \regn, the apodisation scale \apo, the binning scheme \deltal, the multipole ranges \lswitch~and \lmaxabs, and the correction applied to $M_{\ell\ell^\prime}$;
    \item[(ii)] resource-limited settings, whose value is set by the number of simulations one can afford to produce, and which thus carry a concrete cost, \Npp~and \Nll;
    \item[(iii)] validation-only settings, which exist because this analysis is performed on simulations and will not represent an actual choice when analysing real data, \Nval.
\end{enumerate}
This classification is relevant for the interpretation of our results, since a recommendation on a class-(i) setting can be adopted directly, while one on a class-(ii) setting is a statement on how many simulations a future analysis needs. Finally, a dependence on a class-(iii) setting quantifies the precision of the validation itself.

\subsection{Map-level settings: the beam}\label{sec:set_beam}

As anticipated in \autoref{sec:qml_implementation}, the low-resolution maps fed to both the QML and the PB likelihood are obtained by smoothing the \Nside$=64$ MC-NILC maps with a beam \bl~before downgrading them to \Nside$=16$. Since the same maps feed both the large-scale power spectrum estimator and the large-scale benchmark likelihood, the beam is not a QML-specific choice, and we study the dependence of the results on six different beams.

\begin{figure}[t]
    \centering
    \includegraphics[width=.5\textwidth]{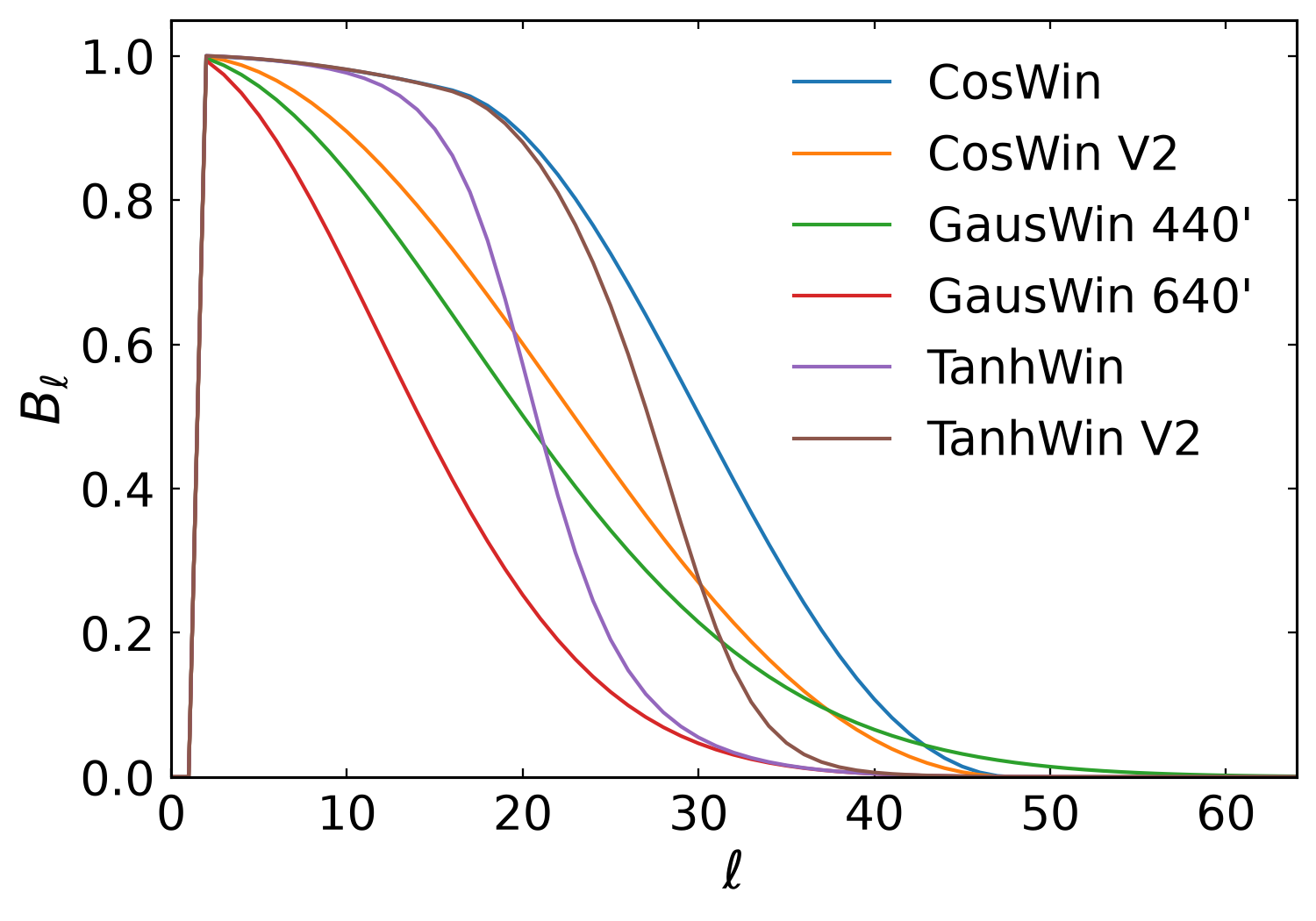}
    \caption{Beams used to downgrade maps and regularise the behaviour of the covariance matrix of pixels on large scales.}
    \label{fig:beams}
\end{figure}

We show these beams in \autoref{fig:beams}: we use two versions of cosine, Gaussian, and $\tanh$ windows. In particular, the cosine windows are defined from
\begin{equation}
    B_\ell = \begin{cases}
        1 \qq{for}\ell\leq\ell_{\rm trans} \\
        \frac{1}{2}\qty[1 + \cos\qty(\frac{\pi(\ell - \ell_{\rm trans})}{\ell_{\rm cut} - \ell_{\rm trans}})] \qq{for}\ell_{\rm trans}<\ell\leq\ell_{\rm cut} \\
        0 \qq{for}\ell>\ell_{\rm cut}
    \end{cases},
\end{equation}
where $\ell_{\rm trans}$ and $\ell_{\rm cut}$ control the regions where the beam is 1 and 0, respectively. After setting $\ell_{\rm cut} = 48$ to avoid harmonic scales greater than $3\Nside$, we test both $\ell_{\rm trans} = 16$ and $1$, named CosWin and CosWin V2, which is more aggressive on the large scales. These kinds of beams were introduced in refs.~\cite{collaborationPlanck2018Results2020, benabed2009teasing} and were widely used in the Planck Collaboration.

In addition, we test two choices for Gaussian window functions (GausWin) obtained using \texttt{healpy}. We use two versions with a FWHM of 440 and 640 arcmin.

Since CosWin and GausWin 640$^\prime$ are the least and most aggressive beams, respectively, and we will show below that the choice of the beam has a significant impact on our analysis, we also define two additional ones to better characterise their differences. In particular, we use a tanh function to weight the CosWin and GausWin, transitioning from one to the other as

\begin{equation}
    B_\ell = \qty(1-f_\ell)\times \text{CosWin} + f_\ell\times \text{GausWin 640$^\prime$}\ ,
\end{equation}
where the weights $f_\ell$ are defined as
\begin{equation}
    f_\ell = \frac{1 + \tanh\qty(\ell - \ell_{\rm trans})}{2}\ .
\end{equation}
Here, we assume two different transition multipoles $\ell_{\rm trans} = 20$ and $30$ to define TanhWin and TanhWin V2.

\subsection{Power spectrum estimation settings}\label{sec:set_cov}

The noise covariance matrix in pixel-space $\mathbf{N}$ must be empirically estimated from simulations. For this reason, we explore this dependence by choosing different values of \Npp, i.e., the number of simulations used to obtain $\mathbf{N}$. In particular, we assume \Npp~$\in \{1,\allowbreak 2,\allowbreak 4,\allowbreak 6,\allowbreak 8,\allowbreak 10,\allowbreak 12,\allowbreak 14,\allowbreak 16,\allowbreak 17\} \times 10^3$ for the QML computation, and \Npp~$\in \{1,\allowbreak 2,\allowbreak 4,\allowbreak 6,\allowbreak 8,\allowbreak 10,\allowbreak 12,\allowbreak 14,\allowbreak 16,\allowbreak 18,\allowbreak 19\} \times 10^3$ for the PB likelihood. The two upper limits reflect how many simulations are left once the validation set has been set aside and, for the QML, once 2000 further simulations are reserved for the harmonic covariance of the oHL.

In order to regularise $\mathbf{N}$, we also add a small isotropic white-noise contribution, which is needed to avoid very small or null eigenvalues. We also study this dependency, assuming \regn~$\in \{0.01,\allowbreak 0.05,\allowbreak 0.1,\allowbreak 0.5,\allowbreak 1,\allowbreak 2,\allowbreak 3,\allowbreak 5,\allowbreak 10,\allowbreak 15\}$~nK-arcmin, the same values for the QML and the PB.

Moving to the small scales, the pseudo-$C_\ell$ estimation requires a choice of mask apodisation and binning scheme. To apodise the mask, we apply Gaussian smoothing with varying FWHM, exploring \apo~$\in \{0,\allowbreak 0.5,\allowbreak 1,\allowbreak 2,\allowbreak 3,\allowbreak 4,\allowbreak 5,\allowbreak 7,\allowbreak 10,\allowbreak 15\}^\circ$. As for the binning choice, we know \textit{a priori} that we will use the QML to estimate the large scales ($\ell \leq 32$); thus, we are interested in assessing the effect of binning the subsequent multipoles.

\begin{figure}[t]
    \centering
    \includegraphics[width=.5\textwidth]{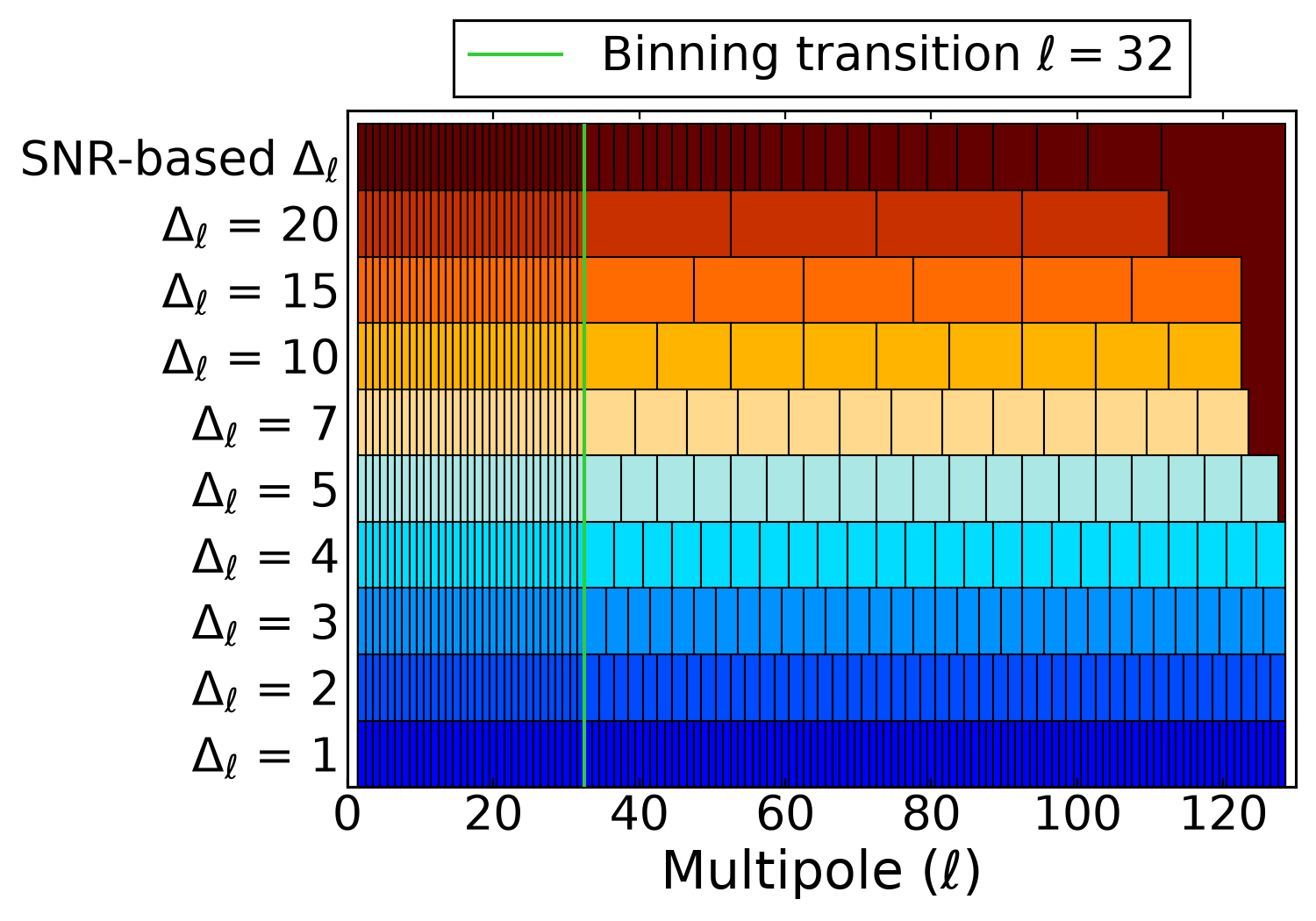}
    \caption{Binning schemes explored. The vertical green line marks the transition from unbinned to binned at $\ell = 32$.}
    \label{fig:binnings}
\end{figure}

Figure~\ref{fig:binnings} showcases our choices. We assume \deltal~$\in \{1,\allowbreak 2,\allowbreak 3,\allowbreak 4,\allowbreak 5,\allowbreak 7,\allowbreak 10,\allowbreak 15,\allowbreak 20\}$ after $\ell = 32$. In addition, we define an SNR-based alternative. Indeed, instead of using constant-width bins, we define them dynamically based on the constraining power on $r$. This naturally results in bins that become larger as the multipoles increase.

\subsection{Likelihood settings}\label{sec:set_like}

The PB likelihood requires two multipole ranges to be specified (see \autoref{sec:pixel-based}), namely \lmaxabs, up to which the signal covariance matrix is built, and \lswitch, below which the power spectrum is allowed to vary with the cosmological parameters. We assume \lmaxabs~$\in \{32,\allowbreak 36,\allowbreak 40,\allowbreak 44,\allowbreak 48,\allowbreak 52,\allowbreak 56,\allowbreak 60,\allowbreak 64\}$ and \lswitch~$\in \{12,\allowbreak 16,\allowbreak 20,\allowbreak 24,\allowbreak 28,\allowbreak 32,\allowbreak 36,\allowbreak 40,\allowbreak 44,\allowbreak 48\}$ to characterise their effect on the results.

On the harmonic side, the oHL likelihood needs the covariance matrix $M_{\ell\ell^\prime}$ of eq.~\eqref{eq:hl_likelihood}, which is itself estimated from simulations; thus, we vary \Nll\ over thirteen values between 100 and 5000, namely 100, 200, 400, 600, 800, 1000, 1200, 1500, 1700, 2000, 2500, 3000, and 5000. For each value, we consider three options for the treatment of the resulting estimation error, i.e., no correction, the SH correction, and the Hartlap factor (see \autoref{sec:harmonic-based}).

\subsection{Validation settings}\label{sec:set_val}

Finally, \Nval~is the number of simulations over which we average the $\chi^2$ before forming the posterior of eq.~\eqref{eq:mean_chi2_posterior}. We vary it over ten values between 100 and 5000, namely 100, 200, 500, 750, 1000, 1500, 2000, 3000, 4000, and 5000. As mentioned above, this is a property of the validation exercise: a real analysis has a single data set, and \Nval~only controls how precisely we can measure the performance of a given configuration.

We emphasise that an important aspect of this analysis is to compartmentalise the use of simulations. Thus, regarding the PB likelihood, the \Npp~and the \Nval~simulations are two non-overlapping sets out of the total $N_{\rm sims}$. This ensures that we do not introduce any unwanted correlation or ``recycling'' of the same information, which could distort the results. Similarly, for the oHL, \Npp~, \Nll~, and \Nval~ are disjoint sets. For example, this means that the noise bias $N_\ell$ used at the likelihood level is obtained from the \Npp~set (see \autoref{sec:qml}), while the systematic part from foregrounds $F_\ell$ is obtained from the \Nll~simulations, ensuring that these spectra are completely independent of the \Nval~ones.

\begin{table}[t]
	\begingroup
	\caption{Summary table with the ranges of the analysis settings explored in this work. The ``Class'' column reports the classification introduced in \autoref{sec:settings}: (i) freely optimisable; (ii) resource-limited; and (iii) validation-only.}
	\label{tab:settings}
	\nointerlineskip
	\footnotesize
	\setbox\tablebox=\vbox{
		\newdimen\digitwidth
		\setbox0=\hbox{\rm 0}
		\digitwidth=\wd0
		\catcode`*=\active
		\def*{\kern\digitwidth}
		\newdimen\signwidth
		\setbox0=\hbox{+}
		\signwidth=\wd0
		\catcode`!=\active
		\def!{\kern\signwidth}
		\halign{\hbox to 0.95in{#\leaderfil}\tabskip=0.6em&
			\hfil#\hfil\tabskip=6pt&
			\hfil#\hfil\tabskip=6pt&
			\hfil#\hfil\tabskip=6pt&
			\hfil#\hfil\tabskip=0pt\cr
			\noalign{\doubleline}
			\omit\hfil \hfil& Class \hfil& Pixel-based \hfil& oHL $\ell\leq32$ \hfil& oHL $33\leq\ell\leq128$ \cr
			\noalign{\vskip 3pt\hrule\vskip 5pt}
			\bl & (i) & [CosWin, GaussWin, TanhWin] &  [CosWin, GaussWin, TanhWin] &  --- \cr
			\Npp& (ii) & $[1000, 19\,000]$ &  $[1000, 17\,000]$ &  --- \cr
			  \regn~$\qty[\text{nK-arcmin}]$ & (i) & $[0.01, 15]$ &  $[0.01, 15]$ &  --- \cr
			\apo~$\qty[{}^\circ]$ & (i) & --- &  --- &  $[0, 15]$ \cr
			\deltal & (i) & --- &  --- &  $[1, 20]$ + SNR-based \cr
			\lswitch & (i) & $[12, 48]$ &  --- &  --- \cr
			\lmaxabs & (i) & $[32, 64]$ &  --- &  --- \cr
			\Nll & (ii) & --- & $[100, 5000]$ & $[100, 5000]$ \cr
			$M_{\ell\ell^\prime}$ Corr. & (i) & --- &  [$\varnothing$, SH, Hartlap] &  [$\varnothing$, SH, Hartlap] \cr
			  \Nval & (iii) & $[100, 5000]$ &  $[100, 5000]$ &  $[100, 5000]$ \cr
			\noalign{\vskip 5pt\hrule\vskip 3pt}}}
	\endPlancktable
	\endgroup
\end{table}

Table~\ref{tab:settings} summarises the analysis settings we will explore in this work, in order of appearance (see also \autoref{fig:pipe}). Given the very high dimensionality of the problem, we typically fix all the parameters except one when investigating a specific dependency. Still, we try to characterise the full space as much as possible. For example, many of the results we will show below are repeated for both CosWin and GausWin 640$^\prime$. Indeed, they are the least and most aggressive beams used; thus, they give us an idea of the variability as a function of the smoothing.

\section{Results}\label{sec:results}

In \autoref{sec:settings}, we have summarised the analysis settings explored in this work. By combining them, we compute 522 posteriors in total. This number excludes the hybrid likelihood, which is itself a combination of some of these posteriors, and other cases that are not relevant here. Since reporting all of them individually is impractical, below we present in detail a reference configuration, which we design in \autoref{sec:benchmark}, and in \autoref{sec:selection} we screen the whole set against it with explicit criteria on precision and accuracy.

\subsection{Choosing the benchmark}\label{sec:benchmark}

\begin{figure}[t]
    \centering
    \includegraphics[width=\textwidth]{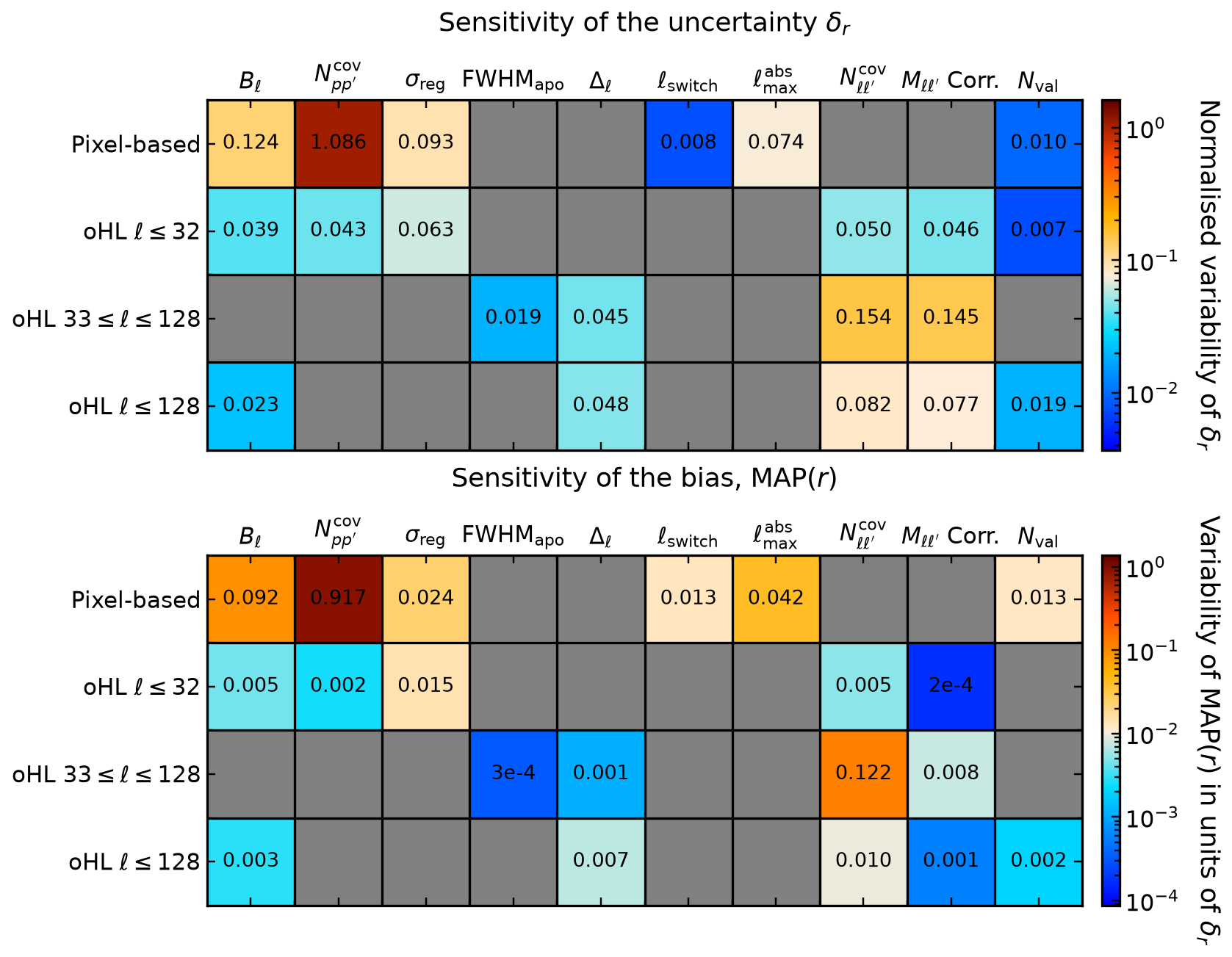}
    \caption{Sensitivity of the two figures of merit to the analysis settings. The upper panel shows the variability of $\delta_r$, normalised within each subgroup by its own mean. The lower one shows the variability of the bias on $r$ about each subgroup's mean in units of the $\delta_r$ of the same subgroup, since the mean of the bias is often exactly zero and cannot be used for the normalisation. Both panels are then dimensionless and share the same scale. Grey pixels indicate that the parameter does not apply to that likelihood, or that we have not explored that dimension of the space of settings. As mentioned in the text, this figure only identifies which settings deserve attention, while \autoref{sec:selection} assesses whether a particular choice of one of them is good.}
    \label{fig:sens_tab}
\end{figure}

In order to assess the ``sensitivity'' of $\delta_r$ to a change in an analysis setting, we compute the standard deviation of $\delta_r$ when a certain setting is varied while the others are fixed. The identification of the subgroup using this logic is not unique; for example, suppose we are interested in the variability of $\delta_r$ when changing \Nll. Then, we have a group of configurations where the correction of the harmonic matrix $M_{\ell\ell^\prime}$ is SH and another where it is the Hartlap factor. Thus, simply taking the standard deviation of these groups is not sufficient, since we are incorrectly including the variability over a parameter different from the one of interest. Thus, before calculating the standard deviation, we normalise the $\delta_r$ values of each case by the mean of the subgroup to which they belong. We repeat the exercise on the bias of $r$, with one change. Since the subgroup mean of the MAP is frequently exactly zero, normalising by it is singular, so we divide instead by the $\delta_r$ of the same subgroup; this is never zero, and it puts the result in the units used by the accuracy criterion of \autoref{sec:selection}. Although this is an empirical way of extracting variability, with no simple interpretation, it provides an idea of which parameters are the most important to consider. We use it only to rank the settings, since the normalised standard deviation has no reference value above which a setting would become unacceptable. It also mixes the values of a setting that a real analysis would consider with those we include only to bracket the range. Finally, it depends on how the subgroups are populated, so two settings with comparable values need not be equally important. We therefore use \autoref{fig:sens_tab} to decide which settings deserve attention, and we leave every statement on whether a particular choice of a setting is good to the explicit criteria of \autoref{sec:selection}.

The results for each likelihood choice and multipole range are shown in \autoref{fig:sens_tab}. It is clear that the most critical parameters are \Npp~and \Nll~for the PB and the oHL, respectively. Thus, overall, the number of simulations available is a crucial aspect of this analysis. This also explains why the correction factor to $M_{\ell\ell^\prime}$ is important, as it allows for the mitigation of this dependence. In addition, the choice of the $B_\ell$ is also relevant, especially for the PB. The bias panel largely follows the same ordering, with two differences. First, \Npp~dominates the bias of the benchmark as strongly as it dominates its width, which makes the accuracy criterion of \autoref{sec:selection} the limiting one for the PB. Second, \lswitch~moves $\delta_r$ by well under a per cent but the bias by nearly twice as much, so a criterion on the width alone would rank it too low. For the oHL at $\ell\leq32$ and over the full range, every entry of the bias panel stays below $0.015\,\delta_r$. However, this value is limited by the clipping discussed in \autoref{sec:selection}: computed on the signed MAP of the grid extended to negative $r$, the corresponding variabilities at $\ell\leq32$ are $0.040$ for \regn, $0.020$ for \bl~and $0.015$ for \Npp, a factor $2.7$ to $4.8$ larger. Hence, the bias of the oHL, although small, does depend on the settings.

Given \autoref{fig:sens_tab}, complemented by appendix~\ref{app: settings} providing more details on the dependency on \bl, \Npp, and \Nll, respectively, in appendix~\ref{app: beam}, appendix~\ref{app: real-space-cov}, and appendix~\ref{app: harm-space-cov}, we are ready to design the reference configuration of our main results. Of course, we have no final guarantee that it will be the absolute truth; however, our analysis allows us to have a very educated guess. 

Since \Npp~is the most critical setting for the PB (see \autoref{fig:sens_tab}), we prioritise maximising it. Under the disjoint-set rule of \autoref{sec:set_val}, \Npp~then takes all the simulations left once the validation set and, for the QML, the harmonic covariance of the oHL are set aside, i.e., 19\,000 for the PB and $17\,000$ for the QML. Indeed, the validation set holds \Nval~$=1000$ simulations, enough to measure the mean of the per-simulation MAP with a sampling error of $0.02\,\delta_r$, while taking as few simulations as possible away from \Npp. For \Nll, instead, we take a value at which the oHL bound has already saturated before applying corrections, \Nll~$=2000$ (see appendix~\ref{app: harm-space-cov}), so that every further simulation goes to \Npp. Since $M_{\ell\ell^\prime}$ is estimated from this finite set, we also correct its estimation error with the SH prescription of \autoref{sec:harmonic-based}, since it marginalises over the estimation error, while the Hartlap factor only rescales the inverse covariance. 

As regards the freely optimisable settings, we keep two windows for \bl, the CosWin and the GausWin 640$^\prime$, which are the least and the most aggressive of the six windows of \autoref{sec:set_beam}; thus, the pair brackets the sensitivity to the smoothing. When a single configuration must be quoted, we take the one with the GausWin 640$^\prime$ beam, since it is the most aggressive of the six windows, and hence the one that most suppresses the power above $4\Nside$ responsible for the drift of the QML estimator (see \autoref{sec:qml_implementation}). We also show that the GausWin 640$^\prime$ beam allows the PB likelihood to converge better to its hypothetical result with \Npp$\to \infty$. Despite this, we show the CosWin results alongside it throughout, since the spread between the two beams is the largest single source of variation in our results. 

The regularisation noise is \regn~$=1$ nK-arcmin for both the PB and the QML, close to the value at which the oHL bound with the GausWin 640$^\prime$ reaches its minimum (see appendix~\ref{app: real-space-cov}). Still, the range of values between 0.5 and 5 nK-arcmin is equally viable. 

For the PB, we set \lmaxabs~$=64$, i.e., $4\Nside$, so that the signal covariance captures essentially all the harmonic content of the smoothed maps, as in the QML (see \autoref{sec:qml_implementation}). Similarly, \lswitch~$=32$ matches the large-scale range $\ell\leq32$ of the oHL. Hence, the PB and the oHL constrain $r$ from the same multipoles, which are also the ones the hybrid likelihood of \autoref{sec:hybrid} takes from the PB. On small scales, we adopt the SNR-based binning of \autoref{sec:set_cov}, whose bins follow the constraining power on $r$ and reduce the number of elements of $M_{\ell\ell^\prime}$ to be estimated from simulations (see appendix~\ref{app: harm-space-cov}). 

Finally, we use the non-apodised mask, \apo~$=0^\circ$. Indeed, the usual motivation for apodising a pseudo-$C_\ell$ mask is the control of the $E$-to-$B$ leakage through purification, which requires a smooth window. In our setup, however, MC-NILC delivers $B$-mode scalar maps directly (see \autoref{sec:compsep}), so no purification is performed and the leakage is already avoided at the component-separation stage. As a consequence, apodising only removes effective sky area and, with it, constraining power on $r$. In the unbinned analysis, $\delta_r$ indeed degrades monotonically as the apodisation scale grows, by $11\%$ between \apo~$=0^\circ$ and $15^\circ$. On the other hand, once the small-scale spectra are binned, the dependence essentially disappears, remaining below $1\%$ across the whole range, consistently with the low sensitivity to \apo~visible in \autoref{fig:sens_tab}. All these settings are listed in \autoref{tab:defaults}, constituting our reference configuration.

\begin{table}[t]
	\begingroup
	\caption{Reference configuration of the analysis settings for the results shown here, designed \textit{a priori} as described in the text.}
	\label{tab:defaults}
	\nointerlineskip
	\footnotesize
	\setbox\tablebox=\vbox{
		\newdimen\digitwidth
		\setbox0=\hbox{\rm 0}
		\digitwidth=\wd0
		\catcode`*=\active
		\def*{\kern\digitwidth}
		\newdimen\signwidth
		\setbox0=\hbox{+}
		\signwidth=\wd0
		\catcode`!=\active
		\def!{\kern\signwidth}
		\halign{\hbox to 0.9in{#\leaderfil}\tabskip=1em&
			\hfil#\hfil\tabskip=10pt&
            \hfil#\hfil\tabskip=10pt&
			\hfil#\hfil\tabskip=10pt&
			\hfil#\hfil\tabskip=0pt\cr
			\noalign{\doubleline}
			\omit\hfil Parameter \hfil& Value \cr
			\noalign{\vskip 3pt\hrule\vskip 5pt}
			\bl & [CosWin, GausWin 640$^\prime$] \cr
			\Npp~(PB)& 19\,000 \cr
			\Npp~(QML)& 17\,000 \cr
            \regn & 1 nK-arcmin \cr
			\apo & $0^\circ$ \cr
            \deltal & SNR-based \cr
			\lswitch & 32 \cr
			\lmaxabs & 64 \cr
			\Nll & 2000 \cr
			$M_{\ell\ell^\prime}$ Corr. & SH \cr
			  \Nval & 1000 \cr
			\noalign{\vskip 5pt\hrule\vskip 3pt}}}
	\endPlancktable
	\endgroup
\end{table}

\subsection{Comparing multipole ranges}\label{sec:res_ranges}

Here, we focus on the results of the various multipole ranges; thus, $\ell \leq 32$ with PB and oHL and $33 \leq \ell \leq 128$ with oHL, together with their ``cross-correlation''. As mentioned above, the standard PB results would be biased by the systematic contribution from foregrounds. As mentioned in \autoref{sec:pixel-based}, in a realistic scenario, it will be extremely challenging to obtain a PB likelihood, which requires a pixel-by-pixel estimate of the systematic signal. However, the PB acts as a benchmark for the large scales in the context of this paper, to which we can compare the oHL. We intend to address in future work whether a pixel-based likelihood can be made viable on real data, and at what cost in accuracy. On small scales, we do not have a benchmark distribution to compare with; however, the oHL likelihood has already been validated in this regime \cite{hamimecheLikelihoodAnalysisCMB2008a}.

\begin{figure}[t]
    \centering
    \includegraphics[width=.49\textwidth]{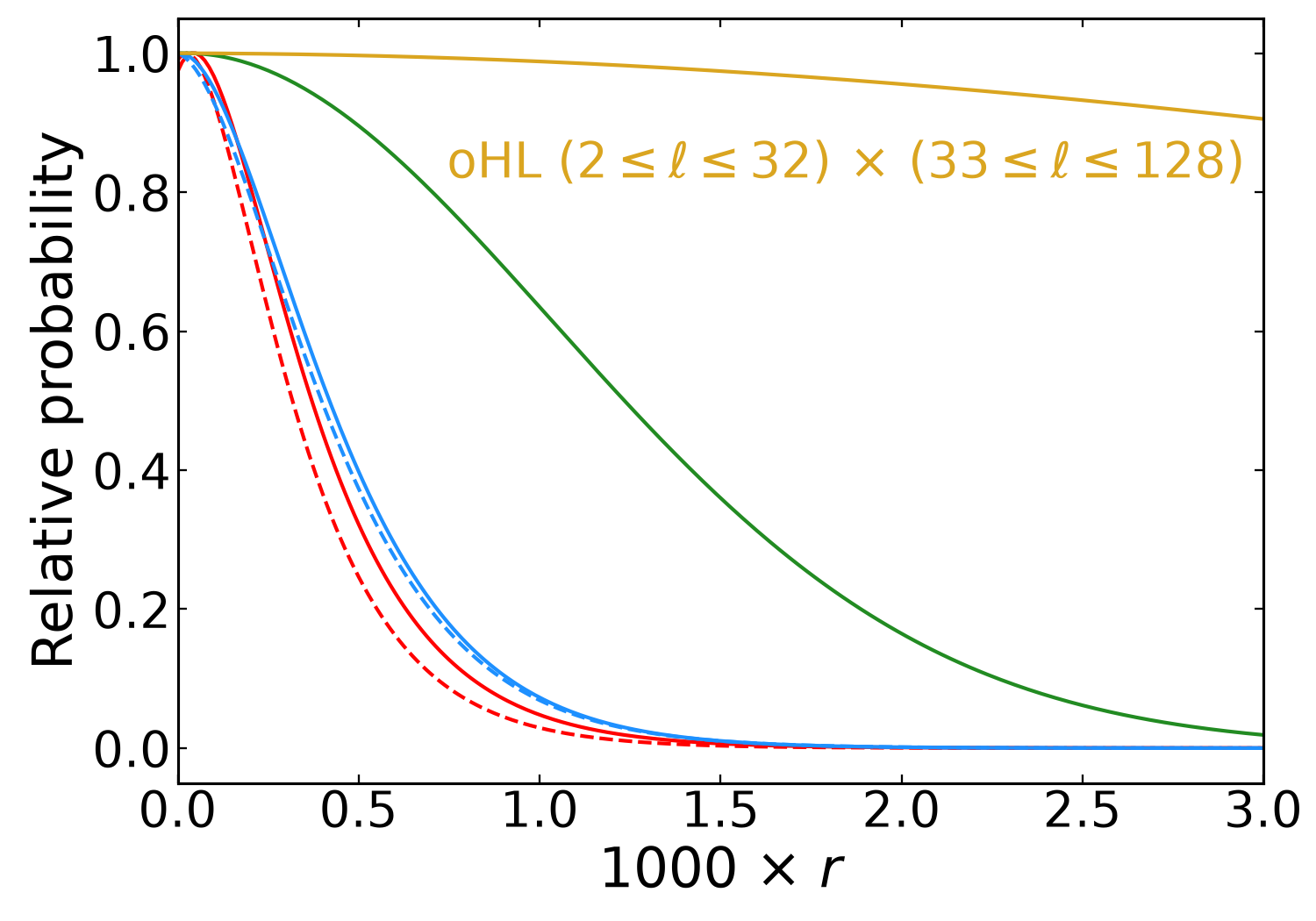}
    \includegraphics[width=.49\textwidth]{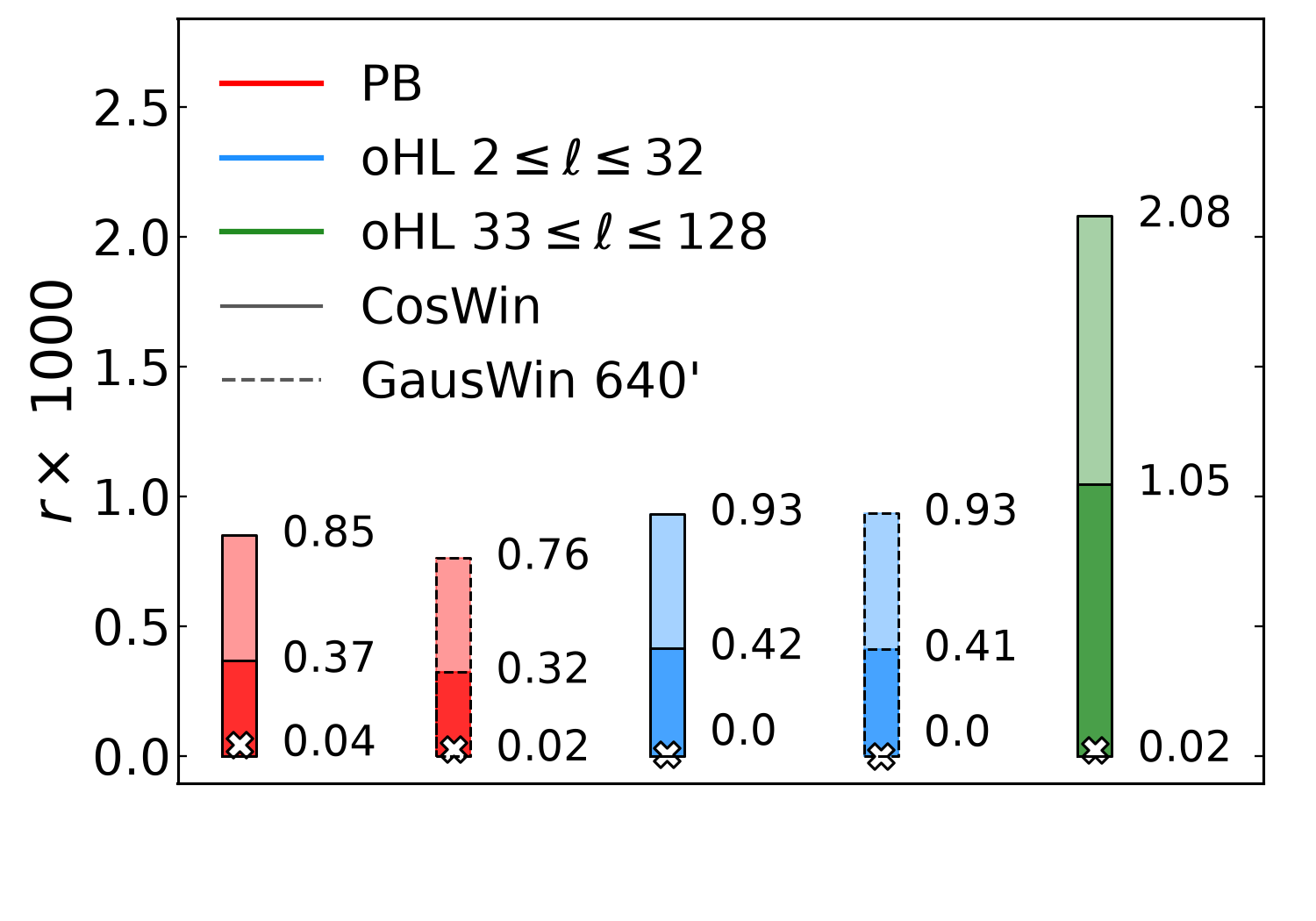}
    \caption{Comparison between PB and oHL on different multipole ranges. Colour identifies the likelihood. Solid (dashed) lines and box outlines refer to the CosWin (GausWin 640$^\prime$) beam, when applicable. The oHL always carries the SH correction to $M_{\ell\ell^\prime}$, which is therefore not repeated in the labels. The left panel shows the posteriors, while the right one shows the MAP together with the 68\% and 95\% HDIs. The yellow curve in the left panel is the cross-correlation between the two multipole blocks, $(2\leq\ell\leq32)\times(33\leq\ell\leq128)$, which is nearly flat over the sampled range of $r$, so its bound is not contained within that range and is not shown in the right panel.}
    \label{fig:comparison_ranges}
\end{figure}

Figure~\ref{fig:comparison_ranges} shows the results for all the ranges of multipoles mentioned above. Starting from the large scale posteriors—the PB and the oHL for $\ell \leq 32$, we note that the assumed \bl~has a stronger impact on the PB rather than the oHL, as expected from \autoref{fig:sens_tab}. All the numerical values of these bounds are reported in the right panel, but let us emphasise that the difference between PB and oHL in $\delta_r$ is 13.5\% for the CosWin case, while using the GausWin 640$^\prime$, it increases to 26.6\% due to increased sensitivity on the PB. Note that this is unexpected, as the Gaussian window is the most aggressive one. We investigated this behaviour, which seems to be due to the scales $20\lesssim\ell\leq32$, where a more aggressive smearing effectively reduces the variance at the pixel level. More details on this can be found in appendix~\ref{app: beam}.

In terms of bias, on the largest scales, the oHL returns ${\rm MAP}(r)=4.5\times10^{-6}$ with the CosWin and a MAP indistinguishable from zero with the GausWin 640$^\prime$, both limited by the $r\geq0$ boundary of the grid, whose effect we quantify below. On the other hand, the benchmark returns ${\rm MAP}(r)=4.4\times10^{-5}$ with the CosWin and $2.5\times10^{-5}$ with the GausWin 640$^\prime$, i.e., $0.12$ and $0.08$ times its own $\delta_r$ (see \autoref{tab:results}). As we show in appendix~\ref{app: real-space-cov}, this residual bias of the PB scales as $1/$\Npp~and extrapolates to a floor of $0.03\,\delta_r$, common to the two windows and left by the map-level debiasing. Hence, most of it is a consequence of estimating $\mathbf{N}$ from a finite set of simulations.

The vanishing MAP of the oHL is set by the boundary of the grid on which the posterior is sampled. Since this grid of $r$ values starts at $r=0$, a simulation that would prefer $r<0$ piles up at the boundary, and eq.~\eqref{eq:mean_chi2_posterior}, being the geometric mean of the individual posteriors, is therefore pulled to zero as soon as most simulations peak there. To measure the bias hidden by the boundary, we extend the grid to negative $r$, as far as the offset keeps the denominator $C_\ell + O_\ell$ of eq.~\eqref{eq:X_vec_ohl} positive, i.e., down to $1.3$--$1.5\,\delta_r$ below zero across our configurations, while the posteriors we quote as results retain the physical $r\geq0$ range. With the extended grid, the pile-up disappears, as the fraction of simulations peaking at its edge drops from $58\%$ to $0.5\%$. The mean of the per-simulation MAP becomes $-0.005$ with the CosWin and $-0.037$ in units of $\delta_r$ with the GausWin 640$^\prime$, against a sampling error of $0.02$ in the same units over our \Nval~$=1000$ simulations; thus, the bias of the oHL on the largest scales is compatible with zero. Moreover, moving the lower edge from $1.0$ to $1.3\,\delta_r$ below zero shifts this mean by $0.004$, a fraction of the sampling error.

In addition, \autoref{fig:comparison_ranges} shows that the vast majority of the constraining power is stored in the large scales, as expected from \autoref{fig:mean_spectra}. Indeed, the small ones alone give $\delta_r = 1.05\times10^{-3}$. Despite this, it is interesting to notice that the cross-correlation does contain some information on $r$, although it is completely negligible with respect to both of the ``auto-correlations'' of the multipole ranges. We do not show the numerical value for the cross-term because our bound is ill-defined in that case. Indeed, our range in $r$ reaches 0.005, but this posterior has a non-negligible volume beyond that point.

\subsection{Full angular range}\label{sec:res_full}

When combining large and small scales, we build the hybrid likelihood to obtain a benchmark for comparing the oHL likelihood (see \autoref{sec:hybrid}). Similarly to the PB, the hybrid likelihood will be particularly sensitive to the systematic contribution of foreground residuals; thus, one needs to consider the debiased PB at large scales when composing $\chi^2_{\rm hyb}$.

\begin{figure}[t]
    \centering
    \includegraphics[width=.49\textwidth]{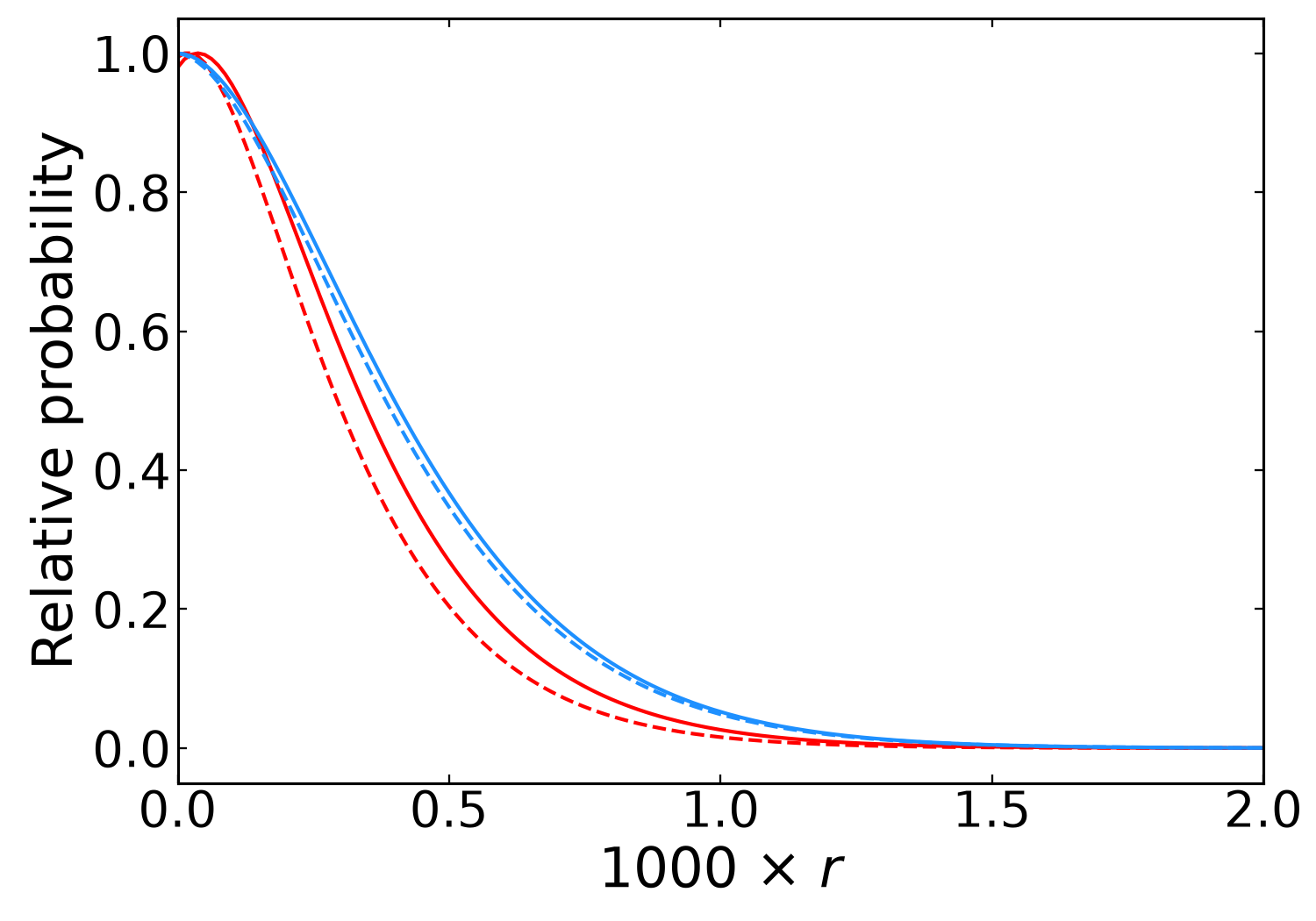}
    \includegraphics[width=.49\textwidth]{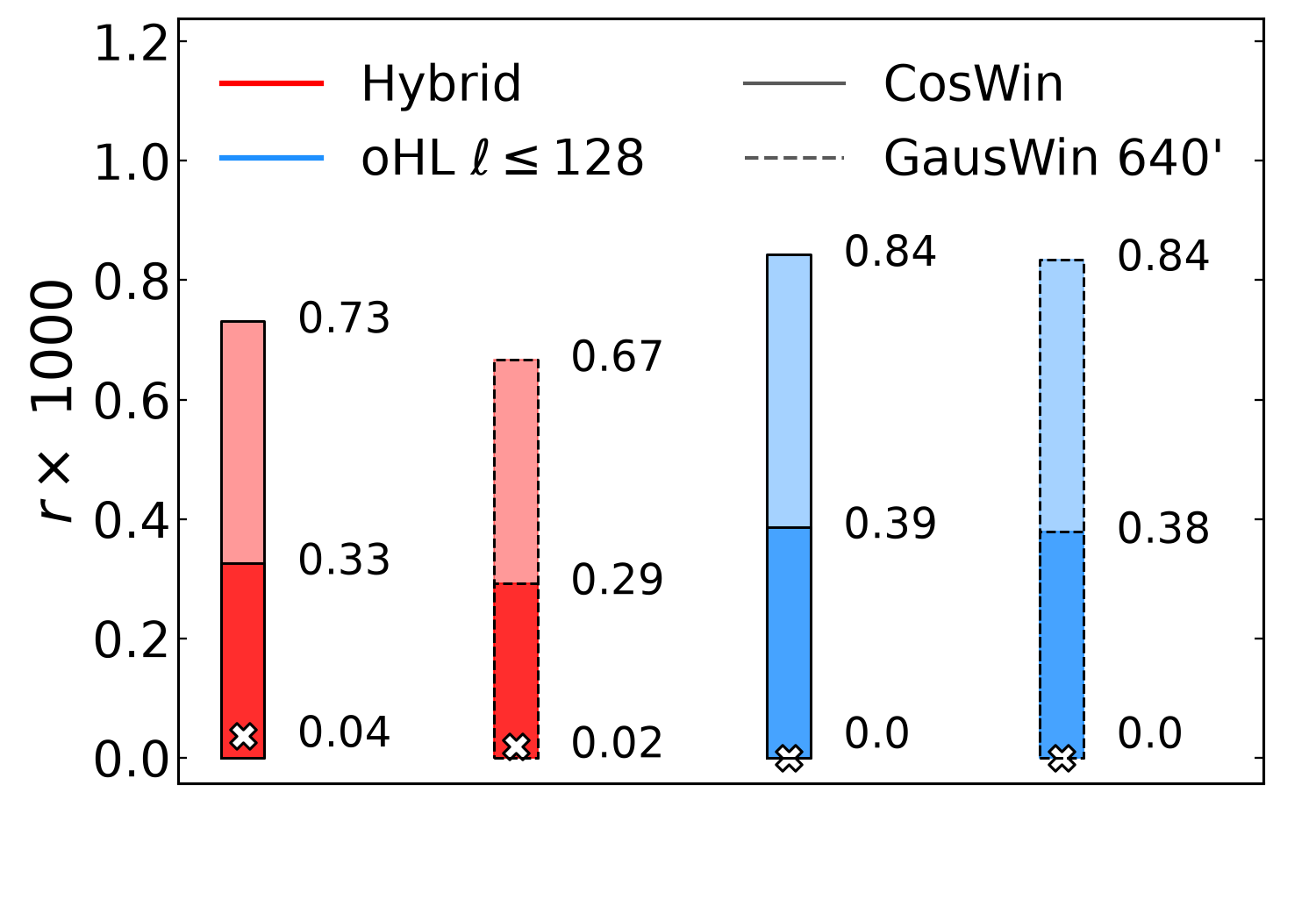}
    \caption{Comparison between hybrid and oHL results on the full angular range $\ell \leq 128$. Colours and line styles are as in \autoref{fig:comparison_ranges}, with colour identifying the likelihood and solid (dashed) lines the CosWin (GausWin 640$^\prime$) beam.}
    \label{fig:comparison_full}
\end{figure}
Figure~\ref{fig:comparison_full} shows the comparison between this hybrid likelihood and the oHL one computed over the full angular range ($\ell \leq 128$), assuming both CosWin and GausWin 640$^\prime$ as the window functions. Note that, similarly to \autoref{sec:res_ranges}, the latter gives better results in terms of $\delta_r$ compared to the former, despite being more aggressive.

Although small, the common constraining power from small scales has helped to shrink both the oHL and our benchmark. In particular, there is a 18.6\% (29.4\%) difference between the two using the CosWin (GausWin 640$^\prime$). This is larger than the 13.5\% (26.6\%) found at $\ell\leq32$, even though the small-scale data is common to both likelihoods. Indeed, adding it lowers the benchmark and oHL bounds by comparable absolute amounts, $0.400$ and $0.290$ with the CosWin and $0.310$ for both with the GausWin 640$^\prime$, in units of $10^{-4}$. Hence, the percentage difference grows mainly because its denominator, the benchmark bound, becomes smaller. On the other hand, the oHL is converged in \Nll~at the adopted value to about $1\%$ over both $\ell\leq32$ and $\ell\leq128$ (see appendix~\ref{app: harm-space-cov}), and dropping the correlation between the ranges from the benchmark changes its bound by $1.4\%$.

The comparison in terms of bias is similar to the one on the largest scales. Indeed, over the full range the oHL gives ${\rm MAP}(r)=0$ for both beams, while the hybrid benchmark returns $3.7\times10^{-5}$ (CosWin) and $1.9\times10^{-5}$ (GausWin 640$^\prime$), that is $0.11$ and $0.07$ times its own $\delta_r$. This is expected: the benchmark inherits the large-scale PB over $\ell\leq32$, and with it the finite-\Npp~bias discussed above. In terms of the per-simulation MAP distribution, the oHL has $57\%$ (CosWin) and $59\%$ (GausWin 640$^\prime$) of the simulations peaking at $r=0$, against $50\%$ and $52\%$ for the hybrid. Therefore, the residual bias of the benchmark sets the floor at which the bias on $r$ of the two likelihoods can be compared. As regards the definition of the hybrid likelihood and the role of the cross-correlation (see \autoref{sec:hybrid}), we find identical results when changing the definition and negligible differences when neglecting the correlation between blocks.

Figure~\ref{fig:comparison_full} represents the main result of this work. Indeed, the 68\% upper bound on $r$ from the oHL, i.e., a harmonic-based likelihood, is 18.6\% (CosWin) and 29.4\% (GausWin 640$^\prime$) larger than that of our benchmark. These numbers shrink to 13.5\% and 26.6\% when the comparison is isolated to its source, i.e., the large scales. Table~\ref{tab:results} reports the numbers behind \autoref{fig:comparison_ranges} and \autoref{fig:comparison_full}, all evaluated on the same posterior grid and with the same estimator of the interval, so the ratios quoted in this section can be reproduced directly from the table. 

\begin{table}[t]
	\begingroup
	\caption{Numerical summary of the results of this section, for the reference configuration of \autoref{tab:defaults} and both beams. The quantity $\delta_r$ is the upper edge of the 68\% highest-density interval of eq.~\eqref{eq:hdi} and $\delta_r^{95}$ that of the 95\% one, while ${\rm MAP}(r)$ measures the bias, since all simulations are generated at $r=0$. The benchmark is the PB likelihood on large scales and the hybrid likelihood over the full range. All values are quoted to three decimal places, which is the precision supported by the sampling of the posterior: the grid on which the intervals are evaluated has a step of $5\times10^{-7}$ in $r$, so a ratio of two entries of this table is meaningful to about $0.3$ percentage points.}
	\label{tab:results}
	\nointerlineskip
	\footnotesize
	\setbox\tablebox=\vbox{
		\newdimen\digitwidth
		\setbox0=\hbox{\rm 0}
		\digitwidth=\wd0
		\catcode`*=\active
		\def*{\kern\digitwidth}
		\newdimen\signwidth
		\setbox0=\hbox{+}
		\signwidth=\wd0
		\catcode`!=\active
		\def!{\kern\signwidth}
		\halign{\hbox to 1.05in{#\leaderfil}\tabskip=0.6em&
			\hfil#\hfil\tabskip=8pt&
			\hfil#\hfil\tabskip=8pt&
			\hfil#\hfil\tabskip=8pt&
			\hfil#\hfil\tabskip=0pt\cr
			\noalign{\doubleline}
			\omit\hfil \hfil& $\delta_r\times10^{4}$ \hfil& $\delta_r^{95}\times10^{4}$ \hfil& ${\rm MAP}(r)\times10^{4}$ \hfil& ${\rm MAP}(r)/\delta_r$ \cr
			\noalign{\vskip 3pt\hrule\vskip 5pt}
			\noalign{\hbox{$\ell\leq32$, CosWin}\vskip 2pt}
			\quad PB & $3.660$ & $8.496$ & $0.435$ & $0.119$ \cr
			\quad oHL & $4.155$ & $9.331$ & $0.045$ & $0.011$ \cr
			\noalign{\vskip 4pt\hbox{$\ell\leq32$, GausWin 640$^\prime$}\vskip 2pt}
			\quad PB & $3.235$ & $7.621$ & $0.245$ & $0.076$ \cr
			\quad oHL & $4.095$ & $9.336$ & $0.000$ & $0.000$ \cr
			\noalign{\vskip 4pt\hbox{$\ell\leq128$, CosWin}\vskip 2pt}
			\quad Hybrid & $3.260$ & $7.311$ & $0.370$ & $0.113$ \cr
			\quad oHL & $3.865$ & $8.426$ & $0.000$ & $0.000$ \cr
			\noalign{\vskip 4pt\hbox{$\ell\leq128$, GausWin 640$^\prime$}\vskip 2pt}
			\quad Hybrid & $2.925$ & $6.666$ & $0.190$ & $0.065$ \cr
			\quad oHL & $3.785$ & $8.351$ & $0.000$ & $0.000$ \cr
			\noalign{\vskip 5pt\hrule\vskip 3pt}}}
	\endPlancktable
	\endgroup
\end{table}

The configuration we simulate is designed for a total $\delta_r<10^{-3}$ \cite{litebirdcollaboration2022ProbingCosmicInflationLiteBIRDa}, a budget that accounts for statistical noise, instrumental systematics, residual foregrounds and lensing. All the values of $\delta_r$ in \autoref{tab:results} lie well below this target. Even though the analysis carried out in this work is quite complex and represents an end-to-end pipeline from simulating maps to the $r$ inference, it is still idealised; thus, the values of $\delta_r$ we obtain cannot be translated into the margin left to \textit{LiteBIRD} for the other contributions. First, our error bars include the scatter of the foreground residuals left by the component separation, since these enter both the simulations and the covariance $\mathbf{N}$ (see \autoref{sec:qml}), but they include no instrumental systematics. Second, our residual foregrounds come from a single sky model, with the tracer of the component-separation method held fixed across the simulations (see \autoref{sec:compsep}); the impact of a more realistic foreground complexity is the subject of a dedicated study in preparation (\textit{LiteBIRD} Collaboration, in prep.). Third, the mission has since been redesigned around a single telescope covering 12 frequency bands, with a revised objective of $\delta_r<2\times10^{-3}$ \cite{LiteBIRD:2026gdf} and a revised instrument whose noise our simulations do not describe. This is the reason why, throughout this work, we used $\delta_r$ to compare the likelihoods with one another, but we purposely place no emphasis on its absolute value. The results above should be read with this in mind.

\subsection{Choosing analysis settings}\label{sec:selection}

The results presented so far correspond to the reference configuration designed in \autoref{sec:benchmark}, i.e., a single point of the space of settings summarised in \autoref{tab:settings} fixed beforehand. In this section we use the whole space, and ask which values of each setting keep the results within tolerance of this reference. These are the values that a future analysis may adopt at no cost for the inference. This requires criteria that are fixed before the answer is known and are expressed in units that are easy to interpret. We adopt two.

The first criterion is on precision: for every configuration we compare its $\delta_r$ with the one obtained from the reference value of the setting being varied, all other settings held fixed, and require
\begin{equation}
    \abs{\frac{\delta_r}{\delta_r^{\rm ref}} - 1} \leq 10\%\ ,
    \label{eq:screen_precision}
\end{equation}
where the reference value is the one of \autoref{tab:defaults}. The tolerance is set by the size of the effect we measure. Indeed, the difference between the oHL and the benchmark measured in this work is $13$--$29\%$ across the two multipole ranges and the two beams of \autoref{tab:results}. A setting able to move $\delta_r$ by more than $10\%$ would therefore act at the same order as the effect under study, while one that moves it by less cannot affect the comparison. Note that eq.~\eqref{eq:screen_precision} is two-sided, since an anomalously tight bound usually signals an underestimated covariance; thus, the corresponding posterior is too narrow. As a consequence, a configuration is retained when it reproduces the reference within the tolerance, and rejected when it departs from it in either direction.

The second criterion is on accuracy: since all our simulations are generated at $r=0$, the MAP is a direct measure of the bias, and we require it to be small compared with the width of the posterior itself,
\begin{equation}
    \frac{\abs{{\rm MAP}(r)}}{\delta_r} \leq \alpha\ , \qq{with} \alpha = 0.2\ .
    \label{eq:screen_accuracy}
\end{equation}
A bias $b$ shifts the estimate while the statistical uncertainty spreads it, and the two combine into a mean squared error $\delta_r^2+b^2$, following the convention adopted to assess the systematic requirements on $r$ for \textit{LiteBIRD} \cite{LiteBIRD:2024eau}. Hence, at $b=0.2\,\delta_r$ the root-mean-square distance from the true value exceeds $\delta_r$ by $2\%$, and we take this as the point beyond which a bias becomes relevant. Both criteria are always evaluated on the highest-density interval of eq.~\eqref{eq:hdi}, never on a bound obtained by integrating the posterior from $r=0$. As mentioned above, the two coincide whenever the posterior is one-sided, but they differ for the configurations whose posterior peaks away from the origin, which are precisely those the screening must measure correctly.

This exercise has two limitations. First, eq.~\eqref{eq:screen_accuracy} uses the MAP of the averaged $\chi^2$, which is clipped at the $r\geq0$ boundary, so the criterion cannot detect a negative bias at all. For this reason we also check it on the grid extended to negative $r$ of \autoref{sec:res_ranges}, where the per-simulation MAP is signed. Over those configurations the signed bias spans $-0.083$ to $+0.082\,\delta_r$ with a median of $-0.018$, so all configurations remain far from $\alpha=0.2$. Second, due to Monte Carlo noise, $\delta_r$ itself scatters at the level of a few per cent between neighbouring values of a resource-limited setting. The $10\%$ tolerance is well above the typical scatter but close to its extremes; hence, we read eq.~\eqref{eq:screen_precision} only as a statement that the effect of a setting is below the ten per cent level.

\begin{figure}[t]
    \centering
    \includegraphics[width=\textwidth]{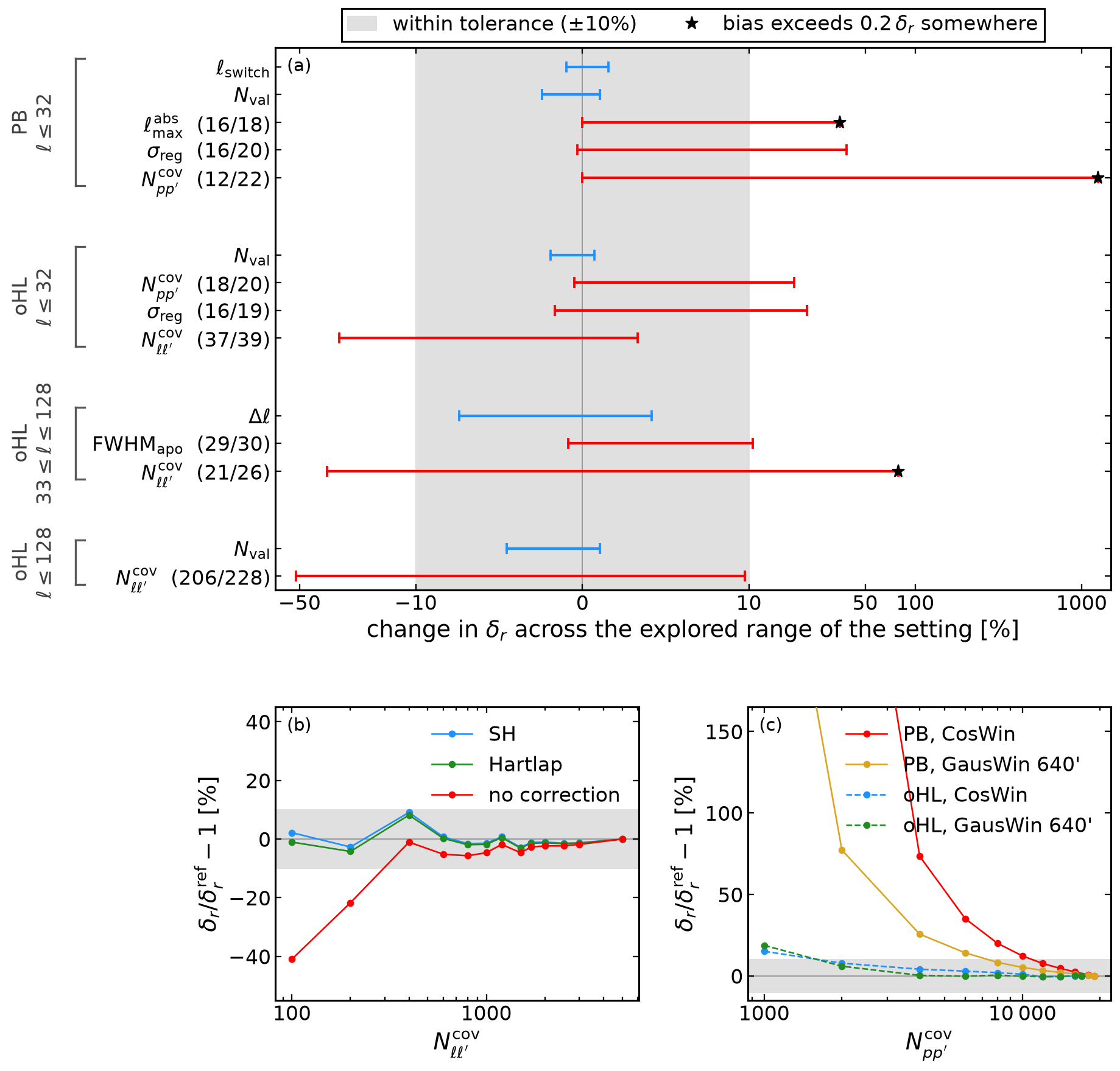}
    \caption{Outcome of the screening. (a) For each setting and each likelihood and multipole range, the range over which $\delta_r$ moves, relative to its reference value, as that setting is varied over all the values we explored. The shaded band is the tolerance of eq.~\eqref{eq:screen_precision}. Bars are blue where every value passes both criteria and red otherwise, with the number that pass given in parentheses, and a star marks the settings for which the bias exceeds $0.2\,\delta_r$ for at least one value, violating eq.~\eqref{eq:screen_accuracy}. The horizontal axis is linear inside the tolerance and logarithmic outside it. (b) The same quantity for \Nll~alone, under the three treatments of the estimation error on $M_{\ell\ell^\prime}$. With the SH or Hartlap correction the bound is stable to a few per cent from a few hundred simulations upward, while without any correction it approaches convergence from below, being $41\%$ too tight at \Nll~$=100$. (c) The same for \Npp. The benchmark approaches its converged value from above and has not flattened at the largest value we can afford, whereas the oHL, which uses \Npp~only through the noise bias of the QML spectra, is already stable at \Npp~$\simeq2000$. \bl~does not appear in (a) since, having no single reference value in \autoref{tab:defaults}, the six windows are ranked against one another, as shown in \autoref{fig:upper_bounds_beams}.}
    \label{fig:screening}
\end{figure}

Figure~\ref{fig:screening} shows the outcome configuration by configuration, and \autoref{tab:screening} summarises it setting by setting. Across the four likelihood and multipole-range combinations we present, the criteria retain the majority of the configurations (469 out of 522, distributed more or less evenly among different likelihoods and different multipole ranges). The distribution of the failures is more informative than their number, and \autoref{fig:screening} shows it clearly. In particular, every failure of the accuracy criterion belongs either to the PB benchmark or to the oHL restricted to small scales at the lowest \Nll. The oHL at $\ell\leq32$ and over $\ell\leq128$ always passes eq.~\eqref{eq:screen_accuracy}, so that in these two cases only the precision criterion rejects configurations. This extends to the whole explored space the conclusion reached in \autoref{sec:res_ranges} and \autoref{sec:res_full} for the default configuration. Panel (a) of \autoref{fig:screening} also shows that the two criteria are complementary, since for most settings the width is the first to exceed the tolerance, while for \Npp, \lmaxabs~and \Nll~at $33\leq\ell\leq128$ it is the bias, marked with a star. Hence, a criterion on $\delta_r$ alone would have accepted these configurations.

Panel (a) of \autoref{fig:screening} leaves out \bl, as explained in its caption. To include \bl~in the totals above, we therefore screen each window against the GausWin 640$^\prime$, which \autoref{sec:res_ranges} designates as the reference whenever a single configuration must be quoted. This screening agrees with the ranking of the windows against one another in \autoref{fig:upper_bounds_beams} and \autoref{tab:screening}: in both cases the GausWin 640$^\prime$ is the best choice, and the CosWin is acceptable wherever $M_{\ell\ell^\prime}$ is corrected.

With this reference, most configurations in which \bl~is varied pass, and the two that fail both use the CosWin and both fail on precision. The first is the PB benchmark, at $+13\%$, which reflects a point already emphasised above. Indeed, the spread between the CosWin and the GausWin 640$^\prime$ is the largest single source of variation in our results, and, as expected, a tolerance calibrated on the $13$--$29\%$ difference between the oHL and the benchmark flags it. The second is the oHL at $\ell\leq32$, at $-26\%$, a spuriously tight bound of the kind that the two-sided form of eq.~\eqref{eq:screen_precision} is meant to catch. It only arises where $M_{\ell\ell^\prime}$ is left uncorrected, since the same window gives $+2.6\%$ and $-6.1\%$ in the two scans that apply the SH correction, and $-8.4\%$ with the Hartlap factor. Thus, in this case the screening rejects the treatment of the covariance, on its own terms.

\begin{table}[t]
	\begingroup
	\caption{Outcome of the screening of eq.~\eqref{eq:screen_precision} and eq.~\eqref{eq:screen_accuracy}. The fourth column gives the largest fractional change in $\delta_r$, relative to its reference value, found anywhere in the explored range of that setting; the fifth gives the values that satisfy both criteria. Classes are those of \autoref{sec:settings}: (i) freely optimisable; (ii) resource-limited; and (iii) validation-only.}
	\label{tab:screening}
	\nointerlineskip
	\footnotesize
	\setbox\tablebox=\vbox{
		\newdimen\digitwidth
		\setbox0=\hbox{\rm 0}
		\digitwidth=\wd0
		\catcode`*=\active
		\def*{\kern\digitwidth}
		\newdimen\signwidth
		\setbox0=\hbox{+}
		\signwidth=\wd0
		\catcode`!=\active
		\def!{\kern\signwidth}
		\halign{\hbox to 0.7in{#\leaderfil}\tabskip=0.6em&
			\hfil#\hfil\tabskip=6pt&
			\hfil#\hfil\tabskip=6pt&
			\hfil#\hfil\tabskip=6pt&
			#\hfil\tabskip=0pt\cr
			\noalign{\doubleline}
			\omit\hfil \hfil& Class \hfil& Range \hfil& $\max\abs{\Delta\delta_r}/\delta_r$ \hfil& Values that pass \cr
			\noalign{\vskip 3pt\hrule\vskip 5pt}
			\bl & (i) & PB $\ell\leq32$ & $13\%$ & five of six; GausWin 640$^\prime$ best, CosWin at $13\%$ \cr
			\bl & (i) & oHL $\ell\leq32$ & $26\%$ & all six with SH or Hartlap; CosWin only if corrected \cr
			\Npp & (ii) & PB $\ell\leq32$ & $>10\times$ & $\geq8000$ (GausWin 640$^\prime$), $\geq12\,000$ (CosWin) \cr
			\Npp & (ii) & oHL $\ell\leq32$ & $19\%$ & $\geq2000$ \cr
			\regn & (i) & PB $\ell\leq32$ & $39\%$ & $0.01$--$5$ nK-arcmin \cr
			\regn & (i) & oHL $\ell\leq32$ & $22\%$ & $0.1$--$10$ nK-arcmin \cr
			\apo & (i) & oHL $33\leq\ell\leq128$ & $11\%$ & $0$--$10^\circ$ unbinned; all, if binned \cr
			\deltal & (i) & oHL $33\leq\ell\leq128$ & $7\%$ & all, including SNR-based \cr
			\lswitch & (i) & PB $\ell\leq32$ & $2\%$ & all \cr
			\lmaxabs & (i) & PB $\ell\leq32$ & $35\%$ & $\geq40$ \cr
			\Nll & (ii) & oHL $\ell\leq32$ & $29\%$ & $\geq100$ (SH, Hartlap); $\geq400$ (none) \cr
			\Nll & (ii) & oHL $33\leq\ell\leq128$ & $79\%$ & $\geq400$ (SH); $\geq600$ (none) \cr
			\Nll & (ii) & oHL $\ell\leq128$ & $53\%$ & $\geq400$ (SH, Hartlap); $\geq1000$ (none) \cr
			$M_{\ell\ell^\prime}$ Corr. & (i) & oHL, all ranges & --- & SH or Hartlap \cr
			\Nval & (iii) & all & $5\%$ & all \cr
			\noalign{\vskip 5pt\hrule\vskip 3pt}}}
	\endPlancktable
	\endgroup
\end{table}

Looking at \autoref{tab:screening} setting by setting, we can separate the explored space into three groups. The settings of the first group are irrelevant at our tolerance and can be chosen on other grounds: \lswitch~moves $\delta_r$ by at most $2\%$ anywhere in $\qty[12, 48]$, \Nval~by at most $5\%$, and the binning by at most $7\%$ relative to the SNR-based scheme, so every value we explored passes. However, for the binning this number depends on the reference, since measured against the unbinned analysis the widest constant-width bin costs $12.5\%$; hence, a future analysis should prefer an SNR-based scheme to a large fixed $\Delta\ell$. 

On the other hand, the settings of the second group pass only within a safe interval. In particular, the regularisation noise is tolerable over $0.1$--$5\,$nK-arcmin for both likelihoods, and fails on either side of this interval for opposite reasons: too little regularisation leaves the smallest eigenvalues of $\mathbf{N}$ ill-determined and inflates $\delta_r$ by up to $22\%$ for the oHL, while too much adds real variance and inflates it by up to $39\%$ for the PB. Across the explored range the two figures of merit pull in opposite directions, with the PB bias falling by a factor of 2 from $0.01$ to $15\,$nK-arcmin while $\delta_r$ grows, so the default of $1$ nK-arcmin is a compromise between the two. The oHL behaves in the same way once its bias is measured on the grid extended to negative $r$. Its signed bias decreases monotonically from $+0.078\,\delta_r$ at $0.01\,$nK-arcmin to $-0.071\,\delta_r$ at $15$ and crosses zero near $0.5$. Here, too little regularisation leaves a positive residual, while too much over-regularises the offset and pushes the peak below zero. Therefore, the default of $1\,$nK-arcmin also sits close to the zero of the harmonic bias. Similarly, \lmaxabs~must reach at least $40$. Indeed, truncating the signal covariance at \lmaxabs~$=32$ costs $35\%$ in $\delta_r$ and pushes the PB bias to $0.23\,\delta_r$, i.e., past the accuracy criterion, while all values from $40$ upward are indistinguishable. The apodisation scale also belongs to this group, with the caveat of \autoref{sec:benchmark} that it matters only in the unbinned analysis, where $15^\circ$ fails, while every value passes once the small-scale spectra are binned.

The third group contains the resource-limited settings, and a recommendation on them comes with a cost. The two settings behave in opposite ways. For \Nll, the criteria are met from a few hundred simulations upward, provided that the estimation error on $M_{\ell\ell^\prime}$ is corrected. In particular, with the SH correction the full-range $\delta_r$ is stable to within $3\%$ from \Nll~$=600$ up and passes from $400$, and individual scans remain within tolerance down to $100$. On the other hand, without any correction the same scans fail badly, and in the direction that a one-sided criterion would have missed. Indeed, at \Nll~$=100$ the bound is $41\%$ tighter than the converged one, and it approaches convergence from below as \Nll~grows. Thus, a future analysis could reduce it by a factor of a few at no cost in performance, as long as it corrects $M_{\ell\ell^\prime}$.

The behaviour of \Npp~is the opposite, and it is the only significant cost identified by this analysis. For the PB benchmark, $\delta_r$ approaches its converged value monotonically from above, and the most stringent criterion is the accuracy one. However, even at the largest value we can afford the dependence has not flattened, and the quoted benchmark $\delta_r$ is still inflated with respect to the \Npp~$\to\infty$ extrapolation (see appendix~\ref{app: real-space-cov}). Therefore, converging the real-space covariance for $B$-mode maps at \Nside$=16$ requires more than 19\,000 simulations, and a benchmark of this kind should be planned accordingly. On the other hand, the oHL is much less affected, since \Npp~enters it only through the noise bias of the QML spectra, and its $\delta_r$ is stable to within $6\%$ from \Npp~$=2000$ upward. Independently of the accuracy with which it reproduces the benchmark, this asymmetry is a practical argument for the harmonic-space approach, since the expensive ingredient is the benchmark.

Finally, we turn to the beam. Among the six windows we explored, the GausWin 640$^\prime$ is the best choice on both criteria simultaneously, and we therefore recommend it. For the PB, it gives the tightest bound, $13\%$ better than the CosWin, and the smallest residual bias, $0.076\,\delta_r$ against $0.119\,\delta_r$ for the CosWin. On the other hand, for the oHL at $\ell\leq32$ all six windows agree to within $4\%$ and return ${\rm MAP}(r)\leq0.011\,\delta_r$. Indeed, the constraining power on $r$ resides at $\ell\lesssim10$, where the six windows differ least, while the more aggressive window reduces the pixel-level variance over $20\lesssim\ell\leq32$, as discussed in appendix~\ref{app: beam}.

We note that this screening inherits the one-setting-at-a-time structure of \autoref{tab:settings}. Hence, it bounds the effect of each setting around the reference configuration, while it leaves untested the interactions between settings that are jointly displaced from it.

\section{Conclusions}\label{sec:conclusions}

In this work, we have presented a comprehensive validation of harmonic-space likelihood methods for constraining the tensor-to-scalar ratio $r$ with \textit{LiteBIRD}. Using 20\,000 end-to-end simulations that include CMB, instrumental noise, and Galactic foreground residuals processed through the MC-NILC component-separation pipeline, we have systematically compared the offset Hamimeche \& Lewis (oHL) likelihood against a debiased pixel-based (PB) likelihood serving as our large-scale benchmark.

Our main result is that the 68\% upper bound on $r$ from the oHL likelihood is $19$--$29\%$ larger than that of the benchmark over the full multipole range $\ell \leq 128$, depending on the beam window function. Isolated on the large scales defined by the PB, the difference shrinks to $13.5$--$26.6\%$. Given the scientific goal of \textit{LiteBIRD} of observing $r$ with a statistical uncertainty $\delta_r<0.001$, including foreground residuals and lensing \cite{LiteBIRD:2026gdf}, this level of agreement validates the use of computationally efficient harmonic-space methods as a baseline for \textit{LiteBIRD} parameter estimation. The residual differences arise primarily from the likelihood approximation itself, with the choice of beam window function \bl~and the number of simulations available for covariance estimation (\Npp~and \Nll) identified as the most critical analysis settings.

In terms of bias, the oHL returns a ${\rm MAP}(r)$ of at most $0.011\,\delta_r$ on the multipole ranges of \autoref{tab:results}. However, this value is limited by the lower edge of the grid at $r=0$; once the grid is extended to negative $r$, the mean of the per-simulation MAP at $\ell\leq32$ is $-0.005$ and $-0.037\,\delta_r$ for the two reference beams, against a sampling error of $0.02$. Thus, the bias of the harmonic approximation is compatible with zero at the precision allowed by our validation set. On the other hand, the benchmark performs worse in this respect, with a residual ${\rm MAP}(r)$ of $0.07$--$0.12\,\delta_r$ that we trace to the finite number of simulations available for the real-space covariance.

These results also allow us to provide recommendations for a future \textit{LiteBIRD} analysis, within the idealised setup of this work. Based on the screening of the explored space against a reference configuration serving as educated guess of the ground truth (see \autoref{sec:benchmark} and \autoref{sec:selection}), we recommend the GausWin 640$^\prime$ beam, which is the best choice on both criteria at once. For the regularisation noise we recommend the range $0.1$--$5$ nK-arcmin, whose boundaries fail for opposite reasons and which also contains the zero of the harmonic bias. We also recommend \lmaxabs~$\geq40$ and the SNR-based binning of the small-scale spectra. Finally, we recommend the SH correction to $M_{\ell\ell^\prime}$, without which the harmonic bound is spuriously tight. All these settings are freely optimisable, and can be adopted as they stand. Despite this, we measure the relative change with respect to the reference value to be small for the vast majority of the considered cases. Thus, deviations from these guidelines will have little-to-no impact at the sensitivity of this work. 

On the other hand, the two recommendations that carry a real cost point in opposite directions. On the harmonic side the requirement is mild. Indeed, with the SH correction in place, \Nll~of a few hundred already meets our criteria, so the $2000$ simulations used here are more than the method needs. On the real-space side, instead, the PB benchmark satisfies both criteria only above \Npp~$\simeq12\,000$ with the CosWin, and this threshold is set by the accuracy criterion. Even at 19\,000 the dependence has not flattened, with the quoted $\delta_r$ still inflated by $6$--$15\%$ relative to the extrapolated value (appendix~\ref{app: real-space-cov}); thus, a future analysis that needs a pixel-based benchmark of this kind, with a setup similar to ours, should plan for more than $2\times10^4$ simulations. As a consequence, independently of how accurately it reproduces the benchmark, the harmonic-space likelihood is the cheaper of the two, which is a practical argument in its favour. The number of simulations needed to estimate a covariance matrix for constraints on $r$ has been discussed extensively in the literature \cite{tristram2021PlanckConstraintsTensortoscalar, tristram2022ImprovedLimitsTensortoscalarratio, beckBiasTensortoscalarRatio2022, Tristram:2023haj, ilicReconstructingEpochReionisation2025}, and the answer depends on the size of the data vector and on how the estimation error is treated. Our results show the same dependence, since a few hundred simulations suffice for the binned harmonic covariance with the SH correction, while the real-space covariance of the PB requires more than $10^4$. In addition, all our simulations share the same fidelity, while in a real analysis end-to-end simulations with a full instrumental model are far more expensive than simplified ones. How to combine many cheap simulations with a few expensive ones, for example through map-level emulators \cite{campetiFewManyMaps2025}, is left to future work.

Beyond the likelihood comparison, this work establishes a fully operational analysis pipeline capable of propagating uncertainties from multifrequency sky simulations to robust constraints on the tensor-to-scalar ratio. The pipeline integrates component separation (MC-NILC), power spectrum estimation (QML for $\ell \leq 32$ and pseudo-$C_\ell$ for smaller scales), and likelihood evaluation in a coherent framework. The agreement between three independent implementations of the pixel-based likelihood (\texttt{parameter\_estimation}, \texttt{PBLik}, and \texttt{PICSLike}) and two of QML (\texttt{pse\_qml} and \texttt{QUBE}) provides an additional robustness layer in our analysis \cite{CosmoForge}.

Looking ahead, several extensions of this work are warranted. The component-separation stage currently relies on a single foreground model; incorporating more realistic foreground complexity will be essential for \textit{LiteBIRD}'s full analysis (\textit{LiteBIRD} Collaboration, in prep.). Additionally, the oHL likelihood could be further refined, for example, by exploring alternative approximations that better capture the non-Gaussian distribution of power spectrum estimates at the largest scales. This couples with the residual difference of $19$--$29\%$ with respect to the hybrid likelihood mentioned above, which could be improved. Furthermore, in this study, we assumed that the pipeline will work on $B$-mode maps directly, since MC-NILC naturally provides that. However, we must emphasise that other component-separation methods may not have this advantage. Parametric component separation typically works in $Q/U$ space since the foreground model is written in terms of Stokes parameters. Such a scenario poses extra complexities to tackle. For example, working with $Q$ and $U$ means that one would have double the pixels to describe in the real-space covariance matrix, meaning the \Npp~identified for $B$-mode maps is most probably not sufficient in that context. This is left to future work. In this work, we have not considered any delensing. Combining \textit{LiteBIRD} with ground-based experiments is expected to improve delensing \cite{LiteBIRD:2026gdf}, and its impact on the validation presented here will be explored in future work. Finally, the interplay between likelihood accuracy and computational cost will become increasingly important as the collaboration scales up to larger simulation ensembles and more complex sky models.

In summary, this validation demonstrates that harmonic-space likelihood methods provide a viable and computationally efficient path towards \textit{LiteBIRD}'s science goals, with a bound on $r$ that is $19$--$29\%$ larger than that of the benchmark and significant advantages in terms of scalability and ease of implementation across the full angular range.

\acknowledgments

This work is supported in Japan by ISAS/JAXA for Pre-Phase A2 studies, by the acceleration program of JAXA research and development directorate, by the World Premier International Research Center Initiative (WPI) of MEXT, by the JSPS Core-to-Core Program of A. Advanced Research Networks, and by JSPS KAKENHI Grant Numbers JP15H05891, JP17H01115, and JP17H01125. The Canadian contribution is supported by the Canadian Space Agency. The French \textit{LiteBIRD} phase A contribution is supported by the Centre National d'Etudes Spatiale (CNES), by the Centre National de la Recherche Scientifique (CNRS), and by the Commissariat à l'Energie Atomique (CEA). The German participation in \textit{LiteBIRD} is supported in part by the Excellence Cluster ORIGINS, which is funded by the Deutsche Forschungsgemeinschaft (DFG, German Research Foundation) under Germany's Excellence Strategy (Grant No. EXC-2094 -- 390783311). The Italian \textit{LiteBIRD} phase A contribution is supported by the Italian Space Agency (ASI Grants No. 2020-9-HH.0 and 2016-24-H.1-2018), the National Institute for Nuclear Physics (INFN) and the National Institute for Astrophysics (INAF). Norwegian participation in \textit{LiteBIRD} is supported by the Research Council of Norway (Grant No. 263011 and 351037) and has received funding from the European Research Council (ERC) under the Horizon 2020 Research and Innovation Programme (Grant agreement No. 772253, 819478, and 101141621). The Spanish \textit{LiteBIRD} phase A contribution is supported by MCIN/AEI/10.13039/501100011033, project refs. PID2019-110610RB-C21, PID2020-120514GB-I00, PID2022-139223OB-C21, PID2023-150398NB-I00 (funded also by European Union NextGenerationEU/PRTR), and by MCIN/CDTI ICTP20210008 (funded also by EU FEDER funds). Funds that support contributions from Sweden come from the Swedish National Space Agency (SNSA/Rymdstyrelsen) and the Swedish Research Council (Reg. no. 2019-03959). The UK \textit{LiteBIRD} contribution is supported by the UK Space Agency under grant reference ST/Y006003/1 -- \lq\lq LiteBIRD UK: A major UK contribution to the LiteBIRD mission-Phase1 (March 25).'' The US contribution is supported by NASA grant no. 80NSSC18K0132. Some of the results in this paper have been derived using the following packages: {\tt CAMB} \citep{lewis2000EfficientComputationCMB, Howlett2012}, {\tt healpy} \citep{healpix2005, Zonca2019}, {\tt NaMaster} \citep{alonsoUnifiedPseudoC_ellFramework2019}, {\tt Matplotlib} \citep{Hunter2007}, {\tt SciPy} \citep{Virtanen2020} and {\tt NumPy} \citep{Harris2020}. We acknowledge the financial support from the INFN InDark initiative and from the COSMOS network through the ASI (Italian Space Agency) Grant 2016-24-H.0. This work has also received funding by the European Union's Horizon 2020 research and innovation program under grant agreement no.\ 101007633 CMB-Inflate. We acknowledge support from the MUR PRIN2022 Project ``BROWSEPOL: Beyond standaRd mOdel With coSmic microwavE background POLarization''-2022EJNZ53 financed by the European Union -- Next Generation EU. We acknowledge CINECA for the availability of high performance computing resources and support through the CINECA-INFN agreement. Co-funded by the Complementary Plan in Astrophysics and High-Energy Physics (CA25944), project C17.I02.P02.S01.S03 CSIC, supported by the Next Generation EU funds, RRF and PRTR mechanisms, and the Government of the Autonomous Community of Cantabria. We acknowledge funding by the European Union (ERC, RELiCS, project number 101116027). Views and opinions expressed are however those of the authors only and do not necessarily reflect those of the European Union or the European Research Council Executive Agency. Neither the European Union nor the granting authority can be held responsible for them. We acknowledge the support from the RadioForegroundsPlus Project HORIZON-CL4-2023-SPACE-01, GA 101135036 and through the Project SPACE-IT-UP by the Italian Space Agency and Ministry of University and Research, Contract Number 2024-5-E.0. This work is supported by the United Kingdom Space Agency (UKSA) under grant ST/Y005902/1, as well as the Dennis Sciama Junior Research Fellowship at Wolfson College, Oxford. During the preparation of this work, the authors used Anthropic's Claude (via the Claude Code agentic coding interface) to assist with manuscript drafting. All AI-generated content was reviewed, edited, and validated by the authors, who take full responsibility for the contents of this publication.

\begin{appendix} 

\section{Dependence on individual analysis settings}\label{app: settings}

\subsection[A focus on \texorpdfstring{\bl}{Bl}]{A focus on \texorpdfstring{$\boldsymbol{B_\ell}$}{Bl}}\label{app: beam}

\begin{figure}[t]
    \centering
    \includegraphics[width=\textwidth]{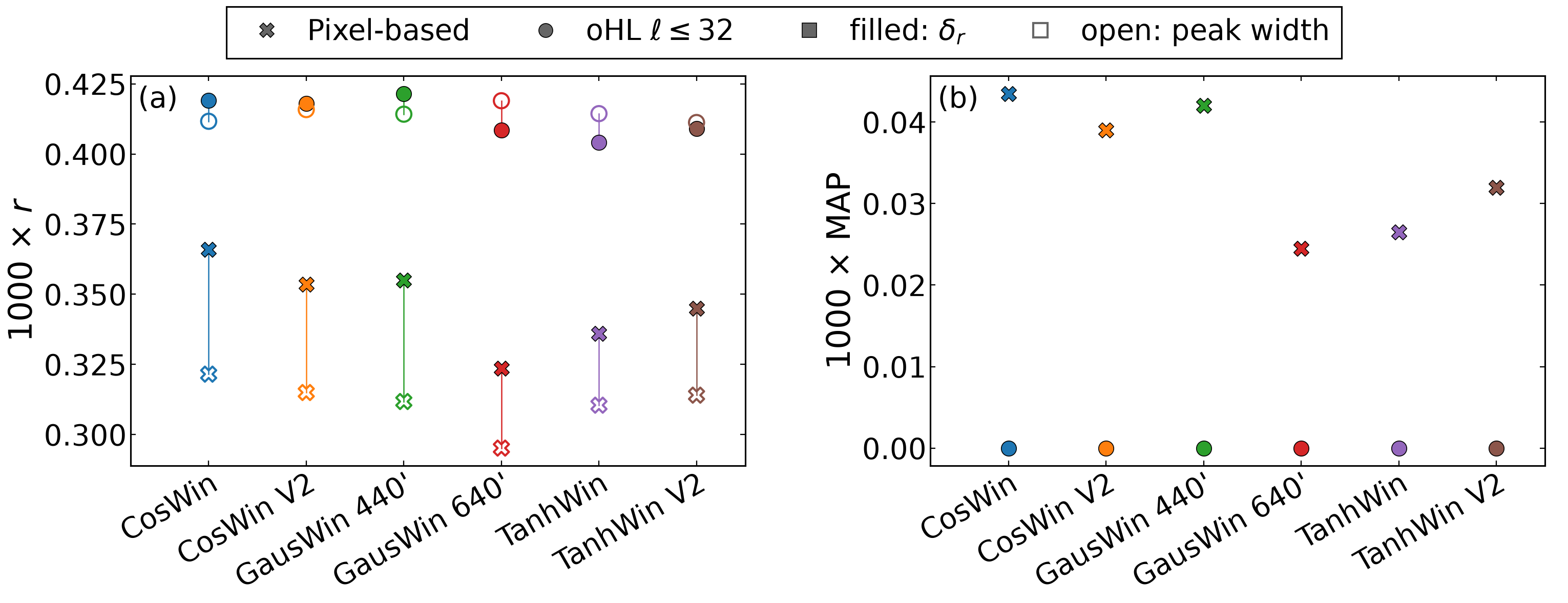}
    \caption{Effect of different beams on the PB and oHL. In panel (a), filled markers show $\delta_r$ and open markers show the local posterior width $\qty(-\partial^2\ln P/\partial r^2)^{-1/2}$ at the MAP, computed from the same averaged $\chi^2$, while panel (b) shows the MAP estimate. Each beam keeps the colour it has in \autoref{fig:beams}, and the marker shape identifies the likelihood, with crosses for the PB and circles for the oHL.}
    \label{fig:upper_bounds_beams}
\end{figure}

As mentioned in \autoref{sec:pixel-based} and \autoref{sec:qml}, to compute the QML algorithm and the PB likelihood on large scales, we need low-resolution maps. These are obtained from the native resolution maps of the MC-NILC at $\Nside=64$ using different window functions (see \autoref{fig:beams}). 
On top of this, in \autoref{sec:results} we show that the GausWin 640$^\prime$ provides more stringent results on $\delta_r$ compared to the CosWin, despite the latter being less aggressive than the former. Panel (a) of \autoref{fig:upper_bounds_beams} shows $\delta_r$ as a function of the chosen beam. Here, this counter-intuitive behaviour is clear. 

It must be emphasised that the vast majority of the SNR is encoded in the very first multipoles ($\ell < 10$), thus in a region where none of the beams are truly cutting the harmonic content. On the other hand, beams are essentially treating the scales smaller than that differently. In fact, the difference between beams seems to stem from $\ell\gtrsim20$: TanhWin and TanhWin V2 allow us to greatly alleviate the difference between CosWin and GausWin 640$^\prime$. We recall that the TanhWin windows are defined simply as a tanh-transition between those two, with a different transition multipole; see \autoref{sec:set_beam}.

As panel (b) of \autoref{fig:upper_bounds_beams} shows, part of this behaviour is due to a shift in the MAP estimate of $r$, while the open markers in panel (a) measure the rest directly. Indeed, they give the local width of the posterior, $\qty(-\partial^2\ln P/\partial r^2)^{-1/2}$, evaluated at the MAP on the same averaged $\chi^2$ from which $\delta_r$ is read. The six windows span $13\%$ in $\delta_r$ but only $8.9\%$ in this quantity; thus, the more aggressive window does return a narrower posterior, and about a third of its advantage in $\delta_r$ comes from the displacement of the peak. For the oHL, on the other hand, the same quantity spans $1.9\%$ against $4.3\%$ in $\delta_r$, so the six windows are indistinguishable in curvature.

As regards the bias, the residual systematic $F_\ell$ of \autoref{fig:mean_spectra} is a fixed and positive contribution to the recovered $B$-mode power, which the likelihood reads as tensor power. It is positive at $30$ of the $31$ multipoles of $\ell\leq32$, and summed over them it amounts to $7\%$ of $C_\ell^{\rm fid}+N_\ell$. To quantify its effect, we fit $F_\ell$ with the same tensor templates used by the likelihood, obtaining an equivalent $r$ of $1.0$--$1.5\times10^{-3}$, and indeed running the PB on the maps before the template subtraction of \autoref{sec:pixel-based} gives, for every window, a posterior peaking at $r\simeq1.0$--$1.2\times10^{-3}$, about three times $\delta_r$. The template subtraction then removes $95$--$98\%$ of this displacement, leaving the $2.5$--$5.0\times10^{-5}$ quoted in \autoref{sec:selection}.

The surviving bias is smaller for the more aggressive window, even though over $2\leq\ell\leq12$ its $F_\ell$ is larger, by a factor $1.6$--$2.2$ with respect to the CosWin, and its pre-subtraction MAP is correspondingly higher, $1.20\times10^{-3}$ against $0.99\times10^{-3}$. The windows differ instead in how well the residual bias can be measured: as shown in appendix~\ref{app: real-space-cov}, the surviving MAP is dominated by a $1/$\Npp~term, whose coefficient is $0.68$ for the CosWin against $0.28$ for the GausWin 640$^\prime$. Indeed, the aggressive window reduces the pixel-level variance over $20\lesssim\ell\leq32$ that $\mathbf{N}$ has to describe. Therefore, at the \Npp~we can afford, the error on the estimate of this covariance and the bias it induces are a factor of $\simeq2.4$ smaller.

\subsection{Estimating real-space covariances}\label{app: real-space-cov}

In \autoref{sec:pixel-based} and \autoref{sec:qml}, we mentioned the need to estimate real-space covariance matrices for the PB likelihood and the QML power spectrum estimation. Also, in \autoref{fig:sens_tab}, we show that \Npp~is indeed one of the critical parameters for this analysis.

\begin{figure}[t]
    \centering
    \includegraphics[width=.49\textwidth]{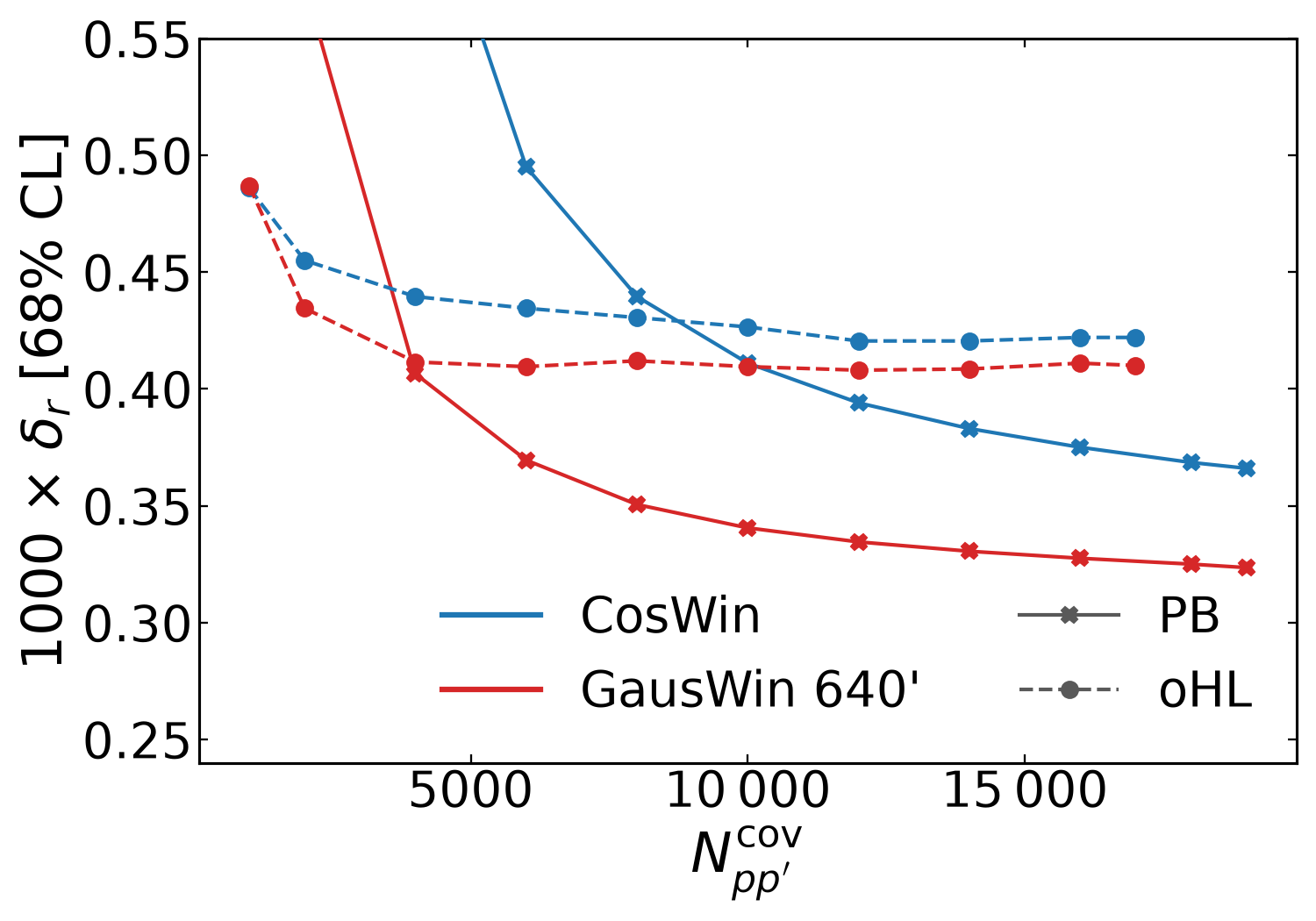}
    \caption{PB and oHL results at $\ell\leq32$ as a function of the number of simulations used for the covariance matrix \Npp. As in \autoref{fig:upper_bounds_beams}, each beam keeps the colour it has in \autoref{fig:beams}, and the likelihood is identified by the marker and line style, with solid crosses for the PB and dashed circles for the oHL.}
    \label{fig:pp_num_cov_sims}
\end{figure}

In \autoref{fig:pp_num_cov_sims}, we show the behaviour of the $\delta_r$ as a function of the number of simulations for PB and oHL at $\ell\leq32$ using both the CosWin and GausWin 640$^\prime$ windows. Having 3072 pixels at the resolution of $\Nside=16$, the number of independent elements of the covariance is very high ($1.1\times10^6$ with 50\% of the sky). Thus, the 20\,000 simulations used for this analysis are relatively low. This is reflected in the behaviour of the PB likelihood in the plot, specifically the two solid lines. These are indeed reaching a plateau; however, the derivative at \Npp~$=19\,000$ is not null yet. Hence, even though 20\,000 simulations are already a non-negligible computational effort, this suggests that even more may be needed to fully characterise the real space covariance matrix. Despite this, it is also important to emphasise that the GausWin 640$^\prime$ case seems to converge faster than the CosWin one; this could indeed explain the residual difference between the two results (see appendix~\ref{app: beam}).

Figure~\ref{fig:pp_num_cov_sims} also shows the same thing for the oHL at large scales, which depends on \Npp~through the QML (see \autoref{sec:qml}). It is clear that this dependence is less important for the harmonic likelihood, and similarly to the PB, the GausWin case converges faster than the CosWin case.

The residual trend affects the two likelihoods differently, as we can quantify by fitting the last five points of each curve with $\delta_r = \delta_r^\infty + b/N_{pp^\prime}^{\rm cov}$. For the PB, we obtain $\delta_r^\infty = 3.18\times10^{-4}$ with the CosWin and $3.05\times10^{-4}$ with the GausWin 640$^\prime$, so that the values quoted at \Npp~$=19\,000$ are overestimated by $15\%$ and $6\%$, respectively. On the other hand, the same fit applied to the oHL gives $1.6\%$ and $-0.4\%$, so the harmonic likelihood is converged in \Npp~within the resolution of our grid, while the benchmark still depends on it. The same fit also tells us how many simulations convergence would require. In particular, bringing the overestimate below $5\%$ requires \Npp~$\simeq23\,000$ with the GausWin 640$^\prime$ and $\simeq57\,000$ with the CosWin, against the 19\,000 available to us. Given the $1/$\Npp~scaling, each further halving of the residual doubles this requirement, and the CosWin needs $2.5$ times as many simulations as the GausWin 640$^\prime$ to reach the same accuracy. The MAP shows this behaviour even more clearly. Indeed, it scales as $1/$\Npp~with a small intercept, and over the last six points of the scan the product ${\rm MAP}\times$\Npp~is constant to within $5\%$ (CosWin) and $12\%$ (GausWin 640$^\prime$). Hence, the bulk of the residual bias discussed in \autoref{sec:res_ranges} is an artefact of the finite simulation set. Fitting ${\rm MAP} = {\rm MAP}^\infty + b/N_{pp^\prime}^{\rm cov}$ over the last four to seven points, which the fit reproduces to better than $1.5\%$, gives $b=0.68$ and $0.28$ for the CosWin and the GausWin 640$^\prime$, respectively, and ${\rm MAP}^\infty\simeq1\times10^{-5}$ in both cases. The intercept is stable to a factor of two against the choice of fit range and is common to the two windows, so we interpret it as the residual left by the map-level debiasing. It amounts to some $0.03\,\delta_r$, a factor of a few below the bias measured at \Npp~$=19\,000$, and well inside the accuracy criterion of eq.~\eqref{eq:screen_accuracy}.

We note that this is a two-parameter extrapolation of a monotonic sequence. However, the residuals of the fit stay below $2\%$ across a factor of three in \Npp, which makes it a reliable estimate. Moreover, since the effect is confined to the benchmark, it propagates to the comparison of \autoref{sec:res_ranges} and \autoref{sec:res_full}, and correcting for it would widen the quoted differences. We believe that part of the difference in bias between the oHL and the benchmark originates here. This is, therefore, a concrete argument for producing a larger set of simulations before the real-data analysis.

The corrections proposed in the literature for the inverse of an empirically estimated covariance, namely the Hartlap factor and the Sellentin \& Heavens prescription \cite{hartlap2007WhyYourModel, sellentin2016ParameterInferenceEstimated}, address the case in which the estimated matrix is inverted on its own and used in a Gaussian $\chi^2$. In the PB likelihood, instead, $\mathbf{N}$ enters as an additive component of $\mathbf{C} = \mathbf{S}(r) + \mathbf{N}$, whose inverse does not follow an inverse Wishart distribution; hence, these corrections are not directly applicable here. For reference, with $p\simeq1.6\times10^3$ observed pixels and $n=$~\Npp~$=19\,000$, the naive Hartlap factor would amount to a $9\%$ distortion of $\mathbf{N}^{-1}$. This is the same order as the $6$--$15\%$ inflation of $\delta_r$ measured above, which we take as a consistency check of our interpretation. On the other hand, these corrections apply to the harmonic covariance $M_{\ell\ell^\prime}$, where we test them explicitly (see appendix~\ref{app: harm-space-cov}).

\begin{figure}[t]
    \centering
    \includegraphics[width=.49\textwidth]{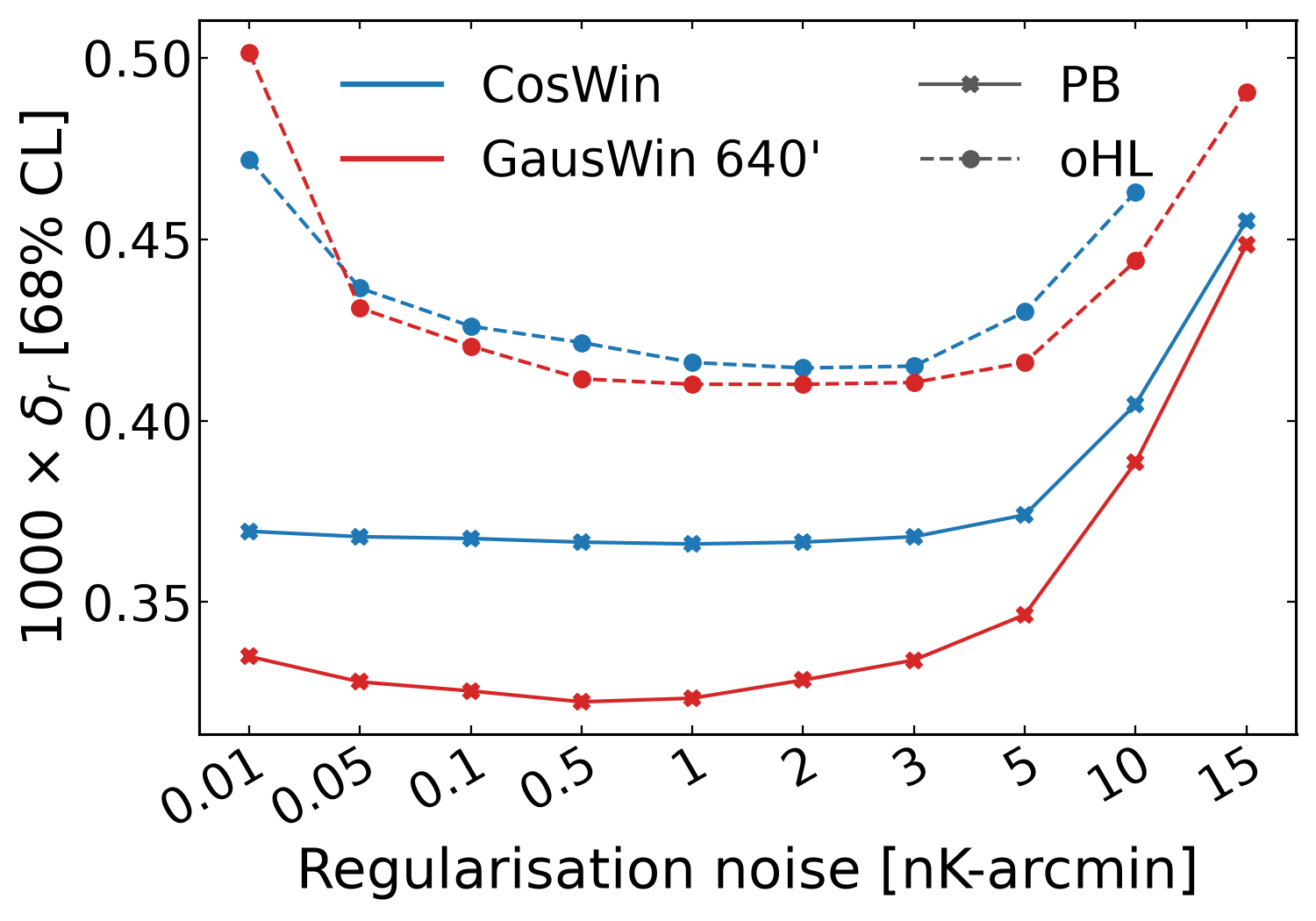}
    \caption{PB and oHL results as a function of the regularisation noise \regn~added to the pixel-space covariance matrix, in nK-arcmin. Colours and markers are as in \autoref{fig:pp_num_cov_sims}.}
    \label{fig:pp_regn}
\end{figure}

When dealing with real-space covariances, it is also important to take care of the regularisation noise. In fact, given the finite number of simulations, we need to add a small regularisation noise to fill the eigenvalues of the matrix. Figure~\ref{fig:pp_regn} shows the behaviour of the same cases as above as a function of \regn~in nK-arcmin.

Here, we find another hint for the fact that the GausWin helps in regularising the behaviour of the pixel covariance matrix. In fact, looking at the oHL results with the GausWin, a minimum is reached for \regn~$\simeq 1$ nK-arcmin; instead, the minimum for the CosWin is at about $3$ nK-arcmin. This indicates that more regularisation is needed in the latter case. The PB, instead, shows a plateau for the CosWin and is minimised at $\simeq 0.5$ nK-arcmin.

\subsection{Estimating harmonic-space covariances}\label{app: harm-space-cov}

Keeping \Npp~fixed, we study the behaviour of the HL posterior as a function of the simulations used to estimate $\eval{M^{-1}_{\ell \ell^\prime}}_{\rm fid}$ and \Nll~(see eq.~\eqref{eq:hl_likelihood}). To mitigate the effect of a low number of simulations, we also consider binning the angular power spectra, the Sellentin \& Heavens (SH) correction, and the Hartlap factor (see \autoref{sec:harmonic-based}).

\begin{figure}[t]
    \centering
    \includegraphics[width=.5\textwidth]{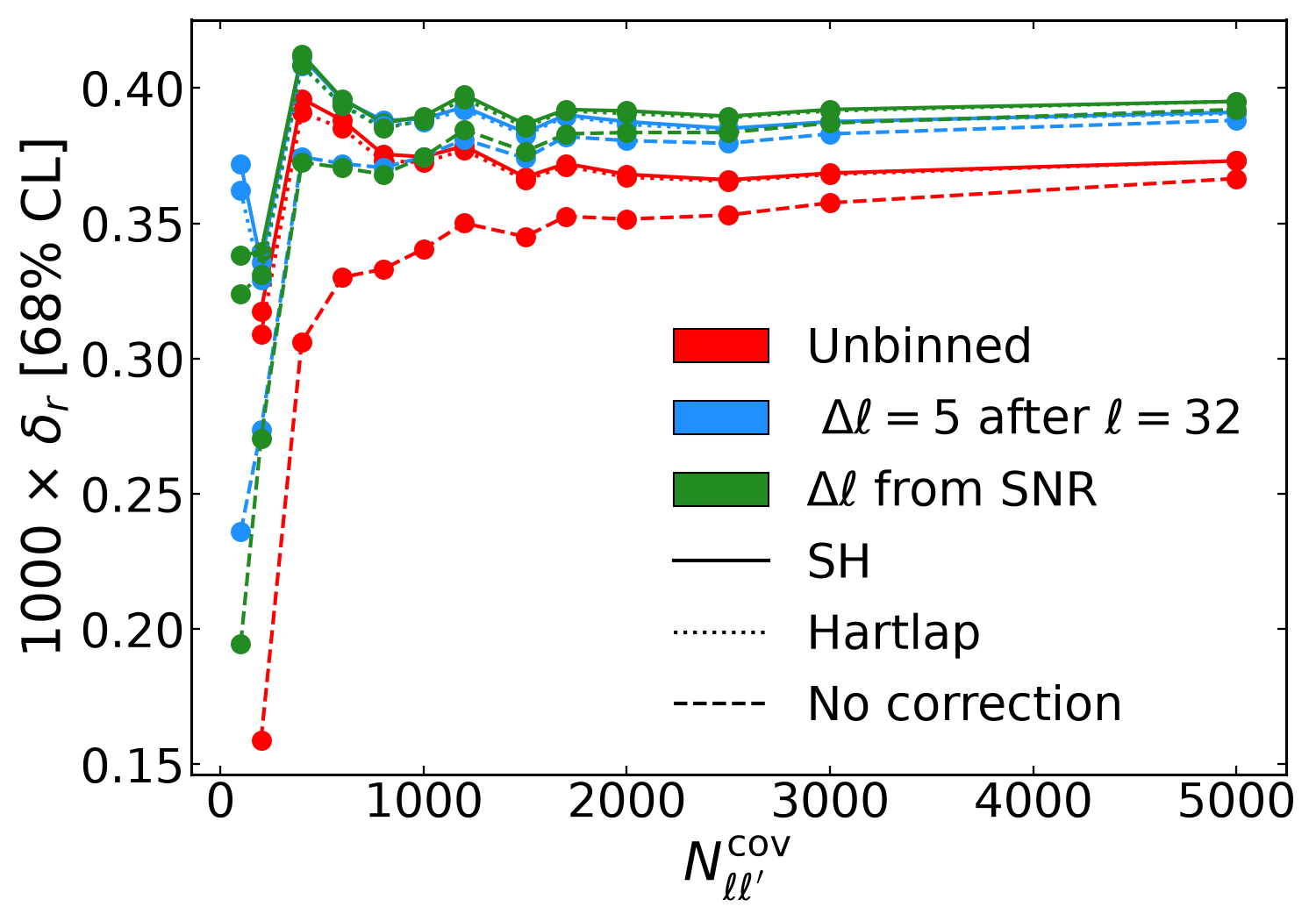}
    \caption{HL results as a function of the number of simulations used to estimate $\eval{M^{-1}_{\ell \ell^\prime}}_{\rm fid}$ (see eq.~\eqref{eq:hl_likelihood}).}
    \label{fig:HL_num_cov_sims}
\end{figure}
Figure~\ref{fig:HL_num_cov_sims} shows the results. Firstly, without considering any mitigation, $\delta_r$ struggles to converge and tends to be underestimated. Indeed, although this is not transparent in the figure, all the unbinned cases show a clear negative bias, likely due to the incorrect estimation of $M^{-1}_{\ell \ell^\prime}$, which has more than 10\,000 independent elements.



The SH and Hartlap corrections yield almost identical results, with the former providing slightly larger $\delta_r$. They both help mitigate this effect; however, it is clear that binning plays a crucial role in decreasing the number of degrees of freedom required for estimation. Even with very few simulations, $\delta_r$ never departs from the converged result. Then, when applying both binning and SH or Hartlap correction, \Nll~$\gtrsim 1000$ is already sufficient to converge to the true answer.

\section{Overview of the \texttt{PBLik} code}\label{app: PBLik}

The \texttt{PBLik} code is a user-friendly Python-based package designed to compute the PB likelihood. The code solves eq.~\eqref{eq:qml_like}, where the data vector $\mathbf{m}$ corresponds either to the observed $Q$ and $U$ pixels ($2N^{\rm{obs}}_{\rm{pix}}$ entries) or to the data represented directly in $B$-mode space. Depending on the chosen representation, the computation of the signal covariance matrix differs. In the $B$-mode case, the signal covariance matrix depends only on $C_{\ell}^{BB}$, whereas for $Q$ and $U$, it depends on both $C_{\ell}^{EE}$ and $C_{\ell}^{BB}$.

The code performs two steps using the MPI standard through \texttt{mpi4py}. First, the signal covariance matrix $\mathbf{S} (C_{\ell})$ is computed, and then the likelihood is evaluated. The construction of the covariance matrix from the theoretical model is parallelised across the available MPI processes and typically requires about 8 seconds when using 50 processors. Since $\mathbf{S}$ is a function of the tensor-to-scalar ratio $r$ (with other cosmological parameters held fixed), a covariance matrix is required for each value of $r$ in the explored grid.

To reduce the computational cost, the code adopts an approximation in which the covariance matrix is computed only once for a reference value $r = 0.01$ and then rescaled as follows: 
\begin{equation}
    \mathbf{S} (C_{\ell}) = \left(\frac{r}{0.01}\right)\cdot \mathbf{S}_{\rm{tensor}}^{r = 0.01}(C_{\ell}) + \mathbf{S}_{\rm{scalar}}(C_{\ell}).
\label{eq:pblik_approx}
\end{equation}
Here, $\mathbf{S}_{\rm{scalar}}$ collects the contributions that do not scale with $r$. In this work the data vector is represented in $B$-mode space, so this term is the lensing $BB$ signal alone. It also includes the $C_\ell^{EE}$ term required to build the polarisation covariance only when the data vector is given as $Q$ and $U$ pixels (see the footnote above). Indeed, $C_{\ell}$ would actually depend nonlinearly on $r$ through the slow-roll consistency relation, which fixes the tensor spectral tilt to $n_{\rm t} = -r/8$ \cite{guzzettiGravitationalWavesInflation2016}. The approximation in eq.~\eqref{eq:pblik_approx} is sufficiently accurate for our purposes, yielding relative errors in $C_{\ell}^{BB}$ well below 1\%, which are negligible compared to the uncertainties in the noise covariance matrix. 

The code is configured through an input \texttt{ini} file\footnote{Full documentation is available in the GitHub repository, \href{https://github.com/ChristianGim/PBLik}{github.com/ChristianGim/PBLik}.}, which specifies, among other things, cosmological parameters and auxiliary settings such as file paths and pre-processing options. In particular, four parameters control the multipole ranges used internally: \texttt{lmax}, \texttt{lmax\_}, \texttt{lmax\_maps}, and \texttt{lswitch}. The \texttt{lmax} defines the maximum multipole considered in the \texttt{camb} call, while \texttt{lmax\_} sets the maximum multipole considered in the covariance matrix computation (equivalent to \lmaxabs~discussed in \autoref{sec:pixel-based}). The parameter \texttt{lmax\_maps} controls the maximum multipole of the map, and finally, \texttt{lswitch} separates the parameter-dependent and parameter-independent regimes of the likelihood: for $\ell \le$ \lswitch, the power spectrum varies with $r$, whereas for $\ell >$ \lswitch~the $C_{\ell}$ are fixed to a reference model, typically assuming no tensor contribution.

The code also implements two additional features mentioned in \autoref{sec:pixel-based}: debiasing and regularisation noise. Foreground residuals introduce systematic offsets that shift the map mean away from zero, breaking a fundamental assumption of the likelihood. To deal with this, a debiasing mode is available in which a user-provided template is subtracted from the maps prior to likelihood evaluation. This template is usually constructed as the mean of simulations containing the residuals.

Regularisation noise is introduced as an optional free parameter to improve the numerical stability of the covariance matrix and enhance the compatibility between the model and the observed data, at the cost of adding an extra noise contribution. It is controlled by the \texttt{reg\_noise} parameter (equivalent to \regn~discussed in \autoref{sec:set_cov}), which specifies the standard deviation of an added white noise component in $\mu{\rm{K}}$. This noise term is consistently included both in the covariance matrix and in the data vector.

\begin{figure}[t]
    \centering
    \includegraphics[width=.77\textwidth]{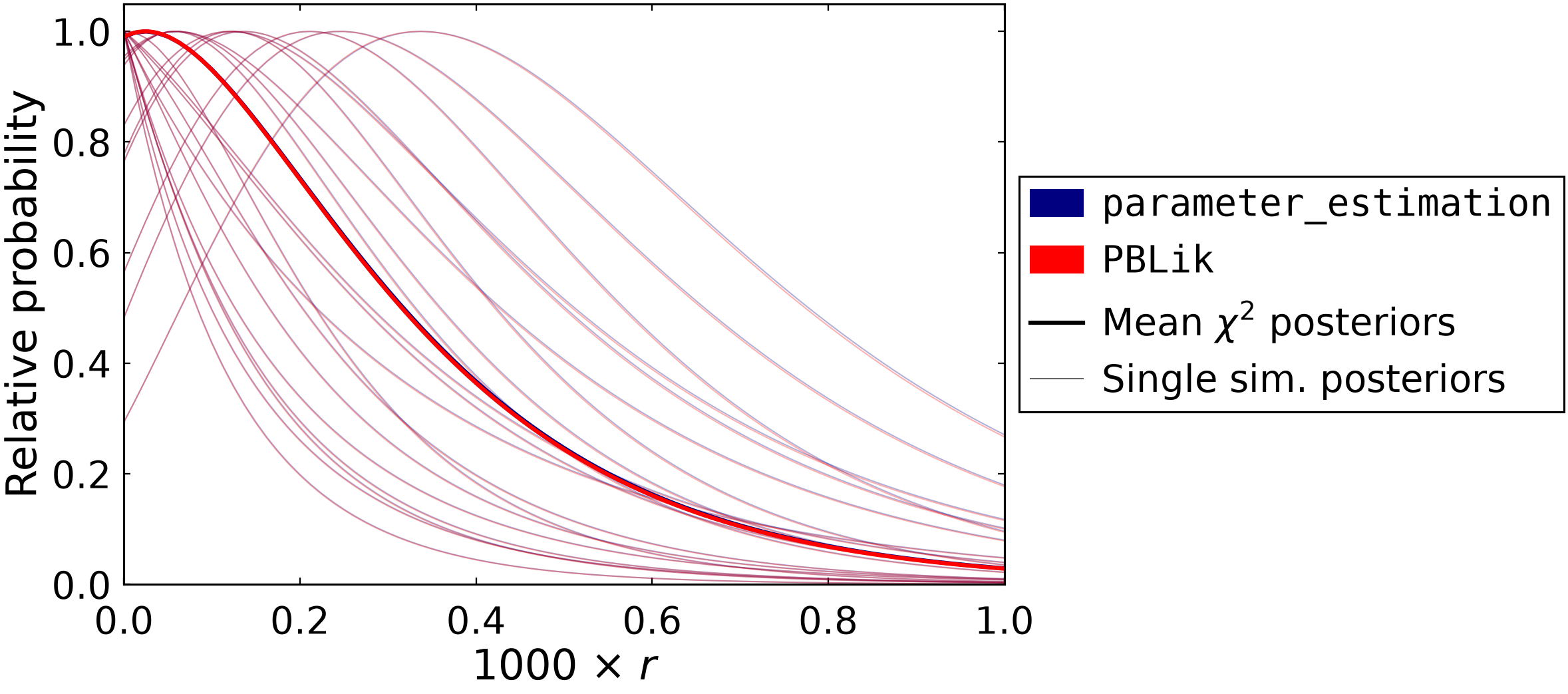}
    \caption{Mean posterior (solid) and single-simulation (shaded) posteriors obtained from \texttt{parameter\_estimation} (navy) and \texttt{PBLik} (red), written respectively in \texttt{Fortran} and \texttt{Python}.}
    \label{fig:pb_comparison}
\end{figure}

In practice, regularisation noise is useful for minimising biases arising from an imperfect knowledge of the noise covariance matrix $\mathbf{N}$. However, its origin is more fundamental. In the absence of noise, the signal covariance matrix computed up to $\ell_{\rm{max}} = 3$\Nside~is often close to being singular and thus not invertible. This issue can be solved by extending the covariance matrix computation to $\ell_{\rm{max}} = 4$\Nside, effectively including sub-pixel scales. This approach, however, introduces a mismatch between the information content of the maps, which typically contain modes up to $\ell_{\rm{max}} = 3$\Nside, and that of the covariance matrix. This is the main reason to distinguish between \texttt{lmax\_} and \texttt{lmax\_maps}. When these parameters differ, with \texttt{lmax\_} > \texttt{lmax\_maps}, the code adds a realisation generated from the theoretical $C_{\ell}$ in the range \texttt{lmax\_maps} $< \ell \le$ \texttt{lmax\_}, ensuring consistency between the data vector and the covariance matrix.

Finally, we compare the results of \texttt{parameter\_estimation} and \texttt{PBLik} under the same settings. Figure~\ref{fig:pb_comparison} shows both the average posterior in solid and 20 random simulations among the \Nval~simulations. It is clear that, simulation-by-simulation, the two implementations of the PB likelihood provide almost identical results. Indeed, this serves as a robustness check of our benchmark—the pixel-based likelihood—since these codes have been developed independently using two different programming languages\footnote{The \texttt{parameter\_estimation} code is written in \texttt{Fortran}, while \texttt{PBLik} is in \texttt{Python}.}. Additionally, the former does not employ the approximation of eq.~\eqref{eq:pblik_approx} and recomputes the signal covariance matrix for each value of $r$. Thus, \autoref{fig:pb_comparison} also proves that this approach is valid in our context.

\end{appendix}

\printbibliography

\end{document}